\documentclass[
prb,
reprint,
twocolumn,
aps,
superscriptaddress,
article
]{revtex4-2}
\usepackage{graphicx}% Include Figure files
\usepackage{adjustbox}
\usepackage{dcolumn}% Align table columns on decimal point
\usepackage{multirow}
\usepackage{hyperref}
\usepackage{ulem}
\usepackage{bm}% bold math
\usepackage{latexsym}
\usepackage{amsmath}
\usepackage{amssymb}
\DeclareMathOperator{\Tr}{Tr}
\usepackage{cleveref}
\crefname{equation}{Eq.}{Eqs.}
\hypersetup{bookmarks,
            colorlinks,
           citecolor=red,
            filecolor=blue,
           urlcolor=blue,
           pdfpagemode=None}
\newcommand{\expect}[1]{\langle #1 \rangle}

\def \cN{\mathcal{N}}
\def \beq{\begin{eqnarray}}
\def \eeq{\end{eqnarray}}

\def \B{{\bm B}}
\def \G{{\bm G}}
\def \H{{\bm H}}

\def \K{{\bm K}}

\def \j{{\bm j}}

\def \mrm{{\rm m}}
\def \r{{\bm r}}

\def \n{{\bm n}}

\def \k{{\bm k}}
\def \p{{\bm p}}
\def \q{{\bm q}}

\def \u{{\bm u}}
\def \v{{\bm v}}
\def \R{{\bm R}}

\def \A{\mathbf{A}}
\def \rr{\bm{\rho}}
\def \cone{c_{11}}
\def \csix{c_{66}}
\def \T{\mathrm{T}}
\def \L{\mathrm{L}}

\begin{document}

\title{Qubit Noise Spectroscopy of Superconducting Dynamics in a Magnetic Field}
\author{Jiajie Cheng}
\thanks{J.C., J.K. and O.K.D. contributed equally to this work.}
\affiliation{Department of Physics, Carnegie Mellon University, Pittsburgh, Pennsylvania 15213, USA}
\author{Jaewon Kim}
\thanks{J.C., J.K. and O.K.D. contributed equally to this work.}
\affiliation{Department of Physics, University of California, Berkeley, CA 94720, USA}
\affiliation{Department of Physics and Anthony J. Leggett Institute for Condensed Matter Theory, University of Illinois Urbana-Champaign, Urbana, Illinois 61801, USA} 
\author{Oriana K. Diessel}
\thanks{J.C., J.K. and O.K.D. contributed equally to this work.}
\affiliation{ITAMP, Center for Astrophysics, Harvard \& Smithsonian, Cambridge, Massachusetts 02138, USA}
\affiliation{Department of Physics, Harvard University, Cambridge, Massachusetts 02138, USA}
\author{Chong Zu}
\affiliation{Department of Physics, Washington University, St. Louis, Missouri 63130, USA}
\affiliation{Institute of Materials Science and Engineering, Washington University, St. Louis, Missouri 63130, USA}
\author{Shubhayu Chatterjee}
\affiliation{Department of Physics, Carnegie Mellon University, Pittsburgh, Pennsylvania 15213, USA}

\begin{abstract}
An applied magnetic field affects a superconductor in two ways --- by promoting pairing fluctuations, and by inducing topological defects called vortices that carry quantized magnetic flux.  
A quantitative characterization of the resultant field-induced superconducting dynamics with spatio-temporal resolution remains challenging, particularly in two-dimensional materials. 
In this work, we analyze magnetic noise measured by the depolarization rate of a proximate single spin qubit as a non-invasive probe of such dynamical fluctuations. 
We demonstrate that the temperature dependence of the magnetic noise spectrum near $T_c$ deviates from  predictions based on quasiparticle excitations due to critical superconducting fluctuations, which in turn are enhanced by a weak applied field. 
By analyzing the magnetic noise due to vortex dynamics, we further show that noise spectroscopy is not only able to distinguish between different vortex phases, but also extract key physical quantities of interest, such as oscillation frequencies of pinned vortices, phonon dispersion of vortex lattices and vortex diffusivity in a vortex liquid.  
Complementing recent work on noise magnetometry of quasiparticle excitations~\cite{CD2022} and Berezinskii-Kosterlitz-Thouless transitions in two-dimensional superconductors without an applied field~\cite{Curtis_2024}, our work highlights the ability of noise spectroscopy to reveal a wealth of superconducting dynamical phenomena in an applied field.
\end{abstract}

\maketitle

\section{Introduction}
Recent advances in the experimental design of Van der Waals heterostructures \cite{Geim2013} have led to a surge in the discovery of atomically thin two-dimensional (2D) superconductors \cite{saito2016naturereview}. 
Remarkably, superconductivity in some of these materials can survive fairly strong out-of-plane magnetic fields \cite{han2025signatures}, or may emerge only in the presence of an applied field \cite{Zhou_2022}.
However, unlike bulk superconductors, techniques to probe 2D superconductors with both momentum and energy resolution are limited. 
An applied magnetic field introduces an additional layer of complication: scattering experiments typically use electrically charged and/or spin-ful probe particles that are deflected by the applied field or any field gradient, making it challenging to obtain momentum-resolved information.

Beyond affecting the nature and fluctuations of the pair condensate itself \cite{larkin2005theory}, magnetic fields also induce vortices, topological defects in the superconducting order parameter carrying quantized fluxes \cite{tinkham,Sonin_RMP1987,sonin2016dynamics}. 
Since flux flow leads to a finite electrical resistance \cite{BlatterRMP}, a detailed understanding of vortex dynamics would aid the design of superconducting materials.
However, studying vortex dynamics with simultaneous spatio-temporal resolution requires sensitivity at nanometer lengthscales over a wide range of temperatures, making such studies difficult for low-temperature micro-scale probes such as SQUIDs \cite{kirtley:1999}.
Developing the capability to resolve in-field pairing fluctuations and vortex motion has the potential to shed light on questions of fundamental interest, e.g., the pairing mechanism in strong coupling superconductors, as well as issues of practical interest, e.g., the design of robust superconductors by impeding flux flow. 

Isolated spin qubits, occurring naturally as solid-state defects such as nitrogen vacancy (NV) centers in diamond \cite{maletinsky2012robust,Wang2022Science,pelliccione2016scanned,rovny2024nanoscale,casola2018probing} or boron vacancy (V$_\mathrm{B}$) centers in hexagonal boron nitride \cite{gottscholl2020BV,gong2023coherent,stern2022room,gong2023isotope}, can detect fluctuating magnetic fields generated by low-dimensional materials with high sensitivity.
Recently, it has been proposed that characterizing the resultant magnetic noise measured by the spin qubit can act as a non-invasive, wireless table-top probe of superconducting dynamics in 2D/thin film materials \cite{CD2022,DC2022,Curtis_2024,de2025nanoscale,kelly2025superconductivity}. 
To achieve this, we may optically initialize a spin qubit in a polarized state in the proximity of the superconductor.
Magnetic field fluctuations at the qubit location, which may originate from Cooper pair fluctuations or the motion of vortices, induces depolarization of the qubit. 
Studying this depolarization rate, denoted by $1/T_1$, as a function of experimental knobs such as temperature $T$, qubit-probe distance $z_0$, and probe frequency $\Omega$ furnishes a wealth of valuable information about dynamical superconducting correlations. 
For instance, previous works have argued that at zero external magnetic field, studying such dynamics can tell us about the pairing symmetry of the superconductor via the temperature dependence of the magnetic noise \cite{CD2022,DC2022}, diagnose critical behavior, e.g., Berezinskii-Kosterlitz-Thouless (BKT) physics of vortex-unbinding via the distance dependence of noise~\cite{Curtis_2024}, or probe time-reversal symmetry-breaking in chiral superconductors \cite{de2025nanoscale}. 
Indeed, some of these predictions have been verified in recent experiments, e.g., in thin film BSCCO \cite{SCNoise}.
Notably, certain noise spectroscopy experiments have gone beyond zero-field measurements, and have also probed magnetic noise in the presence of an applied magnetic field \cite{SCNoise,jayaram2025probing}, leading to an important question:
what is the nature of magnetic noise due to a proximate superconductor in an external magnetic field?

In this work, we theoretically characterize the magnetic noise spectrum detected by an isolated spin qubit placed near a superconductor. 
To set the stage, we first study the noise at a thermal metal-insulator transition within the time-dependent Ginzburg-Landau (TDGL) approach \cite{landau1980statistical,landau2013statistical,cyrot1973ginzburg,schmid1966time,schuller2006time}, and characterize the magnetic noise due to critical pairing fluctuations in 2D/thin film superconductors at zero field. 
Next, we extend our TDGL equations to include an applied magnetic field, and demonstrate how the magnetic noise can detect the enhancement of critical fluctuations in the presence of an external field.
Finally, we consider the magnetic noise due to superconducting vortices at finite fields.
We show that spin-qubit spectroscopy allows us to probe diverse aspects of vortex dynamics for both 2D and 3D superconductors, including Kelvin oscillations of isolated vortices \cite{Sonin_RMP1987,sonin2016dynamics,minowa2025direct}, elastic moduli and phonon dispersion in a vortex lattice \cite{FHP66,FH67,Brandt95}, and the diffusivity of vortices in a vortex liquid phase \cite{Nelson_VL}.  
Our results complement recent works \cite{Curtis_2024,potts2025spin} on the detection of vortex dynamics near a BKT transition in 2D superfluids without an applied field \cite{halperin1979resistive,Beasley}, as well as quasiparticle-induced magnetic noise away from the critical regime \cite{CD2022,DC2022}.
\begin{figure*}[t]
	\includegraphics[width=1
	\linewidth]{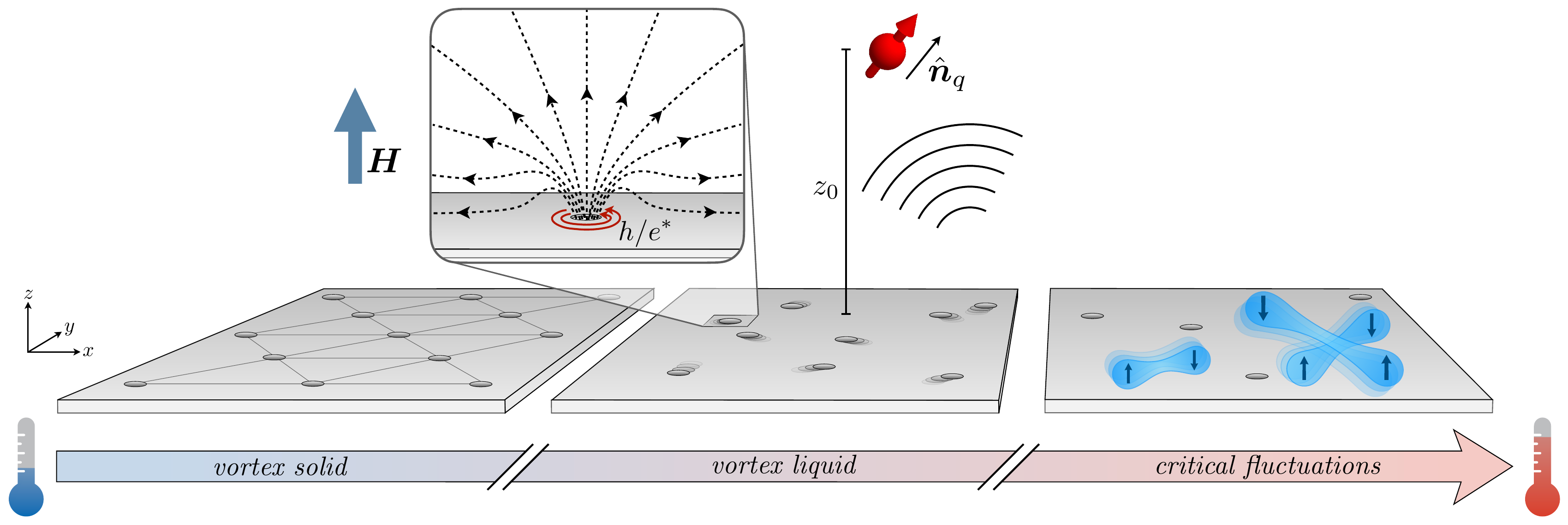}
 \caption{Schematic of the set-up: a qubit sensor with spin-quantization axis $\hat{\bm{n}}_q$ is placed at a distance $z_0$ for a two-dimensional superconducting sample in a perpendicular magnetic field $\bm{H} = H \hat{z}$. Close to the critical temperature $T_c$, the magnetic noise is dominated by critical Cooper-pair fluctuations. Below $T_c$, qualitatively new sources of noise arise in the form of vortex fluctuations: these vortices may be mobile in a liquid phase or freeze into a vortex lattice (or a glass) at lower temperatures.}
 \label{Fig:Schematics}
\end{figure*}

\section{Relating depolarization rate to source fluctuations}

\subsection{Experimental set-up}
We begin by outlining the basic experimental set-up.
A single isolated spin qubit, with quantization axis $\hat{\n}_q$ and intrinsic splitting $\Omega$, is placed at $\r_q = (0,0,z_0)$, i.e., at a distance $z_0$ from the superconducting sample in the $x$-$y$ plane [Fig.~\ref{Fig:Schematics}]. 
The qubit couples to the background static magnetic field $\mu_0 \bm{H}$ as well as the local time-varying magnetic field $\B(\r_q, t)$ generated by pairing fluctuations and/or vortex motion in the superconductor, and is therefore described by the Hamiltonian  
\begin{equation}
\mathcal{H}_q =  \frac{\hbar\Omega}{2}( \hat{\n}_q \cdot \bm{\sigma} )+ g \mu_B \big[\mu_0 \bm{H} + \B(\r_q, t)\big]\cdot \bm{\sigma} .
\label{eq:Hq}
\end{equation}
The component of the static field $\bm{H}$ along the qubit's spin quantization axis $\hat{\n}_q$ modifies the zero-field splitting $\Omega$. 
On the other hand, fluctuations of the magnetic field $\B(\r_q, t)$, arising from current fluctuations and vortex motion in the sample, cause depolarization of a qubit when optically initialized in a fully polarized state. 
The rate of depolarization $1/T_1$ may be evaluated using Fermi's golden rule (e.g., see Refs.~\onlinecite{Agarwal2016,CRD2019}), and relates to the magnetic noise tensor $\mathcal{N}_{\alpha \beta}(\Omega)$ at the qubit location $\r_q$ as 
\begin{align}
\frac{1}{T_1} &= \left(\frac{g \mu_B}{\hbar}\right)^2 \mathcal{N}_{+-}(\Omega), \text{ where } \nonumber \\  
\mathcal{N}_{\alpha \beta}(\Omega) &=  \int_{-\infty}^{\infty} dt\, \, e^{i \Omega t} \langle  B_\alpha(\r_q, t) B_\beta(\r_q, 0) \rangle_T \; .
\label{eq:NoiseTensor}
\end{align}
In Eq.~\eqref{eq:NoiseTensor}, we have used $\langle \ldots \rangle_T$ to denote the thermal expectation value, and defined $B_{\pm} = B_{x^\prime} \pm i B_{y^\prime}$ with $(x^\prime, y^\prime, \hat{\n}_q)$ forming an orthonormal triad~\footnote{We note that in the presence of quantum mechanical sources, the correlation function is promoted to an anti-commutator}. 
Therefore, by orienting the qubit spin-quantization axis $\hat{\n}_q$ appropriately, one may diagnose different components of the noise tensor $\mathcal{N}_{\alpha \beta}(\Omega)$ as a function of the qubit splitting $\Omega$. 
As discussed in Refs.~\onlinecite{Agarwal2016,DC2022}, in systems with additional rotational and mirror symmetries, the magnetic noise tensor is completely characterized by two scalars ~\footnote{Consider setting $\hat{\boldsymbol{n}}_q=(\sin \theta, 0, \cos \theta)$, then the noise $\cN = \left[\mathcal{N}_{\T}\left(1+\cos ^2 \theta\right)/2+\mathcal{N}_{\L} \sin ^2 \theta\right]$.}: (i) the transverse noise $\mathcal{N}_{\T} = \mathcal{N}_{xx} + \mathcal{N}_{yy}$, measured from the depolarization rate for $\hat{\n}_q = \hat{z}$, and (ii) the longitudinal noise $\mathcal{N}_{\L} = \mathcal{N}_{zz}$ which may be detected by setting $\hat{\n}_q = \hat{x}$ and probing the depolarization rate set by $\mathcal{N}_{\L} + \mathcal{N}_{\T}/2$. 
Therefore, the task at hand is to relate the magnetic noise tensor to the fluctuations of its source in the proximate superconducting sample. 

\subsection{Noise tensor and conductivity}
In the vicinity of a thermal phase transition from a metal to a superconductor, the time-varying magnetic field $\B(\r_q, t)$ originates from fluctuations of the current density $\j(\bm{\rho},t)$ via the Biot-Savart law (defining $\bm{\rho} = (x,y)$ to be the in-plane coordinates henceforth)~\footnote{We neglect relativistic retardation effects since typical speeds in solid state systems are much smaller than the speed of light $c$}.
Such current fluctuations are encoded in the non-local dynamical conductivity tensor $\sigma_{\mu \nu}$ of the system, defined via the linear response to an electric field $\bm{E}(\bm{\rho}, t)$ that varies in space and time. 
\begin{equation}
j_\mu(\bm{\rho},t) = \int dt\, \int d^2\bm{\rho} \, \sigma_{\mu \nu}(\bm{\rho} - \bm{\rho}^\prime, t - t^\prime) E_\nu(\bm{\rho}^\prime, t^\prime).
\end{equation}
To account for screening due to mobile charge carriers, it is useful to consider non-local conductivity in momentum space, and decompose it into longitudinal and transverse components (assuming translation and rotational invariance in plane): 
\begin{align}
\sigma_{\mu \nu}(\q, \Omega) & = \int dt\, \, e^{i \Omega t}  \int d^2\bm{\rho} \, e^{-i  \q \cdot \bm{\rho}} \sigma_{\mu \nu}(\r,t) \nonumber \\
& = \hat{q}_\mu \hat{q}_\nu  \sigma_{\L}(q,\Omega) + (\delta_{\mu \nu} - \hat{q}_\mu \hat{q}_\nu ) \sigma_{\T}(q,\Omega).
\label{eq:sigmaRotInv}
\end{align}
As discussed in Ref.~\onlinecite{Agarwal2016}, longitudinal current fluctuations create an imbalance in the charge density and get rapidly screened in the presence of mobile carriers, which results in its suppression by a factor of $(\Omega z_0/c)^2$. 
For typical values of the qubit frequency $\Omega \approx 1$ GHz, qubit-probe distance $z_0 \approx 10$ nm and $c = 3 \times 10^{8}$ m/s, $(\Omega z_0/c)^2 \approx 10^{-15}$, which is negligibly small. 
Therefore, we will focus on transverse current fluctuations, that satisfy $\nabla \cdot \j(\bm{\rho},t) = 0$ or equivalently $\q \cdot \j_\q(t) = 0$ in momentum space, as the primary source of magnetic noise. 
Such current fluctuations are encoded in the non-local transverse conductivity $\sigma_{\T}(\q,\Omega)$, which determines the transverse magnetic noise $\cN_{\T}$ in the experimentally relevant limit $\hbar\Omega \ll k_B T$ as~\cite{Agarwal2016}
\begin{equation}
\mathcal{N}_{\T}(\Omega, z_0) = \frac{\mu_0^2 k_B T}{4 \pi} \int_0^\infty dq \, q \, e^{-2 q z_0} \, \text{Re}\left[ \sigma_{\T}(q,\Omega) \right]  .
\label{eq:Noise}    
\end{equation}
In the vicinity of the critical point $T \approx T_c$, the dominant contribution to $\sigma_{\T}(\q,\Omega)$ comes from fluctuations of the superconducting order parameter $\psi$.
While this temperature window is narrow in 3D superconductors, it may be substantially wider due to enhanced order-parameter fluctuations in quasi-2D superconductors~\cite{SCNoise}.
Accordingly, we will focus on computing the fluctuation conductivity in the vicinity of the critical temperature $T_c$, both without and with an applied out-of-plane magnetic field $\bm{H} = H \hat{z}$. 

We note that our calculation does not account for the contribution of Bogoliubov quasiparticle excitations to the magnetic noise: these were analyzed in Refs.~\onlinecite{CD2022,DC2022}, and are suppressed relative to the noise originating from critical fluctuations when $T \approx T_c$. 
Further, the most general expression for the transverse magnetic noise from a 2D sample is expressed in terms of the reflection coefficient $r_s(\q,\Omega) = \left[1 + 2 i q/\left(\mu_0 \Omega \sigma_{\T}(\q, \Omega)\right)\right]^{-1}$ of s-polarized electromagnetic waves~\cite{Agarwal2016}.
However, the additional corrections are important only in the low-temperature limit when the superfluid density $n_s$ is large~\cite{DC2022}. 
Thus, in the vicinity of the phase transition, $n_s \propto |\psi|^2 \propto T_c - T$ is small, Eq.~\eqref{eq:Noise} provides an accurate description of the magnetic noise.  

\subsection{Noise tensor from vortex dynamics}
An alternate source of magnetic noise at the qubit location is the motion of vortices --- topological defects in the superconducting order parameter $\psi$ with integer winding of its phase around the vortex core.
Each vortex carries an integer number of magnetic flux quantum $\Phi_0 = h/e^*$, with $e^* = 2e$ being the charge of the pair-condensate \cite{tinkham}.
In the absence of an applied magnetic field, the net vortex density remains zero, but the system can nucleate an equal number of vortices and anti-vortices satisfying this constraint.
In fact, the zero-field superconducting transition in two dimensions is described by a vortex-antivortex binding transition --- known as a Berezinskii-Kosterlitz-Thouless (BKT) transition \cite{berezinskii1971destruction,kosterlitz1973ordering,kosterlitz2016kosterlitz}. The signatures of such a transition in magnetic noise, as evidenced by a lengthscale dependent vortex dielectric constant, were analyzed in Ref.~\onlinecite{Curtis_2024}.

At non-zero applied magnetic field, one can instead enter a superconducting phase with a non-zero net vortex density in type-II superconducting materials.
The precise phase of the vortices depends on a complex interplay of thermal fluctuations and disorder \cite{fisherfisherhuse,MFisher89,BlatterRMP}: the vortices may form a liquid, an Abrikosov vortex lattice or a vortex glass phase. 
These phases are characterized by distinct vortex dynamics, which leads to tell-tale imprints of the phase on the magnetic noise detected by the qubit. 
To characterize the noise due to vortex dynamics, we directly compute the correlations of the magnetic field $\B(\r_q, t)$ at the location of the qubit due to vortices.  

To this end, we first evaluate the magnetic field $\B_0(\r)$ in the upper half space ($z > 0$) due to a single vortex at the origin, i.e., at $\r_{\rm vortex} = (\bm{\rho},z) = (0,0,0)$ in a 2D superconductor. 
For analytical progress, we assume the extreme type-II limit, where $\kappa = \lambda_L/\xi \gg 1$, with $\lambda_L$ being the bulk penetration depth and $\xi$ being the order-parameter correlation length that sets the size of the vortex core. 
For a 2D superconductor of thickness $d$, we need to solve the following Maxwell-London equation to find the magnetic field:  
\begin{equation}
\nabla^2 \B_0=\left\{\begin{array}{l}
\dfrac{1}{\lambda_L^2}\left[\B_0-\hat{z}\, \Phi_0 \,\delta^{(2)}(\boldsymbol{\rho})\right], \quad-\dfrac{d}{2}<z<\dfrac{d}{2}\\
0, \qquad |z|>\dfrac{d}{2}   
\end{array}.  \right. 
\end{equation}
Since all the currents are restricted to the 2D sample ($d\to0$), $\nabla \times \B_0 = 0$ for $z > 0$, and accordingly we may write the field outside as the gradient of a scalar potential, i.e., $\B_0 = - \nabla \phi_M$.
Solving for $\phi_M$ with appropriate boundary conditions (see Appendix~\ref{app:C_mag} for details), we find that
\begin{align}
\phi_M(\bm{\rho},z) &= \int \frac{d^2\k}{(2\pi)^2} e^{i \k \cdot \bm{\rho}} e^{-k z} \phi_M^{\rm 2D}(\k), ~ \text{ where } \nonumber \\
\phi_M^{\rm 2D}(\k) &= \frac{\Phi_0}{k(1 + \Lambda k)},  \quad z>0  
\label{eq:PhiM2d}
\end{align}
$\Lambda = 2\lambda_L^2/d$ being the Pearl length \cite{pearl1964current}.
Using Eq.~\eqref{eq:PhiM2d}, we may directly calculate the field at the qubit location as $\B_0(\r_q) =  - \nabla \phi_M|_{\r = \r_q}$.
\begin{equation}
\B_0(\r_q) =  \int \frac{d^2\k}{(2\pi)^2} e^{i \k \cdot \bm{\rho}_q} e^{-k z_0} (- i\k, k) \frac{\Phi_0}{k(1 + \Lambda k)}   .
\label{eq:B02D}
\end{equation}

To contrast our results for magnetic noise with the noise originating from vortex dynamics in bulk superconductors occupying the lower half-space $(z < 0)$, we also calculate the field of a line vortex along the $z$-axis.
In this case, the relevant Maxwell-London equations take the form
\begin{equation}
\nabla^2 \B_0=\left\{\begin{array}{l}
\dfrac{1}{\lambda_L^2}\left[\B_0-\hat{z}\, \Phi_0 \,\delta^{(2)}(\boldsymbol{\rho})\right], \quad z<0 \\
0, \quad z>0  ,
\end{array}\right.
\end{equation}
where we have neglected possible variation of the superfluid density near the surface for simplicity. 
Once again, the magnetic field outside can be characterized by a scalar potential $\phi_M^{\rm 3D}$ as $\B_0 = - \nabla \phi_M^{\rm 3D}$, with
\begin{equation}
\phi_M^{\rm 3D}(\k) = \frac{\Phi_0}{k(1 + k^2 \lambda_L^2)}, \quad z>0
\label{eq:PhiM3D}
\end{equation}
implying that the field is modified as,
\begin{equation}
\B_0(\r_q) =  \int \frac{d^2\k}{(2\pi)^2} e^{i \k \cdot \bm{\rho}_q} e^{-k z_0} (- i\k, k) \frac{\Phi_0}{k(1 + \lambda_L^2 k^2)}   . \label{eq:B03D}  
\end{equation}
Therefore, in either case, we may find the field due to a collection of vortices at positions $\bm{\rho}_i$ by the principle of superposition, in terms of the vortex density $n_v(\bm{\rho},t) = \sum_i \delta^{(2)}(\bm{\rho} - \bm{\rho}_i(t))$, as:
\begin{equation}
\B(\r_q,t) = \int d^2 \rr\, \B_0(\r_q - \bm{\rho})n_v(\bm{\rho},t)   .
\end{equation}
Thus, knowing the field created by a single vortex, we may directly derive the magnetic noise in terms of correlations of the vortex density
\begin{align}
& \mathcal{N}_{\alpha \beta}(\Omega) =  \int d^2 \rr \int d^2 \rr^\prime \, B_{0}^\alpha(\r_q - \bm{\rho}) B_{0}^\beta(\r_q - \bm{\rho}^\prime) C_{\bm{\rho},\bm{\rho}^\prime}(\Omega) ~~~ \nonumber \\
& \text{ where } C_{\bm{\rho},\bm{\rho}^\prime}(\Omega) = \int dt\, \, e^{i \Omega t} \langle n_v(\bm{\rho},t) n_v(\bm{\rho}^\prime,0) \rangle_{T}
\label{eq:Nvortex}
\end{align}
is the vortex density correlation function, and we have assumed that the vortices behave as classical point particles (rods) in 2D (3D). 
For detecting the motion of isolated or pinned vortices, we can simply focus on the field created by a single vortex, while in a vortex solid or liquid phase,
one may use discrete or continuous translation symmetry to further simplify Eq.~\eqref{eq:Nvortex}.
In the latter cases, the transverse noise $\mathcal{N}_{\T}$ and the longitudinal noise $\mathcal{N}_{\L}$ can extracted from the different components of the noise tensor, yielding useful information vortex dynamics.

\section{Zero-field critical magnetic noise}
\label{sec:ZF}
To set the stage for evaluating magnetic noise near criticality at non-zero magnetic fields, we first calculate the magnetic noise arising from critical supercurrent fluctuations near the superconductor to metal phase transition at zero magnetic field.
Strictly speaking, this transition is described by BKT physics in two dimensions due to very weak screening of inter-vortex interactions~\cite{Beasley}, and the magnetic noise originating for the resultant vortex-binding dynamics is discussed in Ref.~\onlinecite{Curtis_2024}.
Nevertheless, in finite-sized samples or in layered quasi-2d materials, a Ginzburg-Landau (mean-field) approach can reasonably describe the conductivity at the onset of local pairing, and thus provide a useful approximation if the pairing temperature is not very distinct from $T_{\rm BKT}$ where phase coherence sets in~\cite{Dorsey91}. 
Therefore, we resort to a calculation of the magnetic noise using a time-dependent Ginzburg-Landau approach, which also has the advantage of being straightforwardly generalizable to finite magnetic fields.

Our analysis shows that measurements of this magnetic noise provide direct access to physically relevant quantities, such as the characteristic lifetime of a critical supercurrent fluctuation, as well as providing an understanding as to how the magnetic noise scales as a function of the qubit-sample distance $z_0$ and coherence length $\xi$.

To make a quantitative analysis of the magnetic noise, our main objective is to find the non-local transverse conductivity $\sigma_{\T}(\q,\Omega)$, from which the magnetic noise can be obtained via Eq.~\eqref{eq:Noise}.
We evaluate $\sigma_{\T}$ with the classical fluctuation-dissipation theorem (FDT), through the transverse current-current correlator \cite{kuboFDT1966}: 
\begin{equation}
\label{eq:sigma_current_correlator}
    \sigma_{\T}(\q,\Omega) = \frac{1}{2k_B T} \langle
    \j^{\T}_{\q,\Omega}\cdot \j^{\T}_{-\q,-\Omega}\rangle,
\end{equation}
where the in-plane transverse current density is defined as $\j^{\T}_\q=\j_\q-\hat{\q}(\hat{\q}\cdot \j_\q)$, with $\hat{\q}$ the unit momentum vector and $\j(\q, \Omega)$ denotes the Fourier transform of the in-plane current density $\j(\bm{\rho},t)=[\hbar e^*/(2im^*)]\big[\psi^*\nabla \psi-\psi(\nabla \psi^*)\big]$, $\psi(\bm{\rho},t)$ being the superconducting order parameter.

We work in the experimentally relevant regime $\hbar\Omega \ll k_B T$, and therefore set $\Omega = 0$ when computing the transverse conductivity.
To evaluate the transverse current correlator we use the time-dependent Ginzburg-Landau framework.
In this approach, the dynamics of the $\psi$ are governed by the following Langevin equations \cite{schmid1966time}, 
\begin{equation}
 \label{eq:Langevin_TGzL}
    \hbar \partial_{t}\psi = -\Gamma \frac{\delta \mathcal{H}}{\delta\psi^{*}}+\zeta,
\end{equation}
with the Ginzburg-Landau Hamiltonian $\mathcal{H}$ given by
\begin{equation}
    \mathcal{H}=\int d^2\bm{\rho}  \left(\frac{\hbar^2}{2m^*} |\nabla\psi|^{2}+ r|\psi|^{2} + \frac{u}{2}|\psi|^{4} \right).
\label{eq:GLFE}
\end{equation}
$r \propto T - T_c$ is an effective parameter controlling the distance from the critical temperature, such that, at mean-field level, $r < 0$ describes the superconducting side, and $r > 0$ describes the metallic side. $m^*$ denotes the effective mass of the Cooper pairs, and $u>0$ controls the strength of the non-linearity of the order paramater.
$\zeta$ is a zero-mean Gaussian noise with correlations given by the FDT, with local temperature $T$,
\begin{equation}
\label{eq:Gamma_defintion}
    \langle \zeta(\rr, t)\zeta(\rr', t') \rangle = 2\Gamma \hbar k_B T \,\delta(t-t') \delta^{(2)}(\rr-\rr')\,.
\end{equation}
The dimensionless parameter $\Gamma$ in Eq.~\eqref{eq:Langevin_TGzL} is determined by the coupling with the higher energy degrees of freedom, which act as a bath that generates the dissipative dynamics.

\subsection{Metallic side}

\begin{figure*}[t]
    \includegraphics[width=1\linewidth]{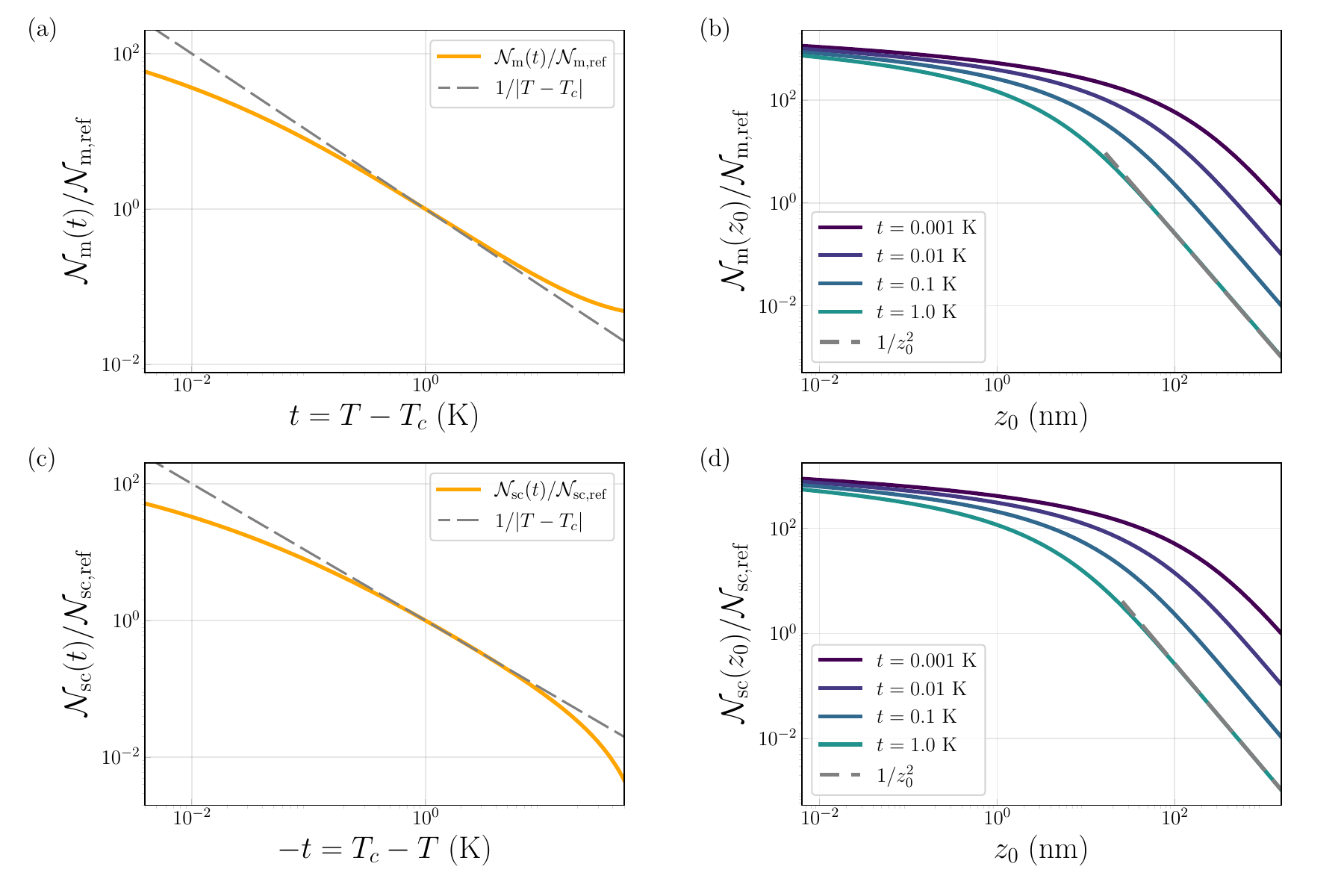}
  \caption{(a)/(c) Rescaled transverse noise on the metallic/superconducting side $\mathcal{N}_{\rm m/sc}/\mathcal{N}_{\rm ref}$ at $H=0$ as a function of $t=T-T_c$; the dashed line indicates the scaling $\sim 1/t$. (b)/(d) Rescaled transverse noise as a function of probe distance $z_0$ for different $t$; the dashed line indicates the scaling $\sim 1/z_0^{2}$. In each panel, the reference noise is $\mathcal{N}_{\rm ref}=\mathcal{N}_{\rm m/sc}(1\,\text{K},\,50\,\text{nm})$. The coherence length at zero temperature is chosen as $\xi_0 = 1.6$ nm, $\xi_{\rm sc}(t)=\xi_0\sqrt{T_c/|t|}$, $\xi_{\rm m} = \sqrt{2}\xi_{\rm sc}$, and transition temperature $T_c = 92.75~$K, which are typical values for high-$T_c$ cuprates.}
 \label{Fig:Metal_B=0}
\end{figure*}

Let us first consider the metallic case, corresponding to $r>0$ in Eq.~\eqref{eq:GLFE}.
We work in the Gaussian limit and neglect the quartic term in Eq.~\eqref{eq:GLFE}.
In this approximation, the order-parameter correlation function, $C(\q,t) = \langle \psi^*_\q(t)\psi_\q(0) \rangle$, is then given by:
\begin{equation}
\label{eq:Cqt_B0}
    C(\q,t) = \frac{k_BT}{\frac{\hbar^2 \q^2}{2m^*} +r}e^{-\Gamma\left(\frac{\hbar^2 \q^2}{2m^*} + r\right)|t|/\hbar}.
\end{equation}
Eq.~\eqref{eq:Cqt_B0} shows that the lifetime of the cooper pair fluctuations and their correlation length is given by,
\begin{equation}
    \tau_{\rm m} = \frac{\hbar}{r \Gamma} \,, \ \xi_{\rm m} = \frac{\hbar}{\sqrt{2m^* r}}   .
    \label{eq:tmxim}
\end{equation}
In the Gaussian limit, the current–current correlator can be evaluated using Wick’s theorem, yielding the following expression for the transverse conductivity.
\begin{equation}
\sigma_{\T}^{\rm m}(q) = \frac{1}{2k_B T}
\left( \frac{\hbar e^*}{ m^*}\right)^2
\int_{\k,t} \k_{\perp}^2 C(\k-\q,t)C(\k,t)   . 
\label{main:sigmaTm}
\end{equation}
Here, $\k_\perp = \k - \hat{\q}(\hat{\q}\cdot\k)$ denotes the component of momentum $\k$ perpendicular to $\q$.
Evaluating the integral (see Appendix~\ref{app:ZeroB}), we find that the transverse conductivity takes the scaling form
\begin{equation}
\begin{split}
    &\sigma_{\T}^\mrm(q) = \sigma_\mrm(0) F_\mrm(q\xi_{\rm m}) \,, \textrm{ where } \\
    & \sigma_\mrm(0) = \frac{1}{8\pi}\left(\frac{e^{*}}{\hbar}\right)^2 k_BT\,\tau_{\rm m} \\
    & F_{\mrm}(\tilde{q}) = \frac{4}{\tilde{q}^2} \left[\ln \frac{16}{\left(4+\tilde{q}^2\right)^2}+\frac{4 \tilde{q}}{\sqrt{4+\tilde{q}^2}} \tanh^{-1} \frac{\tilde{q}}{\sqrt{4+\tilde{q}^2}} \right].
\end{split}\label{eq:sigma_Tm}
\end{equation}
$\sigma_{\mrm}(0)$ denotes the zero-momentum conductivity, and $F_{\mrm}(q\xi_{\rm m})$ is a dimensionless scaling function normalized such that $F_{\mrm}(0)=1$.
$\sigma_{\mrm}(0)$ admits a simple physical interpretation: it is set by the density of thermally excited order-parameter fluctuations $\langle |\psi|^2 \rangle \sim k_B T$, multiplied by their characteristic lifetime $\tau_\mrm$.
Since $\tau_{\mrm} = (2m^{*}\Gamma/\hbar) \xi_{\rm m}^{2}$ diverges as $\xi_{\rm m}^{2} \propto (T-T_c)^{-1}$ as the critical point is approached, the uniform ($q = 0$) conductivity also exhibits a divergence, reflecting the critical slowing down of Cooper-pair fluctuations.

Applying Eq.~\eqref{eq:sigma_Tm} on Eq.~\eqref{eq:Noise}, we arrive at the magnetic noise, 
\begin{align}
    \mathcal{N}_{\rm m}&= \frac{\mu_0^2 k_B T}{4\pi} \sigma_{\rm m}(0) \int_0^\infty dq \, q \, e^{- 2 q z_0} F_{\rm m}(q \xi_{\rm m})   .
    \label{eq:Nm}
\end{align}
In Fig.~\ref{Fig:Metal_B=0} we plot the magnetic noise as a function of $T-T_c$, and also as a function of the qubit-sample distance $z_0$. 
However, the asymptotic limits can be readily understood by noting that in Eq.~\eqref{eq:Nm}, the magnetic noise is simply a function of the dimensionless ratio $\tilde z =z_0/\xi_{\rm m}$.

Away from the critical regime, when $\xi_{\rm m} \ll z_0$, we may approximate $F_{{\rm m}}(q \xi_{\rm m}) \approx F_{\rm m}(0) =  1$ and the integral in Eq.~\eqref{eq:Nm} scales as $1/z_0^{2}$, leading to the following expression for the magnetic noise:
\begin{equation}
\cN_{\rm m}  = \frac{\mu_0^2 k_B T}{16 \pi z_0^2} \sigma_{\rm m}(0).
\end{equation}
This expression highlights two distinct scaling behaviors.
First, since $\sigma_{\rm m}(0) \propto \tau_{\rm m} \propto (T - T_c)^{-1}$, the noise diverges as $(T - T_c)^{-1}$ upon approaching the transition.
By fitting the slope of $\cN_{\rm m}(T)$, one can further determine the characteristic lifetime $\tau_{\rm m}$ of critical supercurrent fluctuations \cite{SCNoise}.
Second, the prefactor reveals a spatial dependence, with the noise decaying as $z_0^{-2}$ with the qubit distance.
These trends are clearly reflected in Fig.~\ref{Fig:Metal_B=0}: panel (a) illustrates the temperature dependence, while panel (b) highlights the distance scaling.

As the temperature is lowered to approach the critical point, the coherence length $\xi_{\rm m} \propto (T  - T_c)^{-1/2}$ becomes much larger than the qubit-sample distance $z_0$.
In this regime, we may approximate $F_{\rm m}(q \xi_{\rm m}) \approx (q\xi_{\rm m})^{-2}$ in the integral in Eq.~\eqref{eq:Nm}, leading to an integral that is independent of $z_0$ (up to logarithmic corrections). 
Consequently, the magnetic noise scales as 
\begin{equation}
\cN_{\rm m} \propto \frac{\mu_0^2 k_B T}{\xi_{\rm m}^2} \sigma_{\rm m}(0) \ln \left( \frac{\xi_{\rm m}}{z_0} \right).
\end{equation}
Since $\sigma_{\rm m}(0) \propto \tau_{\rm m} = (2 m^* \Gamma/\hbar)\xi_{\rm m}^2$, the explicit dependence on $T - T_c$ cancels out in $\sigma_{\rm m}(0)/\xi_{\rm m}^2$.
As a result, the noise no longer diverges algebraically with temperature, but instead saturates, up to a weak logarithmic dependence on either $z_0$ or $T - T_c$.
This saturation is evident in Fig.~\ref{Fig:Metal_B=0}, appearing both in the $t \to 0$ limit of panel (a) and in the $z_0 \to 0$ limit of panel (b).

\subsection{Superconducting side}

We next turn to investigate magnetic noise on the superconducting side of the transition. 
While such a transition is typically a BKT transition as discussed previously, there can be a temperature window where order parameter fluctuations about the local pairing amplitude dominates magnetic noise. 
Here, we consider such Gaussian fluctuations in a quasi-long-range ordered phase, and delineate the corresponding noise signatures --- partly motivated by potentially similar observations in a recent experiment on a thin film high-$T_c$ cuprate~\cite{SCNoise}. 
However, we carefully note that non-linear effects or vortex-antivortex fluctuations, as discussed in Ref.~\onlinecite{Curtis_2024}, will determine the magnetic noise at temperatures very close to the transition. 

We express the order parameter in the amplitude–phase representation as,
\begin{align*}
\psi(\bm{\rho},t)=\sqrt{M_{0}+\chi(\bm{\rho},t)}e^{i\theta(\bm{\rho},t)}  ,
\end{align*}
with $M_0 = |r|/u$ denoting the homogeneous average density evaluated at the mean-field level, and $\chi, \theta$ the density and phase fluctuations, respectively. 
Although the mean local pairing amplitude is set by $\sqrt{M_0}$, the order parameter correlations have a power law decay due to strong fluctuations of the phase $\theta(\bm{\rho},t)$.
In this representation, the current is given by,
$\j(\bm{\rho}) = (\hbar e^*/m^*)\,\mathrm{Im} (\psi^* \nabla \psi )= (\hbar e^*/m^*)[M_0+\chi(\bm{\rho})]\nabla\theta(\bm{\rho})$.
Converting to momentum space, we find the transverse current to be,
\begin{equation}
\j^{\T}_{\q,\Omega}=\frac{i\hbar e^*}{ m^*}\int_{\k,\nu} \k_\perp \chi_{\q-\k,\Omega-\nu}\theta_{\k,\nu}.
\end{equation}
This indicates that \textit{both amplitude and phase fluctuations} are required for a non-zero transverse current density, as phase gradients are purely longitudinal. 
Therefore, to compute the transverse conductivity, we first require the correlation functions of the amplitude and phase fluctuations, $\chi$ and $\theta$. 
To this end, we linearize Eq.~\eqref{eq:Langevin_TGzL} in these variables, upon which we obtain the following Langevin equations,
\begin{equation}
\label{eq:LangevinSc}
\begin{split}
    \hbar\dot{\chi} & = -\Gamma \Big(2 uM_0 \chi - \frac{\hbar^2}{2m^*} \nabla^2\chi \Big)+ \zeta_\chi  ,\\
   \hbar\dot{\theta} & = \Gamma \frac{\hbar^2}{2m^*} \nabla^2\theta+\zeta_\theta  ,
\end{split}
\end{equation}
where $\zeta_\chi=2 \sqrt{M_0} \operatorname{Re} \zeta$ and $\zeta_\theta=\operatorname{Im} \zeta / \sqrt{M_0}$ are zero-mean Gaussian noises with correlations:
\begin{equation}\label{ampphase_noise}
\begin{split}
    \langle \zeta_\chi(\bm{\rho}, t)\zeta_\chi(\bm{\rho}', t') \rangle &= 4\hbar M_0 \Gamma k_B T\delta(t-t') \delta^{(2)}(\bm{\rho}-\bm{\rho}'),\\
    \langle \zeta_\theta(\bm{\rho}, t)\zeta_\theta(\bm{\rho}', t') \rangle &= \frac{\hbar \Gamma k_B T}{M_0}\delta(t-t') \delta^{(2)}(\bm{\rho}-\bm{\rho}'). 
\end{split}
\end{equation}
Crucially, we notice that $\chi$ and $\theta$ are decoupled, and $\theta$ is a massless field (the would-be Goldstone mode in case of true spontaneous symmetry breaking), while the density fluctuations $\chi$ are massive.
From Eq.~\eqref{eq:LangevinSc}, we find the correlation functions of $\chi$ and $\theta$ to be 
\begin{equation}
\begin{split}
     C_{\chi\chi} (\q,t) &= \frac{2
     k_B T M_0}{\frac{\hbar^2 \q^2}{2m^*}  + 2u M_0} e^{-\Gamma\big(\frac{\hbar^2 \q^2}{2m^*}  + 2uM_0 \big)|t|/\hbar}  ,\\
     C_{\theta\theta} (\q,t) &= \frac{k_B T}{ \frac{\hbar^2 \q^2}{m^*} M_0 } e^{-\Gamma \frac{\hbar^2 \q^2}{2m^*} |t|/\hbar}  .
\end{split}\label{eq:Cqt_B0_sc}
\end{equation}
Here, since $\chi$ and $\theta$ are decoupled in Eq.~\eqref{eq:LangevinSc}, their cross-correlations vanish.
Furthermore, Eq.~\eqref{eq:Cqt_B0_sc} shows that the amplitude mode has a correlation length, i.e., superconducting coherence length, $\xi_{\rm sc} = \hbar/\sqrt{4 m^* u M_0} = \hbar/\sqrt{4m^*|r|}$, and that its lifetime (for $q=0$) is given by $\tau_{\rm sc} = \hbar/(2 u M_0 \Gamma)$.

From the transverse current-current correlator, we find that the transverse conductivity is given by, 
\begin{multline}
\sigma_{\T}^{\rm sc}(q) = \frac{1}{2k_B T}\left( \frac{\hbar e^*}{ m^*}\right)^2\int_{\k,t} \k_\perp^2 C_{\chi\chi}(\k-\q,t)C_{\theta\theta}(\k,t).
\label{Eq:CurrentCorreltor_sc}
\end{multline}
Applying Eq.~\eqref{eq:Cqt_B0_sc} to Eq.~\eqref{Eq:CurrentCorreltor_sc} and evaluating the integral (see Appendix~\ref{app:ZeroB}), we finally arrive at the scaling form for the transverse conductivity on the superconducting side.
\begin{equation}
\begin{split}
    &\sigma_{\T}^{\rm sc}(q) = \sigma_{\rm sc}(0) F_{\rm sc}(q\xi_{\rm sc}) \,, \textrm{ where } \\
    & \sigma_{\rm sc}(0) = \frac{\ln2}{\pi} \left(\frac{e^*}{\hbar}\right)^2 k_B T \,\tau_{\rm sc} \\
    & F_{\rm sc}(\tilde{q}) = \frac{1}{\tilde{q}^2\ln2}\left[\ln\left(4\frac{1+\tilde{q}^2}{2+\tilde{q}^2}\right) - \frac{\ln(2+\tilde{q}^2)}{1+\tilde{q}^2}\right].
\end{split}\label{eq:sigma_Tsc}
\end{equation}
$\sigma_{\rm sc}(0)$ denotes the zero-momentum paramagnetic conductivity, and $F_{\rm sc}(\tilde{q})$ stands for the scaling function \footnote{Note that the diverging conductance of superconductor come from the diamagnetic contribution to the conductance, and the paragmagnetic contribution $\sigma_{\rm sc}(0)$ -- driven by supercurrent fluctuations -- is finite}.

Applying Eq.~\eqref{eq:sigma_Tsc} to Eq.~\eqref{eq:Noise}, we arrive at the magnetic noise on the superconducting side.
The results are plotted in Fig.~\ref{Fig:Metal_B=0}, where we plot the magnetic noise as a function of $|T-T_c|$ and also as a function of the qubit-sample distance $z_0$.
We find qualitatively similar behaviors to the metallic side of the transition.

Away from criticality, for coherence length $\xi_{\rm sc} \ll z_0$, we find the magnetic noise to scale as,
\begin{equation}
    \cN_{\rm sc} = \frac{\mu_0^2 k_B T}{16\pi z_0^2} \sigma_{\rm sc}(0) \,.
\end{equation}
Since $\sigma_{\rm sc}(0) \propto \tau_{\rm sc}$, the noise diverges as $(T_c - T)^{-1}$ as the transition is approached, as shown in Fig.~\ref{Fig:Metal_B=0}(c).
Furthermore, the prefactor reveals that the noise decays as $z_0^{-2}$ with the qubit-sample distance, as shown in Fig.~\ref{Fig:Metal_B=0}(d).

On the other hand, as we approach the critical point, and the coherence length $\xi_{\rm sc} \gg z_0$, the magnetic noise becomes approximately $z_0$ independent and scales as,
\begin{equation}
    \cN_{\rm sc} \propto \frac{\mu_0^2 k_B T}{\xi_{\rm sc}^2} \sigma_{\rm sc}(0) \ln \left(\frac{\xi_{\rm sc}}{z_0}\right) \,.
\end{equation}
Since $\sigma_{\rm sc}(0)/\xi_{\rm sc}^2$ has no explicit dependence on $T_c-T$, the noise does not diverge algebraically but instead saturates up to a logarithmic dependence, as seen in Fig.~\ref{Fig:Metal_B=0}.

\section{Magnetic noise at criticality in a field}
\label{sec:B_TDGL}

\subsection{Metallic side}
We now evaluate the magnetic noise near criticality in the presence of a finite magnetic field.
A background field $\H = H \hat z$ not only enhances critical fluctuations and lowers $T_c$, but also introduces an additional lengthscale — the magnetic length $\ell = \sqrt{\hbar/(\mu_0 e^{*} H)}$.
Our goal is to determine how the magnetic noise depends on the three lengthscales of the qubit-sample distance $z_0$, the correlation length $\xi$, and the magnetic length $\ell$.
In the limit $H \to 0$, our results reduce smoothly to the zero-field expressions.
As we show below, the magnetic noise acquires a linear-in-$H$ enhancement at small fields, reflecting an amplification of fluctuation-driven noise in the presence of a background magnetic field.

We model the dynamics of the order parameter by extending the time-dependent Ginzburg-Landau approach in Ref.~\onlinecite{DorseyHuse}.
To this end, we introduce the gauge-covariant derivative $D = \nabla - i e^*\mathbf{A}/\hbar$ to the Hamiltonian of Eq.~\eqref{eq:Langevin_TGzL}, with $\nabla \times \A = \mu_0 \H$,
$$\mathcal{H}=\int d^2\bm{\rho}  \left(\frac{\hbar^2}{2m^*} |D\psi|^{2} + r|\psi|^{2} +\frac{u}{2}|\psi|^{4} \right)  .$$
The resulting Hamiltonian is invariant under local $U(1)$ gauge transformations $\psi \to \psi e^{i\theta}$, $\A \to \A + (\hbar/e^*)\nabla\theta$.
In order to facilitate calculations, we work with the Landau gauge for the static background field, $\A=\mu_0H(0,x,0)$.

\begin{figure}[t]
\hspace{-28pt}
\includegraphics[width=1.1\linewidth]{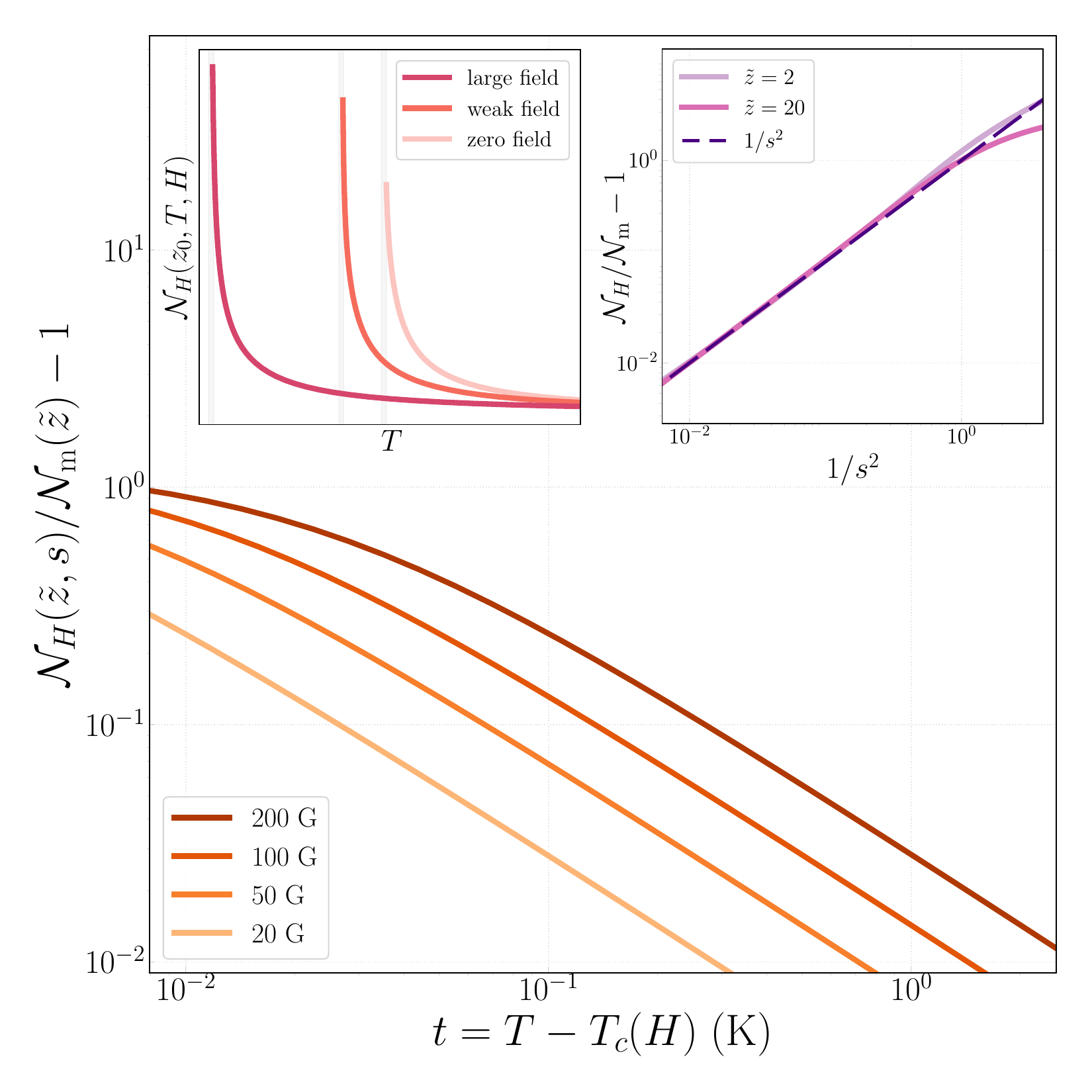}
\caption{
Left inset (schematic): As a magnetic field $H$ is turned on, the peak of the noise $\cN_H(T)$ moves to a lower temperature and increases in magnitude in the weak-field regime.
Main figure: The relative change in noise $\mathcal{N}_H(\tilde{z}, s) / \mathcal{N}_{\rm m}(\tilde{z}) - 1$ as a function of $t = T - T_c(H)$, with $(\tilde{z},s)= (z_0/\xi_H,\ell/\xi_H)$ and $T_c(H) = T_c(0)\sqrt{1 - H/H_{c2}(0)}$. 
The zero-field reference noise $\mathcal{N}_{\rm m}(\tilde{z} = z_0/\xi_{\rm m})$ is evaluated using $\xi_{\rm m} = \xi_0 \sqrt{2 T_c/t}$ with $\xi_0 = 1.6~\rm{nm}$ and $z_0 = 200~\rm{nm}$, while the field-dependent noise $\mathcal{N}_H$ is computed from Eq.~\eqref{eq:NB_int}, using the correlation length $\xi_H$ and relaxation time $\tau_H$ given by Eq.~\eqref{eq:tHxiH}.
As $T$ approaches $T_c(H)$, the relative noise initially exhibits a $1/t$ scaling, but saturates as $t \to 0$.
 Right inset: at weak fields, the relative increase in noise scales as $1/s^2 \propto H$, i.e., linearly with the applied field.}
\label{Fig:Metal_B}
\end{figure}

Solving the Langevin equations, we find that the order parameter correlation function is given by (see Appendix~\ref{app:B_TDGL} for details),
\begin{multline}
    C(\bm{\rho}_1,\bm{\rho}_2,\Omega) = \frac{2\hbar k_BT}{\Gamma}  \int_{q_y}e^{iq_y(y_1-y_2)} \\
    \times\sum_{n=0}^{\infty}\frac{u_n(x_1-x_0)u_n(x_2-x_0)}{(\hbar\Omega/\Gamma)^2 + \epsilon_n^2} \,.
    \label{eq:C_H}
\end{multline}
Here, $\epsilon_n = r + \hbar\omega_c(n+ 1/2)$ denote the Landau level energies with $\omega_c = \mu_0 e^*H/m^*$ is the cyclotron frequency, and $x_0 = q_y\ell^2$.
Last, $u_n(x)$ denotes the Landau level eigenfunctions, which is given by,
\begin{equation*}
\label{eq:Hermite}
u_n(x) = \left( \frac{1}{2^n n!\sqrt{\pi} \ell} \right)^{\frac{1}{2}} e^{-\frac{x^2}{2\ell^2}}H_n\left(\frac{x}{\ell}\right),
\end{equation*}
with $H_n$ the Hermite polynomials.

From Eq.~\eqref{eq:C_H}, we can extract both the correlation length $\xi_H$ by considering the equal-time correlation function $C(\bm{\rho},0;t=0) \sim e^{-\rho/\xi_H}$ at distances $\rho \gg \ell$, and the lifetime $\tau_H$ of zero-momentum Cooper pair fluctuations (see Appendix~\ref{app:B_TDGL} for details).
\begin{equation} 
\xi_H =\frac{\hbar}{\sqrt{2 m^*\left(r+\frac{1}{2} \hbar \omega_c\right)}}, \; \tau_H = \frac{\hbar}{\Gamma (r + \hbar\omega_c/2)}.
\label{eq:tHxiH}
\end{equation}
Note that in the limit of $H \rightarrow 0$, $\tau_H$ and $\xi_H$ reduce to the zero field values of Eq.~\eqref{eq:tmxim}.
Eq.~\eqref{eq:tHxiH} shows that the critical point, indicated by a divergence in the correlation length $\xi_H$ (as well as $\tau_H$), is shifted from $r = 0$ to $r = -\hbar \omega_c/2 = - \hbar^2/(2 m^* \ell^2)$, consistent with our expectations that an applied field will lower the critical temperature $T_c$.
We emphasize, however, that magnetic-field-induced pair breaking, e.g., due to Zeeman splitting of up- and down-spin electrons, is not included in our analysis, and would be expected to further suppress $T_c$.

Using the correlation functions of the order parameter, we compute the transverse current-current correlator and thereby obtain the in-field transverse conductivity $\sigma^H_{\T}$.
A useful simplification follows from rotational invariance, which constrains the general form of the conductivity tensor (see Eq.~\eqref{eq:sigmaRotInv}) and hence allows us to evaluate $\sigma^H_{\T}(q)$ by simply computing $\sigma_{yy}(q\hat{x})$ at a generic field $H$.
\begin{multline}
    \sigma_{\T}^{H}(q) = \frac{-1}{2k_B T} \left (\frac{\hbar e^*}{2m^*} \right)^2 \int_{\Omega, \rr_1-\rr_2} \!\!e^{i q(x_1-x_2)} \\ 
    \times(D_{y_1} \!-\! D^*_{y_3})(D_{y_2} \!-\! D^*_{y_4})C(\rr_2,\rr_3,\Omega)C(\rr_1,\rr_4,\Omega) \Bigg|_{\substack{3=1\\4=2}},
    \label{Eq:CurrentCorreltor_H}
\end{multline}
where we have defined the conjugate differential operator $D^* = \nabla + i e^* \A /\hbar$.
Performing the integration, we find the in-field transverse conductivity $\sigma^H_{\T}$ to be given by (see Appendix~\ref{app:B_TDGL}),
\begin{equation}
\label{eq:sigmaTB}
\begin{split}
    & \sigma_{\T}^H(q) = \sigma_H(0, s) F_H(q\xi_H, s) \,, \textrm{ where} \\
    & \sigma_H(0, s) = \frac{g(s)}{4\pi} \left(\frac{e^*}{\hbar} \right)^2 k_B T \tau_H \,, \\
    & F_H(\tilde{q}, s) = \frac{s^2}{g(s)} \sum_{n,m = 0}^\infty \frac{|A_{mn}(\tilde{q}s)|^2}{\tilde \epsilon_n(s) \tilde \epsilon_m(s) [\tilde \epsilon_n(s) + \tilde \epsilon_m(s)]} \,.
\end{split}
\end{equation}
Here, for notational convenience, we have introduced the dimensionless ratio $s = \ell/\xi_H$, where the limit $s \to 0$ corresponds to strong magnetic fields and $s \to \infty$ recovers the zero-field regime.
$\sigma_H(0,s)$ denotes the zero-momentum conductivity at finite field, and $F_H(\tilde q, s)$ is the corresponding two-parameter scaling function, normalized such that $F_H(0,s)=1$.
The rescaled Landau-level energies are given by $\tilde\epsilon_n(s) = \epsilon_n/(\hbar\omega_c) = n + s^2 /2$, while the dimensionless normalization function $g(s)$ and the matrix function $A_{mn}(x)$ take the form

\begin{align*}
    & g(s) = s^2\left\{1+\left(s^2-1\right)\left[\psi\left(\frac{s^2}{2}\right)-\psi\left(\frac{s^2+1}{2}\right)\right]\right\} ,\\
    & A_{mn}(\tilde{q}s) = \ell\int_{-\infty}^{+\infty} dX\, {u}_m^*(X\ell)\left( X e^{-i\tilde{q}s X} \right) {u}_n(X\ell) .
\end{align*}
An explicit expression for the dimensionless function $A_{mn}(x)$ in terms of associated Laguerre polynomials is provided in Appendix~\ref{app:B_TDGL}. 

Finally, applying Eq.~\eqref{eq:sigmaTB} to Eq.~\eqref{eq:Noise}, we arrive at the magnetic noise in a field,
\begin{equation}
\begin{aligned}
\mathcal{N}_H&=\frac{\mu_0^2 k_B T}{16 \pi} \frac{\sigma_H(0, \ell/\xi_H)}{z_0^2} \int_0^{\infty} d x\, x e^{-x} F_H\left( \frac{x}{2 z_0}\xi_H, \frac{\ell}{\xi_H}\right) \\
&=\frac{\mu_0^2(k_B T)^2}{8\pi^2}\left(\frac{e^*}{\hbar}\right)^2 \frac{m^*}{\hbar\Gamma} g(s) \int_0^{\infty}  d \tilde{q}\, \tilde{q} e^{-2\tilde{q}\tilde{z}} F_H\left(\tilde{q}, s\right).
\label{eq:NB_int}    
\end{aligned}
\end{equation}
Importantly, we find from Eq.~\eqref{eq:NB_int} that the noise is a function of the dimensionless ratios, $\tilde{z}=z_0/\xi_H$ and $s = \ell/\xi_H$.

In the limit $H \rightarrow 0$, $s \rightarrow \infty$, and both $\sigma_H(0,s)$ and $F_H(\tilde q, s)$ reduce to $\sigma_{\rm m}(0)$ and $F_{\rm m}(\tilde q)$ of Eq.~\eqref{eq:sigma_Tm}.
Consequently, the magnetic noise $\cN_H$ approaches its zero-field value $\cN_\mrm$.
In the limit of large $\tilde z$, to leading order in $1/s$, the deviation between the two is $\delta \cN_H = \cN_H-\cN_{\rm m}$,
\begin{subequations}\begin{align}
    &\delta \cN_H = \frac{\mu_0^2(k_B T)^2}{8\pi^2}\left(\frac{e^*}{\hbar}\right)^2 \frac{m^*}{\hbar\Gamma} \left( \frac{1}{8 s^2\tilde{z}^2 } \right) \,,  \label{eq:deltaNH} \\
    &\delta \cN_H/\cN_{\rm m}= \left(\frac{\xi_H}{s \xi_{\rm m}}\right)^2 = \left(\frac{\xi_H^2}{\ell\xi_{\rm m}}\right)^2 .
    \label{eq:deltaNH_ratio}
\end{align}\end{subequations}
It is instructive to study the ratio $\delta\cN_H/\cN_{\rm m}$, which captures the relative change of magnetic noise as a function of the applied field $H$. 
Eq.~\eqref{eq:deltaNH_ratio} shows that for weak magnetic fields when $\xi_H \approx \xi_{\rm m}$, the deviation of the magnetic noise from that at zero field scales (to leading order in $H$) as $\xi_{\rm m}^2/\ell^2 \propto H$, and is therefore a linear increase in the magnetic field. 
We numerically confirm such $H$-linear increase in Fig.~\ref{Fig:Metal_B}~\footnote{In practice, to achieve numerical convergence, we limit the $\tilde{q}$ integral from 0 to 1, which is accurate when $\tilde{z} \gtrsim 1$ }.
This enhancement of magnetic noise can be attributed to the increase in the fluctuations as a result of the background magnetic field.

We note that at stronger magnetic fields, the magnetic length $\ell$ decreases and becomes comparable to the order parameter correlation length $\xi$ even if we are not close to the critical temperature. 
In this regime, we expect the orbital effects of the field to strongly affect Cooper pair propagation, which is not well-captured by our mean-field treatment of order parameter dynamics. 

\subsection{Superconducting side}
Analogous to the metallic side, one can carry out a computation of the noise on the superconducting side arising from Gaussian order parameter fluctuations on top of the mean-field pairing amplitude $\langle |\psi_0| \rangle$. 
However, 2D superconductors cannot effectively screen a three dimensional magnetic field via screening currents restricted to the plane. 
Thus, at a finite magnetic field, the system necessarily nucleates vortices, whose dynamics introduce qualitatively new sources of magnetic noise that are not captured by the linearized time-dependent Ginzburg–Landau theory. 
Therefore, to describe the experimentally relevant behavior for a 2D superconducting material in a magnetic field, we next develop a complementary framework that incorporates vortex motion and evaluates the magnetic noise arising from their dynamics.

\section{Magnetic noise from vortex dynamics}
We now consider type-II superconductors placed in an external field such that $H_{c1} < H < H_{c2}$, where the magnetic flux penetrates the superconducting material in quanta of $\Phi_0 = h/e^*$ via vortices in the superconducting order parameter. 
As discussed previously, we will consider the vortices as classical point defects for analytical convenience. 
Our approximation is accurate in the extreme type-II limit where the Ginzburg-Landau parameter $\kappa$ --- the  ratio of the London penetration depth $\lambda_L$ (Pearl length $\Lambda$ in 2D~\cite{pearl1964current}) to the superconducting coherence length $\xi$ is large. 
Within this limit, we will consider three distinct regimes of vortex dynamics: (i) a single pinned vortex due to impurities (e.g., in a vortex glass phase), (ii) collective phonon modes of an Abrikosov vortex lattice~\cite{abrikosov1957magnetic,abrikosovRMP}, and (iii) diffusive motion of vortices in a vortex liquid phase obtained by melting the aforementioned vortex lattice. 
In each case, we will demonstrate that the magnetic noise detected by the spin qubit encodes unique signatures of the relevant dynamics, thereby making it possible to extract physical quantities such as the phonon dispersion and the vortex diffusivity from a careful characterization of the magnetic noise spectrum. 

\begin{figure*}[!htbp] 
\includegraphics[width=1.0\textwidth]{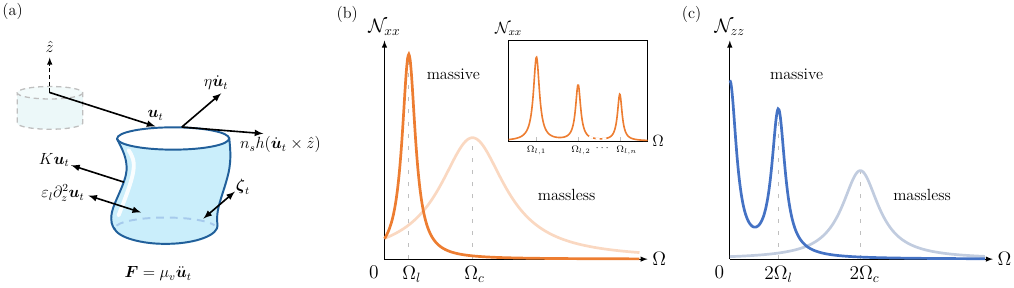}
\caption{Schematics of magnetic noise due to thermal fluctuations of a single pinned vortex in various regimes. (a) Langevin dynamics of a single line vortex, showing the different forces acting on it \cite{BlatterRMP}. (b) Transverse noise response $\mathcal{N}_{xx}(\Omega)$ for a given $k_z$ mode. In the weak damping limit, the response peaks at $\Omega_c \approx \overline{K}(k_z) / (n_sh)$ for a massless vortex, and at $\Omega_l=(\overline{K}(k_z) / \mu_v)^{\frac12}$ for a massive vortex.
The linewidth of the peak decreases for heavier vortex mass.
Inset shows the resonant peaks in $\cN_{xx}$ at ($\Omega_{l,1},\Omega_{l,2},\ldots,\Omega_{l,n}$), expected when multiple $k_z$ modes are thermally populated. (c) Longitudinal noise response $\mathcal{N}_{zz}(\Omega)$: the response peaks at $2\Omega_c$ for a massless vortex and peaks at $\Omega = 0$ and $2\Omega_l$ for a massive vortex.}
\label{fig:SVD}
\end{figure*}

\subsection{Fluctuations of a pinned vortex}
To build intuition, we first consider magnetic noise due to a single vortex pinned by an impurity potential. 
This is physically relevant in a vortex glass phase~\cite{MFisher89,fisherfisherhuse} when the qubit-sample distance $z_0$ is much smaller than the inter-vortex spacing, so that the qubit-probe has only one pinned vortex within its field of view.
For simplicity, we will assume that the pinning potential is harmonic, i.e., $V_{\rm pin}(\rho) = K \rho^2/2$, and that the vortex is in local thermal equilibrium with its surroundings. 
Thermal fluctuations induce vortex motion, leading to a time-varying magnetic field directly above it and resulting in magnetic noise. 
If the qubit-sample distance $z_0$ is much smaller than the superconducting coherence length $\xi$, then these fluctuations are mostly sourced by metallic currents in the core and reflect the non-local conductivity of the metallic state, considered in Ref.~\onlinecite{Agarwal2016}. 
Therefore, we will focus on the opposite limit $z_0 \gg \xi$, where the magnetic fluctuations are instead sourced by circulating super-currents that get displaced due to the motion of the vortex. 

To evaluate the magnetic noise from the motion of a single vortex, we will make a few physically motivated assumptions. 
First, we will assume that the strength of the pinning potential $K$ is larger than the thermal energy scale $k_B T$. 
Consequently, the amplitude of thermal oscillations of the vortex, set by $\sqrt{k_B T/K}$~\footnote{This is a consequence of the equipartition theorem, which continues to hold for the potential energy even in the presence of a Magnus force.}, is much smaller than the relevant screening lengthscale $\lambda_{\rm scr}$ (penetration depth $\lambda_L$ or Pearl length $\Lambda$).
This allows us to recast the magnetic noise tensor in terms of correlations of the vortex displacement $\u_t$ via a Taylor series expansion of the fluctuating magnetic field $\B(\r_q,t)$ at the qubit location in the small parameter $\u_t/\lambda_{\rm scr}$.
Second, we will consider the spin qubit to be located directly above the equilibrium position of the vortex (assumed to be the origin), i.e., $\r_q = (0,0,z_0)$. 
This leads to a different qualitative distance ($z_0$) dependence of $\mathcal{N}_{zz}$ and $\mathcal{N}_{xx}$ as the $z$-component of the field has a maxima when the vortex is at the origin and therefore varies more slowly than the radial components. 
Finally, we will also assume that the vortex motion is entirely characterized by its in-plane displacement, i.e., $\u_t$ does not depend on the $z$-coordinate. 
This is true by definition for 2D Pearl vortices, and holds for rigid (rod-like) Abrikosov vortices when the $z$-axis elastic constant $\varepsilon_l$ is large. 
Later, we will study the consequences of relaxing this assumption by including undulations of the vortex axis. 

Using these assumptions, we may write the time-varying magnetic field at the qubit as
\begin{align}
    \B(\r_q, t) &= \B_0(|\r_q - \u_{t}|) \nonumber \\ & \simeq \B_{0} - (\partial_{\rho} B_0^\rho)_{\rho=0}\; \u_t + \frac{1}{2}(\partial_{\rho}^2 B_0^z)_{\rho=0} \; |\u_t|^2 \hat{z} \,, \nonumber \\
   \text{ where  } & \partial_{\rho} B_0^\rho|_{\rho=0} = \frac{1}{2} \int \frac{d^2\k}{(2\pi)^2} \,  k^2 e^{- k z_0} \phi_M(\k) \, , \text{ and } \nonumber \\ 
    & \partial_{\rho}^2 B_0^z|_{\rho=0} = \frac{1}{2} \int \frac{d^2\k}{(2\pi)^2} \,  k^3 e^{- k z_0} \phi_M(\k).
\label{eq:Bseries}
\end{align}
In Eq.~\eqref{eq:Bseries}, $B_0^\rho$ and $B_0^z$ denote the in-plane and out-of-plane components of the magnetic field respectively, defined for a Abrikosov (rigid 3D) or Pearl (2D) vortex via its magnetostatic potential $\phi_M$ in Eq.~\eqref{eq:PhiM3D} or Eq.~\eqref{eq:PhiM2d} (See Appendix~\ref{app:C_mag} for derivation).
Accordingly, the magnetic noise tensor components are given by
\begin{equation}
\begin{split}
    & \cN_{zz}(\Omega) \simeq \frac{1}{4}(\partial_{\rho}^2 B_{0}^{z})_{\rho=0}^2 \int_{-\infty}^\infty dt\, \ e^{i\Omega t} \expect{|\u_t|^2 |\u_{0}|^2} \, ,  \\
    & \cN_{xx}(\Omega) = \cN_{yy}(\Omega) \simeq \frac{1}{2} (\partial_{\rho} B_{0}^{\rho})_{\rho=0}^2 \int_{-\infty}^\infty dt\, \ e^{i\Omega t}  \expect{\u_t \cdot \u_{0}} \,.
\end{split}\label{eq:Npinned}
\end{equation}
The distance ($z_0$) dependence of the noise tensor is therefore determined by the radial derivatives of $\B_0$ at the origin, while the frequency dependence encodes the dynamical correlations of the vortex displacement. 

Next, we relax the assumption of rigid vortices in the 3D case, and solve the Maxwell-London equations in the bulk without assuming translation invariance in the $z$-direction. 
Remarkably, we can still find an analytical expression for the fluctuating magnetic field due to an undulating vortex with $\u_t(z) = \u_t \, e^{i k_z z}$ as a power series (see Appendix~\ref{app:C_mag} for a derivation):
\begin{align}\label{mag_k_z_modes}
&\B(\r_q,t) = \int \frac{d^2\k}{(2\pi)^2} (- i \k, k) e^{i \k \cdot \bm{\rho}_q} e^{-k z_0} \phi(\k, k_z), \text{ where } \nonumber \\
& \phi(\k, k_z) = \sum_{m=0}^{\infty} \frac{[- i \k \cdot \u_t(0)]^m}{m!} \frac{\Phi_0 }{k \left(1 + \lambda_L^2[k^2 + (mk_z)^2] \right)}  ,
\end{align}
and $\rr_q$ is the in-plane coordinate of the qubit chosen as the $(0,0)$ henceforth, 
and we have chosen periodic boundary condition for the axial oscillations with wave-vector $k_z \in 2\pi \mathbb{Z}/d$, with $d \gg z_0$ being the sample thickness. 
When $k_z = 0$, the series can be summed exactly, i.e., $\phi(\k,0) = e^{-i \k \cdot \u_t}$, such that
\begin{align}
\B(\r_q,t) &= \Phi_0  \int \frac{d^2\k}{(2\pi)^2} (- i \k, k) \frac{e^{i \k \cdot (\bm{\rho}_q - \u_t )} e^{-k z_0}}{k(1 + \lambda_L^2 k^2)} \nonumber \\ &= \B_0(|\r_q - \u_t|),
\end{align}
recovering the magnetic field for rigid Abrikosov vortices. 
Further, we note that the magnetic potential $\phi(\k,k_z)$ has Fourier components at precisely the harmonics of $k_z$, the oscillation wave-vector of the vortex in the $z$-direction.
These harmonics serve to suppress the magnetic field above the vortex, which may be intuitively understood as partial cancellation of the field arising from different segments of the vortex which are slightly misaligned along its axis.  

Once again, to make analytical progress, we assume that we can expand in powers of $\u_t(z=0)/\lambda_L$, and consider only terms up to the lowest order time-dependent harmonics for $B^\rho$ ($m$ = 1) and $B^z$ ($m$ = 2),  
\begin{align}
&\B(\r_q,t) = \B_0(\r_q) - \left[ \frac{\Phi_0}{2} \int \frac{d^2\k}{(2\pi)^2} \frac{ k \, e^{- k z_0}}{1 +  \lambda_L^2(k^2 + k_z^2)} \right] \u_t(0) \nonumber \\
& ~~~~~~~~~~~~+ \hat{z} \left[ \frac{\Phi_0}{4} \int \frac{d^2\k}{(2\pi)^2} \frac{ k^2 e^{- k z_0}}{1 +  \lambda_L^2(k^2 + 4 k_z^2)} \right] \u_t^2(0)  + \ldots \nonumber \\
 &\equiv \B_0(\r_q) - (\partial_{\rho} B^{\rho})_{\rho=0}\; \u_t(0) + \frac{1}{2}(\partial_{\rho}^2 B^{z})_{\rho=0} \; |\u_t(0)|^2 \hat{z} \,.
\label{eq:Bkz}
\end{align}
The first term in Eq.~\eqref{eq:Bkz} is simply the static magnetic field of the rigid 3D vortex, while the latter two terms physically capture the effect of finite $k_z$ by accounting for the suppression of $B^{\rho}$ and $B^z$ respectively. 
The final expression for the magnetic field takes the same form as Eq.~\eqref{eq:Bseries}, where $\u_t(0)$ now denotes the coordinate of the end-point of the vortex at the surface ($z=0$), with the caveat that we need to account for the suppression of the field at non-zero $k_z$, particularly if $k_z \lambda_L$ is large.
Therefore, in both cases, to determine the magnetic noise due to the thermal motion of a pinned vortex, the task at hand is to calculate the correlation functions of its in-plane displacement $\u_t$, which we turn to next.

The Lagrangian for a point-like vortex with mass $\mu_v$ (per unit length in 3D) is given by~\cite{BlatterRMP}:
\begin{align}
\mathcal{L} &= \frac{\mu_v}{2} (\dot{\u}_t)^2 + \bm{f}_{\rm M} \cdot \u_t - \mathcal{H}, \nonumber \\ 
\mathcal{H} &= \mathcal{H}_{\rm tension} + \mathcal{H}_{\rm pinning} = \frac{\varepsilon_l}{2} \left( \frac{\partial \u_t}{\partial z} \right)^2 + \frac{K \u_t^2}{2} ,
\end{align}
where $\bm{f}_{\rm M} = n_s h (\dot{\u}_t \times \hat{z})$ corresponds to the Magnus force on the vortex, $n_s$ is the superfluid density, $\varepsilon_l$ is the line tension of the vortex in the $z$-direction, (see Appendix~\ref{app:E_self_energy} for a derivation of $\varepsilon_l$), and we have assumed that the spring constant $K$ of the pinning potential is independent of $z$ for simplicity. 
One can subsequently derive the following equation of motion for the vortex displacement $\u_t(z)$:
\begin{equation}
    \mu_v \ddot{\u}_t = n_s h (\dot{\u}_t \times \hat{z}) -K \u_t + \varepsilon_l \partial_z^2 \u_t - \eta \dot{\u}_t + \bm{\zeta}_t \,,
    \label{eq:eom_pinned}
\end{equation}
where we have added a phenomenological damping term characterized by the drag coefficient $\eta$.
In the Bardeen-Stephen model~\cite{BardeenStephen65} which considers dissipation arising from the normal metallic core of a vortex in a fully gapped $s$-wave superconductor, $\eta =\Phi_0 \mu_0 H_{c 2}/\rho_n$ (where $\rho_n$ is the normal-state resistivity), but here we will simply treat $\eta$ as a phenomenological drag coefficient.
Since the dissipative drag would damp the motion of the vortex, we have also added a stochastic force $\bm{\zeta}_t$ which provides thermal kicks to keep it in motion, as required for non-zero magnetic field fluctuations at long times. 
For simplicity, we assume the stochastic force to be uncorrelated in time, with an amplitude determined by the FDT: 
\begin{equation}
\langle{\zeta^\alpha_t \zeta^\beta_{t'}\rangle} = 2\eta k_B T \, \delta^{\alpha \beta } \, \delta(t - t')  .
\label{eq:WN}
\end{equation}
In 3D, the stochastic force per unit length $\zeta_t(z)$ depends on $z$. 
In this case, we further assume that $\zeta_t(z)$ is uncorrelated in the $z$-direction, implying that Eq.~\eqref{eq:WN} gets modified to:
\begin{equation}
\langle{\zeta^\alpha_t(z) \zeta^\beta_{t’}(z')\rangle} = 2 \eta k_B T \, \delta^{\alpha \beta } \, \left(  \frac{\delta^{z,z'} }{a_c} \right) \delta(t - t’)  ,
\label{eq:WN3D}
\end{equation}
where we have introduced the spacing $a_c$ between superconducting planes in the $z$-direction, and used a lattice discretization of the Dirac delta function $\delta^{z,z'} = a_c \, \delta(z - z')$ such that Eq.~\eqref{eq:WN3D} is dimensionally consistent.
A schematic of the vortex motion, with all the different forces acting on it, is shown in Fig.~\ref{fig:SVD}(a).

To solve the equation of motion and find the desired correlations, we focus on the more general case of an Abrikosov (3D) vortex which can undulate in the $z$-direction.
Specifically, we consider the following Fourier-space ansatz for the displacement
\begin{equation}\label{eq:u_Fourier}
\u_{t}(z) = \u_{\Omega,k_z} e^{i ( k_z z- \Omega t)}    .
\end{equation}
Eq.~\eqref{eq:eom_pinned} can then be recast as a matrix equation 
\begin{align}\label{eq:eom_frequency}
& \left[M_{\Omega, k_z}^{-1}\right]_{\alpha \beta} u^\beta_{\Omega,k_z} = \zeta^\beta_{\Omega,k_z}, \text{ where }  M_{\Omega,k_z}^{-1} = \nonumber \\
& \begin{pmatrix}
        - \mu_v \Omega^2 - i\eta\Omega + \overline{K}(k_z) & i n_s h \Omega \\
        -i n_s h \Omega & - \mu_v \Omega^2 - i\eta\Omega + \overline{K}(k_z)
    \end{pmatrix}  ,
\end{align}
and $\overline{K}(k_z) = K + \varepsilon_l k_z^2$ is the effective spring constant. Thereafter, we can write down the correlation function by inverting the matrix equation and using Eq.~\eqref{eq:WN} (see Appendix~\ref{app:pinned} for details): 
\begin{equation}
\langle u^\alpha_{\Omega,k_z} u^\beta_{-\Omega,-k_z} \rangle = 2 (\eta/a_c) k_B T [M_{\Omega, k_z} M^\dagger_{\Omega, k_z}]^{\alpha \beta}.
\label{eq:uuCorr}
\end{equation}
Using Eqs.~\eqref{eq:Npinned} and \eqref{eq:uuCorr} and substituting $\partial_{\rho} B_{0}^{\rho} \to \partial_{\rho} B^{\rho}$ to account for $k_z \neq 0$, we arrive at the following expression for $\cN_{xx}$:
\begin{equation}
\cN_{xx}(\Omega) =  (\partial_{\rho} B^{\rho})_{\rho=0}^2 \, (\eta/a_c) k_B T  \, \Tr[M_{\Omega, k_z} M^\dagger_{\Omega, k_z}]  .
\end{equation}
Evaluating the trace leads to the following expression for $\cN_{xx}(\Omega)$ (see Appendix~\ref{app:pinned} for derivation):
\begin{widetext}
    \begin{equation}
    \cN_{xx}(\Omega) = \left(\partial_\rho B_{0}^{\rho}\right)_{\rho=0}^2 \left(\frac{2 \eta k_B T}{a_c}\right)\frac{[\overline{K}(k_z)-\mu_v\Omega^2]^2+[\eta^2 + (n_s h)^2] \Omega^2 }{\left\{[\overline{K}(k_z)-\mu_v\Omega^2]^2-[\eta^2 + (n_s h)^2]\Omega^2 \right\}^2+4 \eta^2\Omega^2 [\overline{K}(k_z)-\mu_v\Omega^2]^2}.
    \label{eq:Nxx_pinned}
\end{equation}
\end{widetext}
Since the full expression for $\cN_{xx}(\Omega)$ is cumbersome to interpret, we next focus on certain simple physical limits [depicted in Fig.~\ref{fig:SVD}(b)].

First, in the low-frequency limit $\Omega \to 0$ or the nearly massless vortex limit $\mu_v \to 0$, the inertial $\mu_v \Omega^2$ term may be neglected, and the Magnus force dominates vortex motion. Thus, the resonant modes are circularly polarized with $\Omega_{c} = \overline{K}(k_z)/(n_sh)$ for weak damping $\eta$.
Consequently, $\cN_{xx}(\Omega)$ takes the form of a Lorentzian peaked at $\Omega = \pm \Omega_{c}$, with a linewidth set by the viscous damping $\eta$ (which is assumed to be small),  
\begin{equation}
\cN_{xx}(\Omega) =  (\partial_{\rho} B^{\rho})_{\rho=0}^2 \, \left[ \frac{2 (\eta/a_c) k_B T (\Omega_c^2 + \Omega^2)}{(n_s h)^2 (\Omega^2 - \Omega_c^2)^2 + 4 \eta^2 \Omega^2 \Omega_c^2 } \right]   .
\end{equation}
We note that as $\eta \to 0$, the Lorentzian approaches a Dirac delta function. Accordingly, the spin qubit is able to find the resonant mode frequencies at different $k_z$, by showing a sharp response when its frequency $\Omega$ approaches $\Omega_{c}$.
This information may be used to extract the strength $K$ of the pinning potential, the vortex-line tension $\varepsilon_l$, and the superfluid density $n_s$.

Second, in the massive vortex limit or if the superfluid density $n_s$ is low, we may neglect the Magnus force on the vortices, and keep only the inertial term. 
Thus, the resonant mode is linearly polarized and features a distinct dispersion $\Omega_l = (\overline{K}(k_z)/\mu_v)^{\frac12}$, and $\cN_{xx}(\Omega)$ remains Lorentzian but peaks at different resonance frequencies, 
\begin{equation}
\cN_{xx}(\Omega) =  (\partial_{\rho} B^{\rho})_{\rho=0}^2 \, \left[ \frac{2 (\eta/a_c) k_B T }{\mu_v^2 (\Omega^2 - \Omega_l^2)^2 + \eta^2 \Omega^2  } \right]  .
\end{equation}
In this case, the sharp increase in the relaxation rate of the spin qubit on tuning its frequency $\Omega \to \Omega_l$ may be used to identify the strength $K$ of the pinning potential, the vortex-line tension $\varepsilon_l$, and the vortex mass $\mu_v$.

We now consider the limit when the viscous damping $\eta$ is large, such that we may neglect both the Magnus force $n_s h \Omega$ and the inertial term $\mu_v \Omega^2$. 
In this limit, $\cN_{xx}$ takes a particularly simple form:
\begin{equation}
\cN_{xx}(\Omega) =  (\partial_{\rho} B^{\rho})_{\rho=0}^2 \, \left[ \frac{2 (\eta/a_c) k_B T}{ \overline{K}(k_z)^2  + \eta^2 \Omega^2} \right]  .
\end{equation}
We note that the resonant peaks are washed out by strong viscous damping.
Nevertheless, even in this limit, we may use the magnetic noise to extract $K$ and $\varepsilon_l$, provided one can excite the different $k_z$ modes. 

We conclude our discussion of $\cN_{xx}$ for an isolated pinned vortex by noting that our results for a line vortex modulated by a single longitudinal mode $k_z$ can be generalized straightforwardly to a superposition of modes. 
Assuming periodic boundary conditions in the $z$-direction for simplicity, we expand the displacement $\u_t(z)$ in terms of a complete set of eigenmodes $\u_{tM}$, where the $M^{\rm th}$ mode has $k_z= 2\pi M/d$:
\begin{equation*}
\u_t(z)=\sum_M \u_{tM} e^{i k_M z}.
\end{equation*}
Then, to leading order in the displacement, we may write the deviation of the in-plane component of the magnetic field (which contributes to the transverse noise) from its equilibrium value as
\begin{align}
\delta \B_{\|}(\r_q,t) &= -\sum_M \left[ \frac{\Phi_0}{2} \int \frac{d^2\k}{(2\pi)^2}  \frac{k \, e^{-k z_0}}{1+\lambda_L^2\left(k^2+k_M^2\right)} \right] \u_{tM}.
\label{eq:Bkz_2}
\end{align}
Within the Langevin dynamics described in Eqs.~\eqref{eq:eom_pinned} and \eqref{eq:WN}, the modes $\u_{tM}$ are decoupled from one another for different $k_M$.
As a result, $\langle{u^\alpha_{t M} u^\beta_{t' M'} \rangle} \propto \delta_{MM'}$.
The total magnetic noise therefore decomposes into a sum of independent modal contributions, with each mode contributing a noise spectrum given by Eq.~\eqref{eq:Nxx_pinned}.
In particular, this implies that $\cN_{xx}(\Omega)$ features a series of resonant peaks corresponding to the different $k_z$ modes as the qubit frequency $\Omega$ is tuned, provided the viscous damping $\eta$ is weak [Fig.~\ref{fig:SVD}(b), inset] (for further details, we refer the reader to Appendix~\ref{app:pinned}, specifically Eq.~\eqref{app:noise_multi_modes}).

So far, our discussion has focused primarily on the noise spectrum $\cN_{xx}(\Omega)$.
Analogous information about vortex dynamics can also be extracted from $\cN_{zz}(\Omega)$.
Unlike $\cN_{xx}$, however, $\cN_{zz}$ depends on the correlation function of the squared vortex displacements, $\expect{|\u_t|^2 |\u_0|^2}$ according to Eq.~\eqref{eq:Npinned}.
This quantity is not independent: it can be expressed entirely in terms of the displacement correlators in Eq.~\eqref{eq:uuCorr}.
Explicitly,
\begin{equation}
\expect{|\u_t|^2 |\u_0|^2} = \expect{|\u|^2}^2 + \sum_{\alpha, \beta} 2\expect{u^{\alpha}_t u^\beta_0}^2 \,.
\label{eq:u2u2corr_wick}
\end{equation}
In Eq.~\eqref{eq:u2u2corr_wick}, the first term on the RHS is independent of time, and has no spectral weight at any non-zero frequency. 
The second term, on the other hand, is the square of the two-point correlator and has spectral weight concentrated on twice the resonant frequency in the weak damping limit.
Thus, $\cN_{zz}(\Omega)$ probes the same vortex dynamics through higher-order correlations, yielding peaks at twice the resonant frequencies with approximately double the linewidth of the peaks of $\cN_{xx}(\Omega)$.

To make this connection explicit, we again focus on three distinct analytically tractable regimes of $\cN_{zz}(\Omega)$ [Fig.~\ref{fig:SVD}(c)] (for the full integral expression, see Eq.~\eqref{app:N_zz_general} in Appendix~\ref{app:pinned}).
We first consider the massless vortex limit ($\mu_v \to 0$).
In this regime, the magnetic noise takes the form
\begin{widetext}
\begin{equation}
\label{eq:Nzz_pinned_massless}
\cN_{zz}(\Omega) =  (\partial_{\rho}^2 B^{z})_{\rho=0}^2 \frac{2\eta}{\tilde\eta^2\overline{K}(k_z)}\left(\frac{k_B T}{a_c}\right)^2 \left[\frac{1}{(\Omega-2\Omega_c)^2+\left(2 \eta\overline{K}(k_z)/\tilde\eta^2\right)^2} + \frac{1}{(\Omega+2\Omega_c)^2+\left(2 \eta\overline{K}(k_z)/\tilde\eta^2\right)^2}  \right] \,,
\end{equation}
revealing resonant peaks at $\Omega=\pm 2\Omega_c$, where $\Omega_c=\overline{K}(k_z)/(n_s h)$ for weak damping ($\eta \ll n_s h$) denotes the resonant frequency of the circularly polarized vortex modes.

Second, we consider the massive vortex limit, corresponding to low superfluid density $n_s$, where the Magnus force may be neglected.
In this regime, the noise spectrum can be approximated by the following analytical expression with resonant features appearing at $\Omega=\pm 2\Omega_l$, where $\Omega_l=(\overline{K}(k_z) / \mu_v)^{\frac12}$, in the limit of weak damping ($\eta \ll \mu_v \Omega_l$):
\begin{equation}
\label{eq:Nzz_pinned_massive}
\cN_{zz}(\Omega) \approx  (\partial_{\rho}^2 B^{z})_{\rho=0}^2 \frac{\eta}{2\mu_v}\left(\frac{k_B T}{a_c\overline{K}(k_z)}\right)^2\left[\frac{1}{\left(\Omega-2 \Omega_l\right)^2+(\eta/\mu_v)^2}+\frac{1}{\left(\Omega+2 \Omega_l\right)^2+(\eta/\mu_v)^2}+\frac{2}{\Omega^2+(\eta/\mu_v)^2}\right] .
\end{equation}

Finally, in the damping dominated regime ($\eta \gg n_s h, \mu_v \Omega$), the noise spectrum simplifies to
\begin{equation}
\label{eq:Nzz_pinned_damping}
\cN_{zz}(\Omega) =  (\partial_{\rho}^2 B^{z})_{\rho=0}^2 \frac{4}{\eta\overline{K}(k_z)}\left(\frac{k_B T}{a_c}\right)^2 \frac{1}{\Omega^2+(2 \overline{K}(k_z)/\eta)^2} \,,
\end{equation}
which shows that in the limit of strong damping, the resonant peaks are washed out.
\end{widetext}

In analogy with $\cN_{xx}(\Omega)$, we observe that the resonant mode frequencies and associated physical quantities, such as the strength of the pinning potential $K$, the vortex line tension $\varepsilon_l$, the vortex mass $\mu_v$ and the superfluid density $n_s$ may also be extracted from $\cN_{zz}(\Omega)$ in the weak damping regime. 
We further note from \cref{eq:Nzz_pinned_massless,eq:Nzz_pinned_massive,eq:Nzz_pinned_damping} that the explicit temperature dependence of $\cN_{zz}$ is markedly different from $\cN_{xx}$, which may be used to isolate these two components.  
This distinct temperature dependence arises from the fact that the $z$-component of the magnetic field is maximum directly above the vortex, where the spin qubit is placed.
As a result, $\cN_{zz}$ probes the correlations of the square of the vortex displacement $\u_t^2$, and scales as $\langle u^2 \rangle^2 \propto (k_B T)^2$ via equipartition theorem. 
Additionally, there are implicit temperature dependences via the drag $\eta$, the vortex line tension $\varepsilon_l$ and possibly the pinning potential $K$, but these are expected to be relatively weak at low temperatures. 

\begin{figure*}[t]
    \centering
    \includegraphics[width=1\linewidth]{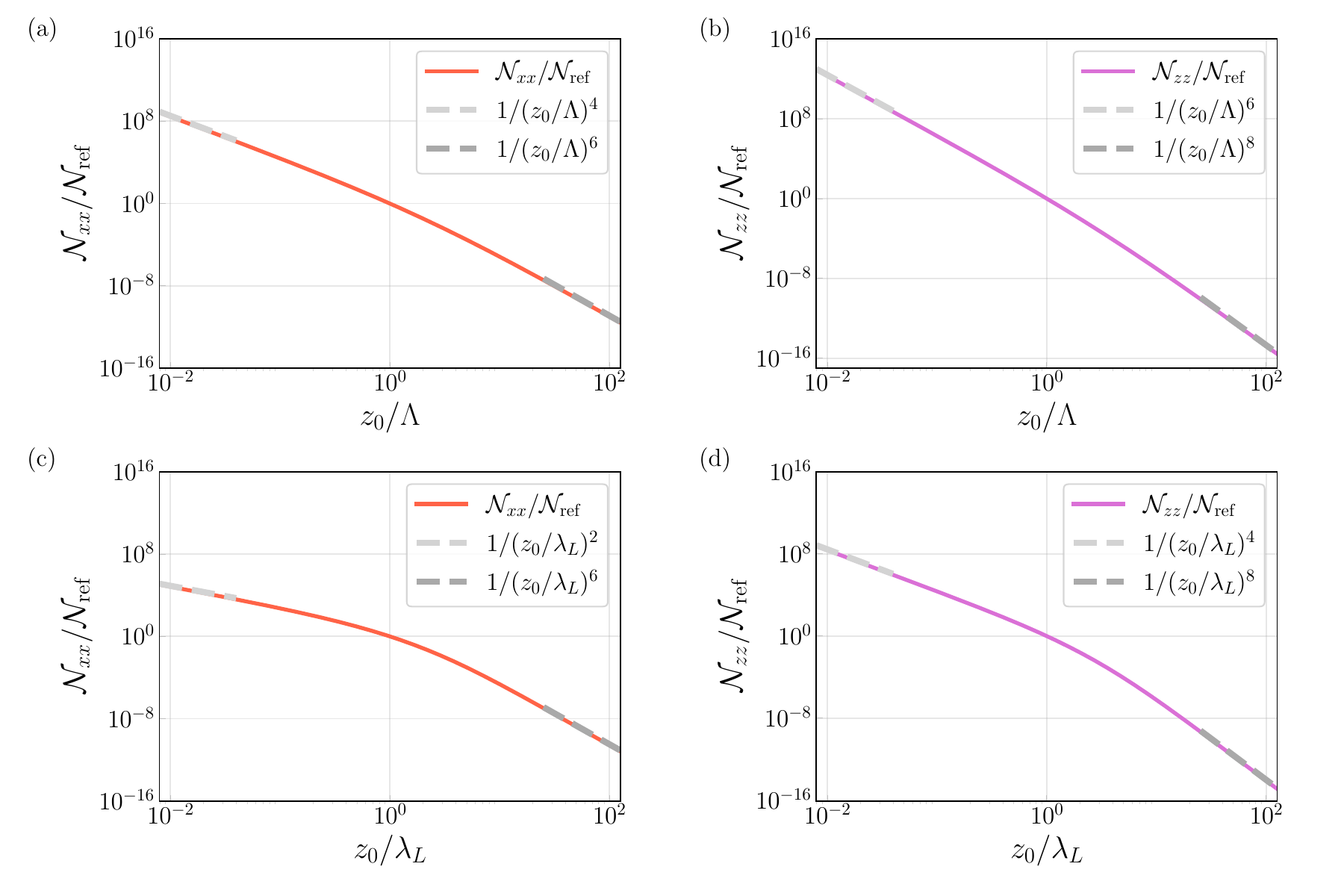}
    \caption{Transverse (longitudinal) noise $\mathcal{N}_{xx}$ ($\mathcal{N}_{zz}$) as a function of qubit-sample distance $z_0$ for a single fluctuating vortex. 
    (a),~(b) Characteristic noise from a Pearl (point) vortex, with asymptotes following Eq.~\eqref{eq:N2DpinnedNFLim} and Eq.~\eqref{eq:N2DpinnedFFLim}. (c),~(d) Characteristic noise from an Abrikosov (line) vortex, with asymptotes following Eq.~\eqref{eq:N3DpinnedNFLim} and Eq.~\eqref{eq:N3DpinnedFFLim}. In each panel, the reference noise is chosen as $\mathcal{N}_{\rm ref}=\mathcal{N}(z_0 = \lambda_{\rm scr})$. }
    \label{fig:Noise_z_dependence}
\end{figure*}

To conclude this section, we discuss the dependence of the noise on the qubit-sample distance $z_0$.
To this end, we explictly evaluate $\partial_{\rho} B_{0}^{\rho}$ and $\partial_{\rho}^2 B_0^{z}$ at the coordinate origin $\bm{\rho} = 0$.
For 2D (Pearl) vortices, $\B = \B_0$ by definition, as the vortex is a point defect with no notion of $k_z$. 
The dependence of $\cN_{xx}$ and $\cN_{zz}$ on $z_0$ can be found by evaluating the integrals in Eq.~\eqref{eq:Bseries} and subsequently using Eq.~\eqref{eq:Npinned}: here we focus on two relevant limits (for the full expressions, see Appendix~\ref{derivative_z_dependence}).
If the qubit-sample distance $z_0$ is much smaller than the Pearl length $\Lambda$, then 
\begin{align}
\cN_{xx}(z_0) & \propto (\partial_{\rho} B_{0}^{\rho})_{\rho = 0}^2 \xrightarrow[]{z_0/\Lambda \ll 1} \left( \frac{\Phi_0}{4 \pi \Lambda } \right)^2 \frac{1}{z_0^4}, \text{ and } \nonumber \\
\cN_{zz}(z_0) & \propto (\partial_{\rho}^2 B_0^{z})_{\rho = 0}^2 \xrightarrow[]{z_0/\Lambda \ll 1}   \left( \frac{\Phi_0}{2 \pi \Lambda}  \right)^2 \frac{1}{z_0^6}.
\label{eq:N2DpinnedNFLim}
\end{align}
By contrast, if the qubit-sample distance $z_0$ is much larger than the Pearl length $\Lambda$, then 
\begin{align}
\cN_{xx}(z_0) & \propto (\partial_{\rho} B_{0}^{\rho})_{\rho = 0}^2 \xrightarrow[]{z_0/\Lambda \gg 1} \left( \frac{\Phi_0}{2 \pi} \right)^2 \frac{1}{z_0^6}, \text{ and } \nonumber \\
\cN_{zz}(z_0) & \propto (\partial_{\rho}^2 B_0^{z})_{\rho = 0}^2 \xrightarrow[]{z_0/\Lambda \gg 1}   \left( \frac{3\Phi_0}{2 \pi} \right)^2 \frac{1}{z_0^8}.
\label{eq:N2DpinnedFFLim}
\end{align}
We note that the distinct $z_0$ dependence of $\cN_{xx}$ and $\cN_{zz}$ can be used to detect an oscillating pinned vortex, even when it cannot be resolved using static magnetic field measurements by the qubit sensor. 
Further, $\cN_{xx}$ decays more slowly with qubit-sample distance $z_0$ than $\cN_{zz}$ in both the near-field and far-field scenarios, potentially making $\cN_{xx}$ easier to measure in experiments.
We confirm these asymptotic scalings by an exact numerical evaluation of the magnetic noise for a fluctuating Pearl vortex in Fig.~\ref{fig:Noise_z_dependence}(a), (b).

Next, we consider Abrikosov vortices, which are line defects, and find the general expression for vortex undulations along its axis ($k_z \neq 0$).
Since the analytical expressions for the derivatives of the field in Eq.~\eqref{eq:Bkz} are cumbersome, we relegate these expressions to Appendix~\ref{derivative_z_dependence}, and discuss the limiting cases for the noise tensor components. 
If the qubit-sample distance $z_0$ is much smaller than the London penetration A few comments are in order.depth $\lambda_L$, then 
\begin{align}
 \cN_{xx}(z_0) & \propto (\partial_{\rho} B_0^{\rho})_{\rho = 0}^2 \xrightarrow[]{z_0/\lambda_L \ll 1} \left( \frac{\Phi_0}{4 \pi \lambda_L^2} \right)^2 \frac{1}{z_0^2}, \text{ and } \nonumber \\
\cN_{zz}(z_0) & \propto (\partial_{\rho}^2 B_0^{z})_{\rho = 0}^2 \xrightarrow[]{z_0/\lambda_L \ll 1}   \left( \frac{\Phi_0}{4 \pi \lambda_L^2} \right)^2 \frac{1}{z_0^4}.
\label{eq:N3DpinnedNFLim}
\end{align}
On the other hand, if the qubit-sample distance $z_0$ is much larger than the London penetration depth $\lambda_L$, then 
\begin{equation}
    \begin{split}
        \cN_{xx}(z_0) & \propto (\partial_{\rho} B_0^{\rho})_{\rho = 0}^2 \xrightarrow[]{z_0/\lambda_L \gg 1} \left( \frac{\Phi_0}{2 \pi[1 + (\lambda_L k_z)^2]} \right)^2 \frac{1}{z_0^6}, \\
\cN_{zz}(z_0) & \propto (\partial_{\rho}^2 B_0^{z})_{\rho = 0}^2 \xrightarrow[]{z_0/\lambda_L \gg 1}   \left( \frac{3\Phi_0}{2 \pi[1 + (2\lambda_L k_z)^2]} \right)^2 \frac{1}{z_0^8}.  \\
    \end{split}\label{eq:N3DpinnedFFLim}
\end{equation}
Similar to the 2D case, $\cN_{zz}$ is suppressed relative to $\cN_{xx}$ as a function of $z_0$, although the specific power laws differ in the near-field limit $z_0 \ll \lambda_L, \Lambda$ owing to the weaker screening of electromagnetic fields in 2D. 
Once again, these asymptotic scalings for a fluctuating Abrikosov vortex are confirmed via exact numerical evaluations of the noise in Fig.~\ref{fig:Noise_z_dependence}(c), (d).

A few comments are in order. 
First we note from Eq.~\eqref{eq:N3DpinnedNFLim} that the finite $k_z$ effects are irrelevant when the qubit-sample distance $z_0$ is much smaller than the London penetration depth $\lambda_L$.
This fits our expectations that electromagnetic screening effects do not play an important role in determining the magnetic noise when $z_0$ is much smaller than $\lambda_L$, since the qubit approximately probes the field fluctuations within an area $\sim z_0^2$ on the sample. 
Second, we observe that in the far-field limit $z_0 \gg \lambda_L$ (Eq.~\eqref{eq:N3DpinnedFFLim}), non-zero $k_z$ reduces the magnetic noise due to an overall reduction of the strength of the field, as intuitively expected. 
This reduction is particularly important if $k_z \lambda_L \gg 1$, i.e., the vortex undulates several times along the $z$-axis within the penetration depth, leading to an efficient cancellation of the field from different segments of the vortex. 
Finally, comparing Eqs.~\eqref{eq:N2DpinnedFFLim} and \eqref{eq:N3DpinnedFFLim}, we find that in the far-field limit, the magnetic noise is identical for Pearl vortices and rigid Abrikosov vortices ($k_z = 0$), and does not depend on the electromagnetic screening length $\lambda_{\rm scr}$. 
This happens because at distances $z_0 \gg \lambda_{\rm scr}$, the magnetic field of a vortex resembles that of a point magnetic monopole at the origin, i.e,. $\B_0(\r) = \Phi_0 \hat{\r}/(2\pi r^2)$ in both cases, so its variation as the vortex moves is simply set by this Coulomb-like magnetic field. 
By contrast, comparing Eqs.~\eqref{eq:N2DpinnedNFLim} and \eqref{eq:N3DpinnedNFLim} tells us that in the near-field limit $z_0 \ll \lambda_{\rm scr}$, the field fluctuations scale differently for 2D and rigid Abrikosov vortices, leading to distinct $z_0$ dependencies and potentially providing a route to identifying the dimensionality of the pinned vortex. 

\subsection{Collective modes: Vortex lattice phonons}
Having discussed noise signatures of a single fluctuating vortex, we now consider the noise signatures of the collective modes of a vortex lattice --- the vortex phonon modes. 
To this end, we assume that the vortices form a two-dimensional triangular lattice of lattice constant $a_\Delta$ with primitive vectors $\{\bm{a}_1, \bm{a}_2\}$, such that their equilibrium positions are given by $\R_i = m \bm{a}_1 + n \bm{a}_2$ ($m, n \in \mathbb{Z}$).
The low-energy fluctuations of the vortex lattice correspond to gapless Goldstone modes of the broken continuous translation symmetry. 
To describe these modes, we write the time-dependent displacement of the vortex cores as $\r_i(t) = \R_i + \u_i(t)$, where we again assume that $|\u_i|$ is small compared to the lattice spacing $a_\Delta$.
These fluctuations of the vortex positions lead to a time-dependent magnetic field at the location of the qubit, and contribute to magnetic noise that leads to its depolarization. 

Intuitively, we expect the magnetic noise to depend on the longitudinal fluctuations of vortices (i.e., $\u_\k \,  \| \, \hat{\k}$). 
The simplest way to understand this dependence is to consider the limit of large qubit-sample distance $z_0 \gg \Lambda, \lambda_L, a_\Delta$, where the magnetic  field $\B_0(\r) = \Phi_0 \hat{\r}/(2\pi r^2)$ due to a vortex at the origin resembles that of a magnetic monopole ($z_0 > 0$). 
At large distances, the qubit-sensor is not able to resolve individual vortices, so it senses a sheet of uniform \textit{magnetic} sheet charge, which creates a uniform magnetic field (this is precisely the applied field $\bm{H} = H \hat{z}$). 
Transverse vortex fluctuations with $\u_\k \perp \hat{\k}$ do not change the \textit{magnetic sheet charge density} on the surface of the superconductor.
As a result, the total magnetic field sensed by the qubit does not change with time, and there is no magnetic noise. 
By contrast, longitudinal fluctuations with $\u_\k \, \| \, \hat{\k}$ lead to local variations of the \textit{magnetic charge density} and lead to a time-varying magnetic field. 
Therefore, the magnetic noise should be determined solely by dynamical correlators of longitudinal vortex displacements. 

To substantiate the above argument, we now explicitly compute the magnetic field due to the collective modes of a vortex lattice. 
We assume that we are deep inside the superconducting phase, so that $z_0 \gg \Lambda, \lambda_L$ and further that the field $H$ is large enough so that $z_0 \gg \ell \sim a_\Delta$. 
For simplicity, we choose the qubit to be at $\r_q = (0,0,z_0)$, as we do not expect the transverse coordinates to affect the magnetic noise in the thermodynamic limit. 
Then, the magnetic field (both for 2D Pearl vortices and rigid Abrikosov vortices, we will study the consequences of allowing $z$-axis undulations later) is simply given by the sum of Coulombic fields of each individual vortex: 
\begin{align}
\B_{\rm lat, 0}(\r_q, t) = \frac{\Phi_0}{2\pi} \sum_{i} \frac{\r_q - \r_i(t)}{|\r_q - \r_i(t)|^3},
\end{align}
where the sum runs over all lattice sites $\R_i$. 
To make analytical progress, we now expand $\B(\r_q, t)$ to linear order in $\u_i(t)$, in exact analogy to our previous calculation for the single fluctuating vortex: 
\begin{align}
 \B(\r_q,t) &= \B_{\rm lat, 0}(\r_q) + \delta B_z(\r_q,t) \hat{z} + \delta \B_\|(\r_q,t), \text{ where } \nonumber \\
 \B_{\rm lat, 0}(\r_q) &=\frac{\Phi_0}{2\pi} \left( \sum_i    \frac{z_0}{(z_0^2 + R_i^2)^{3/2}} \right) \hat{z}  ,  \nonumber \\
 \delta B_z(\r_q,t) &= - \frac{\Phi_0}{2\pi} \sum_i  \frac{3 z_0 \R_i \cdot \u_i(t)}{(z_0^2 + R_i^2)^{5/2}}, \, \text{ and }  \nonumber \\
 \delta \B_\|(\r_q,t) &= \frac{\Phi_0}{2\pi} \sum_i  \frac{3 \R_i (\R_i \cdot \u_i(t)) - \u_i(t)\left(z_0^2+R_i^2\right)}{(z_0^2 + R_i^2)^{5/2}}   .
 \label{eq:Blat}
\end{align}
The first term in Eq.~\eqref{eq:Blat} is the static field due to the vortex lattice. 
When $z_0$ is large, we may take the continuum limit $\sum_i \to n_v \int d^2 \rr$ where $n_v = 1/A_{\rm uc} =  2/(\sqrt{3} a_\Delta^2)$ is the vortex density, and explicitly evaluate the integrals.
In this limit, we find that $\B_{\rm lat, 0}(\r_q) = n_v \Phi_0 \hat{z} = \mu_0 \H$, precisely corresponding to the applied external field. 

The second and third terms in Eq.~\eqref{eq:Blat} correspond to the out-of-plane and in-plane fluctuating fields respectively due to vortex motion (to linear order in  displacement). 
These can be conveniently written in Fourier space as (fixing $\r_q = (0,0,z_0)$ for clarity)
\begin{align}
& \delta B_z(t)  = \frac{1}{\sqrt{N_v}} \sum_\k G_z^\mu(\k) \u_{\mu}(\k,t), \nonumber \\ 
& ~~~~~~~~~~~~~~~ G_z^\mu(\k) \approx n_v \Phi_0 (- i k_\mu) e^{- k z_0} ,\nonumber \\
& [\delta \B_\|(t)]^\mu  = \frac{1}{\sqrt{N_v}} \sum_\k G_\|^{\mu \nu}(\k) \u_\nu(\k,t), \nonumber \\ 
& ~~~~~~~~~~~~~~~ G_\|^{\mu \nu}(\k) \approx  n_v \Phi_0 \left( \frac{k_{\mu} k_{\nu}}{k} \right) e^{- k z_0} ,
\label{eq:BlatFourier}
\end{align} 
where $N_v$ is the total number of vortices, and the propagators $G_\|^{\mu \nu}(\k)$ and $G_z^{\mu}(\k)$ have again been evaluated in the same continuum approximation as $\B_{\rm lat, 0}$ (see Appendix~\ref{app:solid} for a detailed derivation). 
Crucially, Eq.~\eqref{eq:BlatFourier} makes explicit our earlier claim: $\delta \B(t)$ depends only on the longitudinal component of the displacement $\hat{\k} \cdot \u_\k$, i.e., transverse vortex motion does not contribute to magnetic noise. 
We note that our approximation allows for analytical progress at the expense of periodicity in the Brillouin zone, i.e., the approximate expressions do not obey the constraint $G_\|^{\mu \nu}(\k) = G_\|^{\mu \nu}(\k + \bm{b}_{\rm rl})$ for any reciprocal lattice vector $\bm{b}_{\rm rl}$ of the vortex lattice.
Nevertheless, at distances $z_0 \gg a_\Delta$, the qubit-sensor probes low-momentum phonon modes with momenta $k a_\Delta \ll 1$: therefore, the deviation of our approximation from the exact value of the propagator at large momenta $|\bm{b}_{\rm rl}| \sim a_\Delta^{-1}$ is not important. 

From Eq.~\eqref{eq:BlatFourier}, one can simply compute the noise tensor components by Fourier transforming in time-domain. 
Extending the summation over $\k$ in the Brillouin Zone to an integral over all $\k$ in the same spirit as the previous continuum approximations, and using translation invariance of the correlation function, we arrive at the following expression for $\cN_{xx}$,
\begin{align}
\cN_{xx}(\Omega) & = \int dt\, \, e^{i \Omega t } \langle \delta B_x(t) \, \delta B_x(0) \rangle  \nonumber \\ 
& = \frac{1}{N_v} \sum_\k G^{x\mu}_\|(\k)  G^{x\nu}_\|(-\k) \int dt\, \, e^{i \Omega t }  \langle u^{\mu}_{\k}(t) u^{\nu}_{-\k}(0) \rangle \nonumber \\
& \to \frac{n_v \Phi_0^2 }{2}  \int \frac{d^2\k}{(2\pi)^2} \langle \k \cdot \u_{\k,\Omega} \; \k \cdot  \u_{-\k,-\Omega} \rangle e^{-2 k z_0}     . 
\label{eq:NxxVL}
\end{align}
Similarly, we can write $\cN_{zz}$ as
\begin{align}
\cN_{zz}(\Omega) & = \int dt\, \, e^{i \Omega t } \langle \delta B_z(t) \, \delta B_z(0) \rangle  \nonumber \\ 
& = \frac{1}{N_v} \sum_\k G^{\mu}_z(\k)  G^{\nu}_z(-\k) \int dt\, \, e^{i \Omega t }  \langle u^{\mu}_{\k}(t) u^{\nu}_{-\k}(0) \rangle \nonumber \\
& \to n_v \Phi_0^2   \int \frac{d^2\k}{(2\pi)^2} \langle \k \cdot \u_{\k,\Omega} \; \k \cdot  \u_{-\k,-\Omega} \rangle e^{-2 k z_0}     . 
\label{eq:NzzVL}
\end{align}
Remarkably, $\cN_{zz} = 2 \cN_{xx}$, so it suffices to study the properties of $\cN_{zz}$.
Just like the single fluctuating vortex discussed in the previous subsection, the frequency dependence of $\cN_{zz}(\Omega)$ is determined by the dynamical correlations of the collective vortex vibrations. 
However, in contrast to single vortex dynamics, the distance dependence of $\cN_{zz}(z_0)$ also depends on these dynamical correlations, which we turn to next. 

The low-energy, long-wavelength Lagrangian density $\mathcal{L}$ for the (2D/rigid 3D) vortex lattice is given by~\cite{BlatterRMP}:
\begin{align}
\mathcal{L} = \sum_\k & \left\{ n_v \left[\frac{\mu_v}{2}|\dot{ \u}_{\k}|^2 + n_s h (\dot{\u}_{\k} \times \hat{z}) \cdot \u_{-\k} \right] \right. \nonumber \\ 
& \left. - \frac{1}{2} u^\mu_\k \phi_{\mu \nu}(\k) u^\nu_{-\k} \right\}, 
\end{align}
where $\phi_{\mu \nu}(\k) = k_\mu k_{\nu} c_{11}(k) + (k^2 \delta_{\mu \nu} - k_\mu k_\nu) c_{66}(k)$ is the elastic matrix of the vortex lattice --- $c_{11}(k)$ being the compression modulus and $c_{66}(k)$ being the shear modulus. 
Subsequently, we derive the equation of motion for the collective displacement mode $\u_\k(t)$ as
\begin{equation}
 \mu_v \ddot{u}_\k^\mu =  n_s h \, \varepsilon_{\mu \nu} \, \dot{u}_{\k}^\nu - n_v^{-1} \phi_{\mu \nu}(\k) u^\nu_{\k} -  \eta \dot{u}_{\k}^\mu +  \zeta^\mu_{\k}  ,
\label{eq:eom_VL}
\end{equation}
where we have again added a phenomenological viscous drag coefficient $\eta$~\cite{BardeenStephen65} and a local stochastic force $\bm{\zeta}(t)$ to account for scattering off high energy modes.
The stochastic force $\bm{\zeta}(t)$ (per unit length in 3D) satisfies the FDT:
\begin{equation}
\langle  \zeta^\mu_{\k}(t) \zeta^\nu_{\k^\prime}(t') \rangle = \begin{cases}
2 \eta k_B T \delta^{\mu \nu} \, \delta(t - t') \, \delta_{\k,-\k^\prime}, \text{ Pearl} \\ 
2 \frac{\eta}{a_c} k_B T \delta^{\mu \nu} \delta(t - t') \delta_{\k,-\k^\prime} \text{ Abrikosov}
\end{cases} .
\label{eq:WNVL}
\end{equation}

To solve Eq.~\eqref{eq:eom_VL}, we move to the frequency domain and recast the equation of motion as a matrix equation (in analogy with the pinned vortex scenario):
\begin{align}
&[M^{-1}_{\k,\Omega}]_{\mu \nu} u^{\nu}_{\k,\Omega} = \zeta^\mu_{\k,\Omega},  \nonumber \\  M^{-1}_{\k,\Omega} =  & \, ( c_{11}(k)k^2/n_v -\mu_v \Omega^2 - i \eta \Omega )P_\L(\k) \nonumber \\  
&+ (c_{66}(k) k^2/n_v - \mu_v \Omega^2 - i \eta \Omega) P_{\T}(\k) + i \Omega n_s h \varepsilon  ,
\label{eq:MinvVL}
\end{align}
where $[P_{\L}(\k)]_{\mu \nu} = k_\mu k_\nu/k^2$ is the longitudinal projector, $[P_{\T}(\k)]_{\mu \nu} = \delta_{\mu \nu} -  k_\mu k_\nu/k^2$ is the transverse projector, and $\varepsilon$ is the fully antisymmetric rank-2 tensor in 2D. 
The correlation function is obtained by inverting the matrix equation and using Eq.~\eqref{eq:WNVL} (see Appendix~\ref{app:solid} for details):
\begin{widetext}
\begin{equation}
\begin{aligned}
    C_{\L}(\k,\Omega) &\equiv \langle \k \cdot \u_{\k, \Omega} \,~ \k \cdot \u_{-\k,-\Omega} \rangle
= 2 \eta k_B T k^2 \Tr[P_{\L}(\k) M_{\k, \Omega} M^\dagger_{\k, \Omega}] \\
&= \frac{2 \eta k_B T k^2 \left[ |c_{66}(k) k^2/n_v -\mu_v \Omega^2 - i \eta \Omega|^2 + (\Omega n_s h)^2 \right]}{|(c_{11}(k) k^2/n_v -\mu_v \Omega^2 - i \eta \Omega)(c_{66}(k) k^2/n_v -\mu_v \Omega^2 - i \eta \Omega) - (\Omega n_s h)^2|^2}, 
\end{aligned}
\label{eq:uuLongCorrVL} 
\end{equation}
\end{widetext}
where we have specified to Pearl vortices (for Abrikosov vortices, we have to divide the RHS of Eq.~\eqref{eq:uuLongCorrVL} by an additional factor of $a_c$, and note that $\eta$ ($n_s$) denotes the viscous drag (superfluid density) per unit length).
In principle, we can find $\cN_{zz}(\Omega)$ by inserting Eq.~\eqref{eq:uuLongCorrVL} into Eq.~\eqref{eq:NzzVL}, and performing a numerical integration over all momenta. 
However, in analogy with our discussion of  magnetic noise due to a single fluctuating vortex, we focus on simple physical limits. 
In our expressions for $C_{\L}(\k,\Omega)$, we will often set the spacing between superconducting layers $a_c = 1$ for Abrikosov vortices, but we will restore them as appropriate while discussing the noise tensor $\cN_{zz}(\Omega)$.

First, in the massless vortex limit, we can set $\mu_v = 0$ and neglect the inertial term $\mu_v \Omega^2$. 
In this limit, the vortex phonons are circularly polarized as a consequence of the Magnus force, with a  dispersion $\Omega_\k = \sqrt{c_{11}(k) c_{66}(k)}k^2/(n_v n_s h)$~\cite{FHP66}. 
Therefore, the longitudinally projected  correlation function takes the form of a Lorentzian with a peak at $\Omega = \Omega_\k$ and a linewidth set by the viscous drag $\eta$ and parametrized by the ratio of the drag to the Magnus force $\alpha_\eta \equiv \eta/(n_s h)$: 
\begin{equation}
C_{\L}(\k,\Omega)  = \frac{2 \eta k_B T k^2[(c_{66}k^2/n_v n_s h)^2 + \tilde{\Omega}^2]}{( n_s h)^2[\Omega_\k^2 - \tilde{\Omega}^2]^2 + \alpha_\eta^2 \Omega^2 [(c_{11} + c_{66})/n_v]^2 k^4},
\end{equation}
where $\tilde{\Omega} = \Omega(1 + \alpha_\eta^2)^{1/2}$.
Crucially, the asympototic behavior of the compression modulus $c_{11}(k)$ differs at small momenta for Pearl and Abrikosov vortices, leading to distinct dispersion for the phonon modes for 2D and 3D superconducting vortex lattices. 
This, in turn, leads to a distinct dependence of $\cN_{zz}(\Omega)$ on the frequency $\Omega$ in the resonant limit. 
\begin{figure*}[!t]
    \includegraphics[width = 1.0\linewidth]{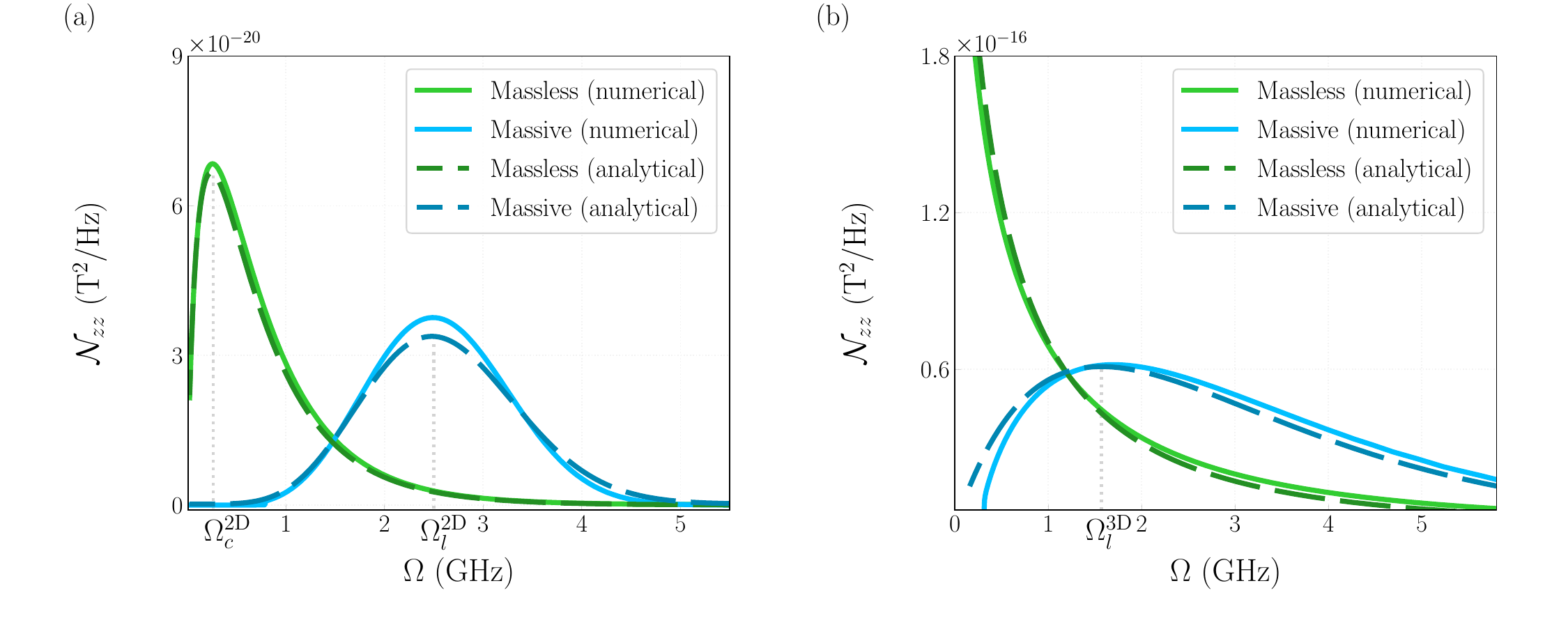}
    \caption{Longitudinal noise spectral density $\cN_{zz}(\Omega)$ due to vortex lattice phonons in the far field limit, i.e., for qubit-sample distance $z_0$ much larger than the relevant screening length $\lambda_{\rm scr}$, evaluated using Eqs.~\eqref{eq:NzzVL} and \eqref{eq:uuLongCorrVL} \cite{numerical_integration}. 
    For both panels, we consider inter-vortex spacing $a_\Delta = 20~\mathrm{nm}$ ($\mu_0 H\approx 6~\T$) and temperature $T = 30~\mathrm{K}$.
    (a) Vortex phonon noise for a lattice of Pearl (point) vortices, evaluated numerically in the massless and massive limits with resonant peaks at $\Omega_{c}^{\rm 2D}$ and $\Omega_l^{\rm 2D}$ respectively.
    The parameters used are $\mu_v=5\times10^{-27}~\mathrm{kg}$ (massless) and $2\times10^{-21}~\mathrm{kg}$ (massive), $\eta = 5\times 10^{-16}~\mathrm{kg/s}$, $n_s h= 5\times 10^{-13}~\mathrm{kg/s}$, $d = 50~\mathrm{nm}$, and $z_0 = 1~\mu\mathrm{m}$, which ensures $z_0 \gtrsim \Lambda$.
    The analytical curves follow Eqs.~\eqref{eq:phonon_P_massless} and \eqref{eq:phonon_P_massive}.
    (b) Vortex phonon noise for a lattice of Abrikosov (line) vortices, evaluated numerically in the massless and massive limits: only the massive case shows a resonant peak at $\Omega_l^{\rm 3D}$ in accordance with our analytical results. 
    The analytical curves follow Eqs.~\eqref{eq:phonon_A_massless} and \eqref{eq:phonon_P_massive}.
    The parameters used are: $\mu_v=10^{-20}~\mathrm{kg/m}$ (massless) and $4\times10^{-15}~\mathrm{kg/m}$ (massive), $\eta = 10^{-8}~\mathrm{kg/(m\cdot s)}$, $n_s h= 10^{-6}~\mathrm{kg/(m\cdot s)}$, $a_c=1~\mathrm{nm}$, and $z_0 = 500~\mathrm{nm}$. 
    Note that in 3D, $\mu_v$ and $\eta$ are considered per unit length, and $n_s$ is the volume density. 
    }
    \label{fig:lattice}
\end{figure*}

To find the field-dependence of $\cone(k)$ and the resulting dispersion of the vortex lattice phonons, we consider intermediate values of the external field $H$, i.e., not too close to $H_{c2}$ when the vortex cores strongly overlap, and not too close to $H_{c1}$ when the superconductor is about to expel magnetic flux and enter the Meissner phase. 
For a lattice of Pearl vortices, the strongly non-local interaction between vortices leads to a divergent Fourier space interaction $\tilde{V}_{\rm int}(k) \sim 1/k$ at small $k$: in turn, the compression modulus $c_{11}(k) \approx n_v^2 \, \tilde{V}_{\rm int}(k)$ also diverges as $k \to 0$ as a compression involves a change of vortex density (see Appendix~\ref{app:solid} for a derivation).
At the same time, the shear modulus $c_{66}(k)$ has a finite limit $c_{66}$ as $k \to 0$ as a shear does not change the vortex density, and the resulting energy change is dominated by lattice scale contributions. 
More precisely, the elastic moduli for a lattice of Pearl vortices are given by~\cite{Brandt95} 
\begin{equation}\label{eq:elastic_moduli_2D}
\cone(k) \approx \frac{(\mu_0 H)^2}{\mu_0 k(1 + \Lambda k)}, ~~ \csix(k) \approx \frac{(\mu_0 H) \Phi_0}{8 \mu_0 \pi \Lambda}   .
\end{equation}

Therefore, the phonon dispersion has a non-analytic momentum dependence for a 2D vortex lattice: 
\begin{align}
\Omega_\k = \frac{\sqrt{c_{11}(k) c_{66}(k) } k^2}{n_v n_s h} &= \frac{\alpha k^{3/2}}{\sqrt{1 + \Lambda k}}, \nonumber   \\
\text{ where }\alpha &= \sqrt{\frac{\Phi_0^3 H}{8 \pi \mu_0 \Lambda (n_s h)^2} } .
\end{align}
By contrast, for a 3D vortex lattice, electromagnetic screening by the bulk superconductor is more effective: hence, $c_{11}(k)$ approaches a finite limit as $k \to 0$ while $\csix$ remains approximately momentum independent~\cite{Brandt95}:
\begin{equation}\label{eq:elastic_moduli_3D}
\cone(k) \approx \frac{(\mu_0 H)^2}{\mu_0(1 + k^2 \lambda_L^2)}, ~~ \csix(k) \approx \frac{(\mu_0 H) \Phi_0}{16 \mu_0 \pi \lambda_L^2}  .
\end{equation}
Hence, the low-energy dispersion for a 3D vortex lattice is quadratic in momentum~\cite{FHP66}: 
\begin{equation}
\Omega_\k = \frac{\sqrt{c_{11}(k) c_{66}(k)} k^2}{n_v n_s h} = \frac{\bar{\alpha} k^2}{\sqrt{1 + k^2\lambda_L^2}}, ~~ \bar{\alpha} = \frac{\sqrt{\pi \Phi_0^3 (\mu_0 H)}}{4 \pi \mu_0 \lambda_L n_s h}  .
\end{equation}
This difference in dispersion leads to a prominent distinction in the frequency dependence of the magnetic noise.
Such distinction is easiest to see in the resonant limit $(\eta \to 0)$, when the longitudinal projection of the correlator takes the form of a Dirac delta function:
\begin{align}
C_{\L}(\k,\Omega) = \frac{2\pi k_B T |\Omega| n_v}{c_{11}(k)} \, \delta\left( \Omega_\k^2 - \Omega^2\right)  .
\end{align}
The resultant magnetic noise is then simply obtained by using this correlator in Eq.~\eqref{eq:NzzVL}. 

For a lattice of Pearl vortices, in the limit of $\Lambda \ll z_0$ such that $\Lambda \, k_{\rm typical} \sim \Lambda/z_0 \ll 1$, we find that the magnetic noise in the resonant limit is given by
\begin{equation}\label{eq:phonon_P_massless}
\cN_{zz}(\Omega) = \frac{\mu_0 k_B T \Omega}{3 \alpha^2} \exp\left[ - 2 \left( \frac{\Omega}{\alpha}\right)^{2/3} z_0 \right]  .
\end{equation}
We observe that resonant peak in $C_{\L}(\k,\Omega)$ at the phonon frequency $\Omega_\k = \alpha k^{3/2}$ translates to a noise peak at $\Omega_c^{\rm 2D} \sim \alpha z_0^{-3/2}$ [Fig.~\ref{fig:lattice}(a)], consistent with our intuition that the resonant modes with $k \sim z_0^{-1}$ dominate the contribution. 
The low-energy dispersion of the phonon modes of the Pearl vortex lattice can thus be extracted from the scaling of the noise peak frequency with $z_0$. 

For rigid Abrikosov vortices, in the limit of $\lambda_L \ll z_0$ such that $k \lambda_L \ll 1$, the magnetic noise is given by
\begin{equation}\label{eq:phonon_A_massless}
\cN_{zz}(\Omega) = \frac{\mu_0 k_B T}{4 a_c \bar{\alpha}} \exp\left[-  2 \left( \frac{\Omega}{\bar{\alpha}}\right)^{1/2} z_0 \right]  .
\end{equation}
In this case, the sharp resonance in $C_{\L}(\k,\Omega)$ is smeared out in the magnetic noise by the the momentum integral in Eq.~\eqref{eq:NzzVL}.
Nevertheless, the noise $\cN_{zz}(\Omega)$ has a pronounced frequency dependence [Fig.~\ref{fig:lattice}(b)] that serves to distinguish between Pearl vortices and Abrikosov vortices. 
Further, noting that $\alpha, \bar{\alpha} \sim \sqrt{H}$, the expressions for $\cN_{zz}$ explicitly show that the magnetic noise takes distinct non-analytic forms as a function of the applied field $H$, depending on the dimensionality: however, they both vanish as $H \to 0$.
Finally, while we assumed explicit expressions for the elastic constants $c_{11}$ and $c_{66}$ to deduce the field dependence, in principle we may extract these elastic moduli by fitting the magnetic noise data with our theoretical model.

Next, we consider the limit when the vortices are massive, i.e., $\mu_v  \Omega^2 \gg n_s h \Omega$. 
Neglecting the Magnus force, or equivalently setting $n_s h \Omega \to 0$ in Eq.~\eqref{eq:uuLongCorrVL}, we find that the dynamical correlation function $C_{\L}(\k,\Omega)$ takes the following form:
\begin{equation}
C_{\L}(\k,\Omega) = \frac{2 \eta k_B T k^2}{(\cone(k) k^2/n_v - \mu_v \Omega^2)^2 + \eta^2 \Omega^2}  .
\end{equation}
Once again, we focus on the resonant limit ($\eta \to 0$), such that the longitudinal correlator $C_{\L}$ is given by
\begin{equation}
C_{\L}(\k,\Omega) = \frac{2 \pi k_B T k^2}{\mu_v |\Omega|} \; \delta\left(\Omega_\k^2 - \Omega^2\right), \text{ where } \Omega_\k^2 = \frac{\cone(k) k^2}{n_v \mu_v}   .
\end{equation}
Once again, the divergent form of  the compression modulus $\cone(k)$ at low momentum in two dimensions leads to distinct, dimension-dependent dispersions $\Omega_\k$ for the longitudinal phonon mode. 
For Pearl vortices, the longitudinal mode disperses as 
\begin{equation}
\Omega_{\k} = \sqrt{\frac{ \Phi_0 H k}{\mu_v(1 + \Lambda k)}} \propto \sqrt{k} \, , \text{ for } k\Lambda \ll 1,
\end{equation}
whereas for Abrikosov vortices,
\begin{equation}
\Omega_{\k} = \sqrt{\frac{\Phi_0 H k^2}{\mu_v(1 + k^2 \lambda_L^2)}} \propto k \, , \text{ for } k \lambda_L \ll 1.
\end{equation}
This distinction shows up in the magnetic noise spectrum. 
For massive Pearl vortices, the noise in the resonant limit is given by
\begin{equation}\label{eq:phonon_P_massive}
\cN_{zz}(\Omega) = \mu_0 k_B T \left( \frac{\mu_v}{\Phi_0 H}\right)^3 \Omega^5 \exp\left[ - \frac{2 \mu_v \Omega^2 z_0}{\Phi_0 H} \right].
\end{equation}
The noise exhibits a distinct peak at the qubit frequency $\Omega = \Omega_l^{\rm 2D} \sim \sqrt{\Phi_0 H/\mu_v z_0}$, corresponding to the resonant mode frequency with $k \sim 1/z_0$ [Fig.~\ref{fig:lattice}(a)].
Further, the peak frequency scales as $1/\sqrt{z_0}$, reflecting the anomalous $\Omega_\k \sim \sqrt{k}$ low-energy dispersion of longitudinal phonons in the Pearl vortex lattice. 

By contrast, for massive Abrikosov vortices in the resonant limit, the magnetic noise is given by
\begin{align} \label{eq:phonon_A_massive}
\cN_{zz}(\Omega) &= \frac{(\mu_0 H)^3 k_B T \mu_v \Omega}{2 a_c \cone^2(0) \Phi_0} \exp\left[ - 2 \sqrt{\frac{n_v \mu_v}{\cone(0)}} \Omega z_0\right] \nonumber  \\
&= \frac{\mu_0 k_B T \mu_v \Omega}{2 a_c H \Phi_0} \exp\left[ - 2 \sqrt{\frac{\mu_v}{\Phi_0 H}} \Omega z_0 \right],
\end{align}
where the peak frequency $\Omega_l^{\rm 3D} \sim \sqrt{\Phi_0 H/\mu_v z_0^2}$ [Fig.~\ref{fig:lattice}(b)] scales as $1/z_0$, reflecting the linear dispersion of the phonons (recall that $\mu_v$ has the dimensions of mass per unit length in 3D).  
Just like the massless limit, either the distinct frequency dependence of $\cN_{zz}(\Omega)$ or the distinct non-analytic field ($H$) dependence can be used to extract low-energy phonon dispersions and distinguish between Pearl vortices and Abrikosov vortices in the massive limit as well. 
Further, only the longitudinal phonon mode, which is sharply defined in the massive vortex limit, contributes to the magnetic noise. 
This implies that only the compressional elastic modulus $\cone$ can be extracted from $\cN_{zz}$ in this limit. 
Finally, we also note that the low density of states of vortex phonons implies that the magnetic noise is suppressed by powers of $\Omega$, which is typically the smallest energy scale in the set-up. 
This suppression is particularly severe for Pearl vortices, owing to the sharp $\Omega_\k \sim \sqrt{k}$ dispersion of the phonon modes.

Lastly, we consider the strongly dissipative limit of large viscous damping relative to the other kinetic terms, i.e., $\eta \gg n_s h, \mu_v \Omega$. 
The longitudinal correlator $C_{\L}(\k, \Omega)$ in Eq.~\eqref{eq:uuLongCorrVL} now takes a simple form:
\begin{equation}
C_{\L}(\k, \Omega) = \frac{2 \eta k_B T k^2}{\left( \cone(k) k^2/n_v \right)^2 + \eta^2\Omega^2}
\end{equation}
For a lattice of Pearl vortices, $\cone(k) \sim k^{-1}$ at small momentum, leading $C_{\L}(\k,\Omega)$ to go approach a finite limit as $\Omega \to 0$. 
Therefore, just like the overdamped fluctuating single vortex, we can simply take the $\Omega \to 0$ limit in the magnetic noise:
\begin{equation}
\cN_{zz}(\Omega \to 0) = \frac{\mu_0 \eta k_B T}{4 \pi H \Phi_0 z_0^2}  .
\end{equation}
By contrast, for a lattice of rigid Abrikosov vortices, we cannot simply set the probe frequency $\Omega$ to zero while evaluating the magnetic noise.
In this case, $\cone(k) $ approaches a constant in the low momentum limit, and hence $C_{\L}(\k,\Omega \to 0) \sim k^{-2}$ is strongly divergent: intuitively this divergence arises due to the enhanced density of states of vortex lattice phonons in 3D. 
The qubit frequency $\Omega$ naturally acts as an IR cutoff, and leads to a logarithmic dependence of the magnetic noise on both $\Omega$ and $z_0$:
\begin{align}
\cN_{zz}(\Omega) = \frac{\mu_0 \eta k_B T}{2 \pi a_c H \Phi_0} \ln\left( \frac{H \Phi_0}{\Omega \eta z_0^2} \right)  .
\end{align}
Remarkably, the magnetic noise is quite strong at weak magnetic fields in the overdamped limit, and the distinct qubit-sample distance dependence can be used to tell apart the vortex lattice dimensionality even when the viscous forces are strong. 
This enhancement stems from the softening of the compressional elastic modulus $\cone$ at weak magnetic fields, leading to enhanced low-energy fluctuations. 
However, at sufficiently weak magnetic fields, the significant suppression of the elastic moduli can cause the vortex lattice to melt into a vortex liquid due to thermal fluctuations, thereby precluding an unphysical divergence in the magnetic noise as $H \to 0$. 
The noise signatures of such a vortex liquid phase will be discussed in the next section.

To conclude this section, we discuss the effect of vertical distortions of Abrikosov vortices, i.e., a non-zero $k_z$, on our results for the 3D vortex lattice.
Drawing from the intuition built by incorporating $k_z \neq 0$ for a single fluctuating vortex, we expect a weaker signal when the vortices are allowed to distort in the $z$-direction.
Additionally, we expect that the magnetic noise also depends on the elastic modulus $c_{44}$~\cite{BlatterRMP, Brandt95} that controls such out-of-plane oscillations. 
The precise expression for the magnetic noise that accounts for such modulations of line vortices is left for future work.

\subsection{Vortex diffusion in a vortex liquid}
Having discussed the signatures of magnetic noise from vortex lattice phonons, we now consider the possibility of a molten vortex lattice due to thermal/quantum fluctuations, leading to a translation invariant vortex liquid phase.  
Since the vortex-antivortex pairs get bound into dipolar pairs below the BKT transition temperature, we can treat the density of free vortices $n_v$ in a vortex liquid as approximately conserved~\cite{MinnhagenRMP}.
While the average free vortex density $n_v$ in the vortex liquid is set by the magnetic field $H$, i.e., $n_v = \mu_0 H/\Phi_0$, the long-wavelength late-time  density fluctuations are expected to be diffusive in nature. 
Such fluctuations lead to magnetic noise that contain discernible signatures of the vortex liquid phase.
Conveniently, the noise tensor can be calculated by rewriting Eq.~\eqref{eq:Nvortex} in momentum space in terms of $\B_0(\k) = e^{- k z_0}(- i \k,k) \phi_M(k)$ obtained by Fourier transforming $\B_0(\r_q)$, and the dynamical vortex correlation function $C_{n_v n_v}(\k,\Omega)$: 
\begin{align}
& \cN_{\alpha \beta}(\Omega) = \int \frac{d^2\k}{(2\pi)^2} B^\alpha_0(\k) B^\beta_0(-\k) C_{n_v n_v}(\k,\Omega), \text{ where }  \nonumber \\
& C_{n_v n_v}(\k,\Omega) = \int dt\, \, e^{i \Omega t} \int d^2 \rr\, e^{- i \k \cdot \bm{\rho}} \langle n_v(\bm{\rho},t) n_v(\bm{0},0) \rangle  .
\end{align}

We start by noting that one can explicitly derive the diffusive dynamical correlation function $C_{n_v n_v}(\k,\Omega)$ for the vortex density by assuming simple Langevin dynamics for point-like vortices:
\begin{equation}
\mu_v \ddot{\u} = n_s h (\dot{\u} \times \hat{z})- \bar{\eta} \dot{\u} + \bar{\bm{\zeta}}_t \,,
\label{eq:eom_Vliquid}
\end{equation}
which follows from the equation of motion of the pinned vortex Eq.~\eqref{eq:eom_pinned} by neglecting restoring force from the pinning potential $(K)$ or elastic deformations ($\varepsilon_\ell$).
Here, the viscous drag $\bar{\eta}$ can have multiple physical origins: it can arise from dissipative electronic excitations in the vortex core~\cite{BardeenStephen65} (as quantified by $\eta$ in the setting of pinned vortices or vortex lattice phonons), and also from dissipative collisions between vortices. 
For simplicity, we restrict to the physically relevant massless limit ($\mu_v = 0$) of Eq.~\eqref{eq:eom_Vliquid}.
By deriving the Fokker-Planck equation corresponding to the Langevin equation, which describes the evolution of the vortex density distribution (see Appendix~\ref{app:liquid} for details), we confirm our intuitive expectation that vortex dynamics is diffusive.
Consequently, the vortex-density correlation function takes the standard form in momentum space:
\begin{equation}\label{eq:n_v_correlator}
C_{n_v n_v}(\k,\Omega) = \langle n_v(\k,\Omega) n_v(-\k,-\Omega) \rangle = \frac{2 k_B T \chi_v D_v k^2}{\Omega^2 + (D_v k^2)^2},
\end{equation}
where $\chi_v$ is the vortex compressibility, $D_v$ is the vortex diffusion constant, and the Einstein relation dictates that the vortex conductivity $\sigma_v = \chi_v D_v$ (assuming unit winding number for each vortex). 
Using the rotational invariance of $C_{n_v n_v}(\k,\Omega)$, the non-zero components of the noise tensor are given by
\begin{align}\label{eq:Noise_liquid}
\cN_{\alpha \alpha}(\Omega) = \int \frac{d^2\k}{(2\pi)^2} |B^\alpha_0(\k)|^2 \left[ \frac{2 k_B T \chi_v D_v k^2}{\Omega^2 + (D_v k^2)^2} \right]  .
\end{align}
Using the explicit form of $B^\alpha_0(\k)$, we again find that $\cN_{zz} = 2 \cN_{xx} = 2 \cN_{yy}$, just as we found for the noise tensor from vortex lattice vibrations. 
Therefore, it suffices to focus on $\cN_{zz}(\Omega)$, the expression for which can be conveniently recast by introducing a diffusive lengthscale $\ell_\Omega = \sqrt{D_v/\Omega}$.
For a liquid of point-like Pearl vortices, $\cN_{zz}(\Omega)$ is given by: 
\begin{equation}
\cN_{zz}(\Omega) = \frac{ k_B T \sigma_v \Phi_0^2}{\pi D_v^2} \int_0^\infty d\tilde{k} \, \frac{\tilde{k}^3  \, e^{- 2 \tilde{k} z_0/\Lambda}}{(1 + \tilde{k})^2 \left[ \left( \frac{\Lambda}{\ell_\Omega} \right)^4 + \tilde{k}^4 \right]}  . ~~~~~~
\end{equation}
On the other hand, for a liquid of 3D rod-like Abrikosov vortices, we find that:
\begin{equation}
 \cN_{zz}(\Omega) = \frac{k_B T \sigma_v \Phi_0^2}{\pi D_v^2} \int_0^\infty d\tilde{k} \, \frac{\tilde{k}^3  \, e^{- 2 \tilde{k} z_0/\lambda_L}}{(1 + \tilde{k}^2)^2 \left[ \left( \frac{\lambda_L}{\ell_\Omega} \right)^4 + \tilde{k}^4 \right]}  .   ~~~~~~
\end{equation}
\begin{table}[!t]
\centering
\resizebox{1.0\linewidth}{!}{ % scales entire table
\renewcommand{\arraystretch}{3.0}
\begin{tabular}{ccc}
\hline\hline\vspace{0.5em}
Vortex type  & ~~ Length-scale hierarchy & $\cN_{zz}(\Omega,z_0)$ \\ \hline
\multirow{2}{5em}{\centering Pearl or Abrikosov} & $\lambda_{\rm scr}, z_0 \ll \ell_\Omega$ &  $\dfrac{\Phi_0^2 k_B T \sigma_v}{\pi D_v^2}\ln\left( \dfrac{\ell_\Omega}{\max(z_0,\lambda_{\rm scr})} \right)$ \\ 
& $\lambda_{\rm scr}, \ell_\Omega \ll z_0$ & $\dfrac{3\Phi_0^2 k_B T \sigma_v}{8 \pi D_v^2} \left( \dfrac{\ell_\Omega}{z_0} \right)^4$  \\ \hline         
\multirow{2}{5em}{\centering Pearl} & $\ell_\Omega \ll z_0 \ll \Lambda$ & $\dfrac{\Phi_0^2 k_B T \sigma_v}{\pi D_v^2} \dfrac{\ell_\Omega^4}{4 z_0^2 \Lambda^2}$ \\
& $z_0 \ll \ell_\Omega \ll \Lambda$ & $\dfrac{\Phi_0^2 k_B T \sigma_v}{4 D_v^2} \left( \dfrac{\ell_\Omega}{\Lambda} \right)^2$ \\ \hline
\multirow{2}{5em}{\centering Abrikosov} & $\ell_\Omega \ll z_0 \ll \lambda_L$ &$\dfrac{\Phi_0^2 k_B T \sigma_v}{2\pi D_v^2} \left(\dfrac{\ell_\Omega}{\lambda_L} \right)^4 \ln\left(\dfrac{\lambda_L}{z_0} \right)$ \\
& $z_0 \ll \ell_\Omega \ll \lambda_L$ & $\dfrac{\Phi_0^2 k_B T \sigma_v}{\pi D_v^2}\left(\dfrac{\ell_\Omega}{\lambda_L} \right)^4 \ln\left(\dfrac{\lambda_L}{\ell_\Omega} \right)$ \\ \hline\hline
\end{tabular}  }
\caption{Dependence of the magnetic noise on the hierarchy of screening length $\lambda_{\rm scr}$, qubit-sample distance $z_0$ and diffusion length $\ell_\Omega = \sqrt{D_v/\Omega}$ in the vortex liquid phase assuming difffusive density correlations.}
\label{tab:VLscalings}  
\end{table}
While the above expressions are quite general, for physically relevant qubits such as the NV center, the frequency $\Omega$ is typically the smallest energy scale, and accordingly we expect $\ell_\Omega \gg z_0$ and the screening lengthscale $\lambda_{\rm scr}$ (where $\lambda_{\rm scr} = \Lambda$ for a liquid of Pearl vortices, and $\lambda_{\rm scr} = \lambda_L$ for a liquid of rigid Abrikosov vortices). 
Therefore we focus on this limit, and find that 
\begin{equation}
\cN_{zz}(\Omega) = \frac{\Phi_0^2 k_B T \sigma_v}{\pi D_v^2}\ln\left( \frac{\ell_\Omega}{\max(z_0,\lambda_{\rm scr})} \right).
\end{equation}
\begin{figure}[!t]
\includegraphics[width = 0.96\linewidth]{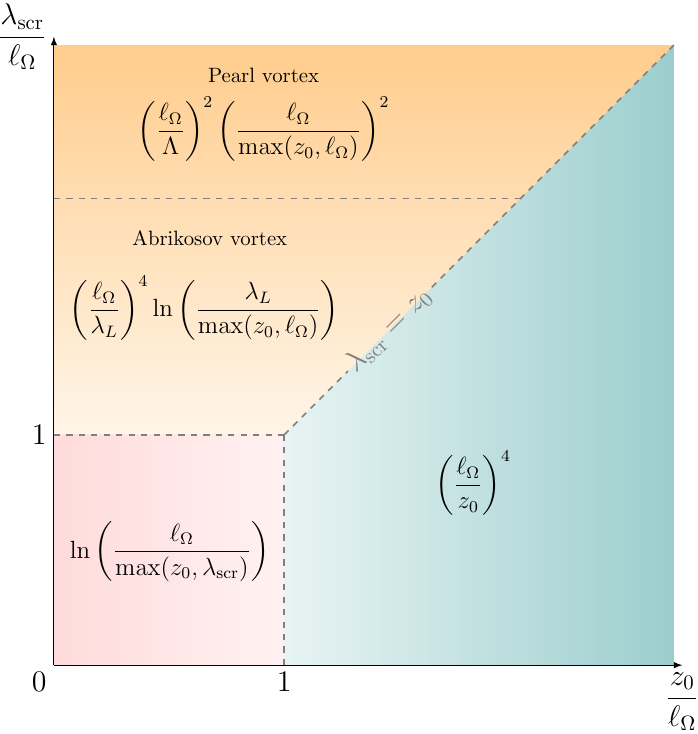}
\caption{Schematic cross-over diagram of scaling behavior of the magnetic noise as a function of qubit-sample distance $z_0$ and the electromagnetic screening lengthscale $\lambda_{\rm scr}$ (both normalized by the diffusive lengthscale $\ell_\Omega = \sqrt{D_v/\Omega}$) in the vortex liquid phase: $\mathcal{N}_{zz} \sim \Phi_0^2 k_B T \sigma_v D_v^{-2} f(z_0, \lambda_{\rm scr}, \ell_\Omega)$ and the expressions for $f(z_0, \lambda_{\rm scr}, \ell_\Omega)$ are given in different regimes.}
\label{fig:VLiquid}
\end{figure}

A few comments are in order. First, we note that in the hierarchy of lengthscales assumed above, the noise depends very weakly on the frequency $\Omega$ (via $\ell_\Omega$) and the sample-qubit distance $z_0$.
Intuitively, this is because the diffusive correlations scale as $1/k^2$, leading to a nearly scale-invariant integral for the noise, with a lower cutoff set by $\Omega$ and an upper cutoff the qubit-sample distance $z_0$ or the screening length $\lambda_{\rm scr}$.

Next, the field dependence of $\cN_{zz}$ contains information about the origin of viscous drag forces on the vortices. 
In the dilute limit, the viscous force is dominated by dissipative electronic quasiparticles in the vortex core, which leads to a drag on the vortex in the classic Bardeen-Stephen picture of flux-flow~\cite{BardeenStephen65}. 
In this limit, inter-vortex scattering can be neglected and the diffusion constant $D_v$ does not depend on the vortex density and hence on $H$.
On the other hand, the Einstein relation dictates that the vortex conductivity $\sigma_v = \chi_v D_v$, with $\chi_v$ being the vortex compressibility. 
For a dilute gas of vortices, the compressibility, given by $\chi_v = n_v/(k_B T)$~\cite{Kardar_2007}, scales linearly with $H$, indicating that $\cN_{zz}(\Omega) \propto n_v \propto H$.
On the other hand, in the dense limit, the interaction between the vortices are expected to be strong, and the viscous drag $\bar\eta$ depends on inter-vortex collisions. 
Further, if we assume, within a simple Drude-like model, that the vortex scattering time $\tau_v$ is inversely proportional to its density $n_v$, then $\sigma_v \propto n_v\tau_v$ is independent of $n_v$. 
On the other hand, the vortex diffusion constant $D_v$ will scale as $\langle v^2 \rangle \tau_v \sim 1/n_v$, where we have assumed that the mean-squared vortex velocity $\langle v^2 \rangle$ is determined by the thermal energy $k_B T$. 
Thus, in this limit, the magnetic noise scales as $D_v^{-2} \propto n_v^2 \sim H^2$.
Note that in both limits, the field-scaling of $\cN_{zz}$ is distinct from the field scaling for vortex lattice phonons, and may be used to distinguish the two phases.

Next, we note that it is possible for the Pearl length $\Lambda$ to exceed both $\ell_\Omega$ and $z_0$ in very thin samples.
This unscreened regime was considered in Ref.~\onlinecite{Curtis_2024}: in this limit the noise scales as $\sigma_v \ell_\Omega^2/D_v^2 \propto \sigma_v/D_v$.
Remarkably, for both the dilute and dense limits that we discussed previously, the noise scales linearly with $H$, with the linear scaling arising from $\sigma_v \propto H$ in the dilute limit, and from $D_v^{-1} \propto H$ in the dense limit. 
However, a more complicated field-dependence of the inter-vortex scattering rate (that determines $D_v$) can potentially alter this scaling on increasing $H$ to reach the dense limit. 

Finally, we also note that for the dynamical vortex density correlations to be appropriately described by a diffusive process, we require the qubit to be able to probe low-energy and low momenta behavior. 
Accordingly, we require the qubit-sample distance $z_0$ to be larger than the ballistic lengthscale, and the qubit frequency $\Omega$ to be much smaller than the typical vortex scattering rate. 
Provided these assumptions hold, we summarize the scalings of magnetic noise for all possible length-scale hierarchies in Table.~\ref{tab:VLscalings} and Fig.~\ref{fig:VLiquid}.

\section{Conclusion and Outlook}
In this work, we presented a comprehensive study of spin-qubit–based spectroscopy of magnetic noise from a superconductor placed in an external magnetic field.  
Specifically, we demonstrated how an applied field, known to lower the critical temperature for the superconducting transition, also significantly increases the magnetic noise near criticality due to enhanced pairing fluctuations. 
Further, we showed that magnetic noise spectroscopy is able to detect fluctuations of superconducting vortices, characterize their low-energy dynamics, and diagnose key quantities of physical interest --- such as the elastic moduli of vortex lattices and the associated phonon dispersion, as well as the diffusivity of vortices in a liquid phase obtained by melting of the vortex lattice. 
Our results complement existing theoretical results on noise spectroscopy of superconductors \textit{in the absence of an applied magnetic field}~\cite{CD2022,DC2022,Curtis_2024,de2025nanoscale,kelly2025superconductivity}, and establishes the importance of this technique for interrogating pairing and vortex dynamics in a superconducting sample placed under a field. 

On the experimental front, we note a couple of very recent experiments~\cite{SCNoise,jayaram2025probing} investigating dynamical processes in superconductors via depolarization ($T_1$) and decoherence ($T_2$) noise spectroscopy with NV centers. 
Ref.~\onlinecite{SCNoise} found that the peak in the magnetic noise in the GHz range, detected by $T_1$ spectroscopy, shifts to a lower temperature relative to zero-field transition temperature $T_c(H = 0)$ in a thin film cuprate superconductor (BSCCO). 
Additionally, the peak noise spectral density increases linearly with $H$. 
While Ref.~\onlinecite{SCNoise} interpreted this $H$-linear increase of noise in terms of a vortex liquid state, our present results show that an alternate interpretation in terms of field-induced enhancement of order-parameter fluctuations can also lead to such a linear increase. 
Future experiments that simultaneously measure transport along with magnetic noise can settle this issue: the vortex liquid phase is expected to have a finite resistivity. 
At lower temperatures, both Ref.~\onlinecite{SCNoise} and Ref.~\onlinecite{jayaram2025probing} uncover evidence of slow motion of vortices in a vortex glass phase, attributed to thermal fluctuation-induced depinning, in two distinct superconductors --- thin film BSCCO and NbSe$_2$. 
Our results on single vortex dynamics can shed light on the nature of the pinning potential as well as the elastic line tension of vortices in such scenarios. 

We conclude by discussing a few future directions of potential interest that are motivated by our results. 
First, our results can be straightforwardly extended to decoherence-based ($T_2$) noise spectroscopy, which is able to detect and characterize noise at an even lower frequency scale (typically MHz for NV centers, as opposed to GHz-scale noise detected by depolarization-based spectroscopy discussed here). 
While such methods necessarily lead to a broadened frequency filter function set by the pulse sequence~\cite{NMRqc_RMP,machado2023quantum} and a resultant smearing of the noise spectral density $\cN(\Omega)$ in frequency, they can prove useful to diagnose universal low-energy dynamics near superconducting phase transitions and slow motion of superconducting vortices, e.g., in a vortex glass phase. 
Second, our theoretical approach for computing critical noise in an applied field can be extended by using non-linear corrections via a self-consistent calculation~\cite{Dorsey91}. 
Likewise, characterizing the noise signatures of vortex dynamics beyond our present considerations, such as thermally assisted flux creep in a vortex-glass phase~\cite{BlatterRMP}, can be useful to shed significant light on glassy dynamics of topological defects.  
Third, given that the qubit sensor measures a stronger signal from dynamical fluctuations in the layers close to the sensor, it is naturally more sensitive to the surface than the bulk. 
This makes it an ideal probe for surface superconductivity and interfacial superconductivity, predicted to occur in a variety of materials~\cite{gozar2008high,gozar2016high,brun2016review,song2021high}. 
Longer term, one can attempt to leverage the ability of the same qubit sensor to detect both magnetic fluctuations and electrical fluctuations~\cite{sahay2025noise} to shed light on superconductivity in low-density superconductors in putative critical ferroelectric materials~\cite{volkov2022superconductivity,VladPRX,VladPRL}.

\textit{Note added:} Near the completion of this work, we became aware of Ref.~\onlinecite{2026detectinghalfquantum}, which proposes applying qubit noise spectroscopy to detect half-quantum vortices in spin-triplet superconductors.

\vspace{8mm}
\acknowledgements
We gratefully acknowledge discussions with Emily Davis, Pavel Dolgirev, Zhongyuan Liu, Ruotian Gong, Mathias S. Scheurer, Norman Y. Yao and Zhipan Wang. 
J. K. acknowledges support from the National Science Foundation, Grant No. DMR-2225920. 
O. K. D. acknowledges support from the NSF through a grant for ITAMP at Harvard University. C. Z. acknowledges support from the National Science Foundation under grant No. 2514391.
S. C. acknowledges support from a PQI Community Collaboration Award.

\bibliography{Noise_SC_in_field}

@book{Kamenev, place={Cambridge}, edition={2}, title={Field Theory of Non-Equilibrium Systems}, publisher={Cambridge University Press}, author={Kamenev, Alex}, year={2023}, url={
https://doi.org/10.1017/9781108769266}}

@article{CD2022,
  title = {Single-spin qubit magnetic spectroscopy of two-dimensional superconductivity},
  author = {Chatterjee, Shubhayu and Dolgirev, Pavel E. and Esterlis, Ilya and Zibrov, Alexander A. and Lukin, Mikhail D. and Yao, Norman Y. and Demler, Eugene},
  journal = {Phys. Rev. Res.},
  volume = {4},
  issue = {1},
  pages = {L012001},
  numpages = {6},
  year = {2022},
  month = {Jan},
  publisher = {American Physical Society},
  doi = {10.1103/PhysRevResearch.4.L012001},
  url = {https://link.aps.org/doi/10.1103/PhysRevResearch.4.L012001}
}

@article{DC2022,
  title = {Characterizing two-dimensional superconductivity via nanoscale noise magnetometry with single-spin qubits},
  author = {Dolgirev, Pavel E. and Chatterjee, Shubhayu and Esterlis, Ilya and Zibrov, Alexander A. and Lukin, Mikhail D. and Yao, Norman Y. and Demler, Eugene},
  journal = {Phys. Rev. B},
  volume = {105},
  issue = {2},
  pages = {024507},
  numpages = {27},
  year = {2022},
  month = {Jan},
  publisher = {American Physical Society},
  doi = {10.1103/PhysRevB.105.024507},
  url = {https://link.aps.org/doi/10.1103/PhysRevB.105.024507}
}

@article{Agarwal2016,
  title = {Magnetic noise spectroscopy as a probe of local electronic correlations in two-dimensional systems},
  author = {Agarwal, Kartiek and Schmidt, Richard and Halperin, Bertrand and Oganesyan, Vadim and Zar\'and, Gergely and Lukin, Mikhail D. and Demler, Eugene},
  journal = {Phys. Rev. B},
  volume = {95},
  issue = {15},
  pages = {155107},
  numpages = {25},
  year = {2017},
  month = {Apr},
  publisher = {American Physical Society},
  doi = {10.1103/PhysRevB.95.155107},
  url = {https://link.aps.org/doi/10.1103/PhysRevB.95.155107}
}

@book{tinkham,
  author = {Tinkham, Michael},
  day = 14,
  edition = 2,
  howpublished = {Paperback},
  isbn = {0486435032},
  month = jun,
  publisher = {Dover Publications},
  title = {Introduction to Superconductivity},
  url = {https://store.doverpublications.com/products/9780486435039},
  year = 2004
}

@article{BlatterRMP,
  title = {Vortices in high-temperature superconductors},
  author = {Blatter, G. and Feigel'man, M. V. and Geshkenbein, V. B. and Larkin, A. I. and Vinokur, V. M.},
  journal = {Rev. Mod. Phys.},
  volume = {66},
  issue = {4},
  pages = {1125--1388},
  numpages = {0},
  year = {1994},
  month = {Oct},
  publisher = {American Physical Society},
  doi = {10.1103/RevModPhys.66.1125},
  url = {https://link.aps.org/doi/10.1103/RevModPhys.66.1125}
}

@article{FHP66,
  title = {Stability of a Lattice of Superfluid Vortices},
  author = {Fetter, A. L. and Hohenberg, P. C. and Pincus, P.},
  journal = {Phys. Rev.},
  volume = {147},
  issue = {1},
  pages = {140--152},
  numpages = {0},
  year = {1966},
  month = {Jul},
  publisher = {American Physical Society},
  doi = {10.1103/PhysRev.147.140},
  url = {https://link.aps.org/doi/10.1103/PhysRev.147.140}
}

@article{FH67,
  title = {The Mixed State of Thin Superconducting Films in Perpendicular Fields},
  author = {Fetter, Alexander L. and Hohenberg, P. C.},
  journal = {Phys. Rev.},
  volume = {159},
  issue = {2},
  pages = {330--343},
  numpages = {0},
  year = {1967},
  month = {Jul},
  publisher = {American Physical Society},
  doi = {10.1103/PhysRev.159.330},
  url = {https://link.aps.org/doi/10.1103/PhysRev.159.330}
}

@ARTICLE{Brandt95,
    title = {The Flux-Line Lattice in Superconductors},
  author = {Brandt, E H},
  year = {1995},
  month = nov,
  journal = {Reports on Progress in Physics},
  volume = {58},
  number = {11},
  pages = {1465--1594},
  issn = {0034-4885, 1361-6633},
  doi = {10.1088/0034-4885/58/11/003},
  urldate = {2025-11-04},
  langid = {english},
}

@article{BardeenStephen65,
  title = {Theory of the Motion of Vortices in Superconductors},
  author = {Bardeen, John and Stephen, M. J.},
  journal = {Phys. Rev.},
  volume = {140},
  issue = {4A},
  pages = {A1197--A1207},
  numpages = {0},
  year = {1965},
  month = {Nov},
  publisher = {American Physical Society},
  doi = {10.1103/PhysRev.140.A1197},
  url = {https://link.aps.org/doi/10.1103/PhysRev.140.A1197}
}

@article{Zhou_2022,
   title={Isospin magnetism and spin-polarized superconductivity in Bernal bilayer graphene},
   volume={375},
   ISSN={1095-9203},
   url={http://dx.doi.org/10.1126/science.abm8386},
   DOI={10.1126/science.abm8386},
   number={6582},
   journal={Science},
   publisher={American Association for the Advancement of Science (AAAS)},
   author={Zhou, Haoxin and Holleis, Ludwig and Saito, Yu and Cohen, Liam and Huynh, William and Patterson, Caitlin L. and Yang, Fangyuan and Taniguchi, Takashi and Watanabe, Kenji and Young, Andrea F.},
   year={2022},
   month=feb, pages={774–778} }

@article{Curtis_2024,
  title = {Probing the Berezinskii-Kosterlitz-Thouless vortex unbinding transition in two-dimensional superconductors using local noise magnetometry},
  author = {Curtis, Jonathan B. and Maksimovic, Nikola and Poniatowski, Nicholas R. and Yacoby, Amir and Halperin, Bertrand and Narang, Prineha and Demler, Eugene},
  journal = {Phys. Rev. B},
  volume = {110},
  issue = {14},
  pages = {144518},
  numpages = {21},
  year = {2024},
  month = {Oct},
  publisher = {American Physical Society},
  doi = {10.1103/PhysRevB.110.144518},
  url = {https://link.aps.org/doi/10.1103/PhysRevB.110.144518}
}

@article{de2025nanoscale,
  title = {Nanoscale Defects as Probes of Time-Reversal Symmetry Breaking},
  author = {De, Suman Jyoti and Pereg-Barnea, T. and Agarwal, Kartiek},
  journal = {Phys. Rev. X},
  volume = {16},
  issue = {1},
  pages = {011001},
  numpages = {16},
  year = {2026},
  month = {Jan},
  publisher = {American Physical Society},
  doi = {10.1103/9h4l-21mt},
  url = {https://link.aps.org/doi/10.1103/9h4l-21mt}
}

@misc{SCNoise,
       author = {{Liu}, Z. and {Gong}, R. and {Kim}, J. and {Diessel}, O. K. and {Xu}, Q. and {Rehfuss}, Z. and {Du}, X. and {He}, G. and {Singh}, A. and {Eo}, Y. S. and {Henriksen}, E. A. and {Gu}, G.~D. and {Yao}, N. Y. and {Machado}, F. and {Ran}, S. and  {Chatterjee}. S and {Zu}, C.},
        title = "{Quantum noise spectroscopy of superconducting dynamics in thin film Bi$_2$Sr$_2$CaCu$_2$O$_{8+\delta}$}",
      journal = {arXiv e-prints},
         year = {2025}, 
        month = feb,
          eid = {arXiv:2502.04439},
        pages = {arXiv:2502.04439},
          doi = {10.48550/arXiv.2502.04439},
archivePrefix = {arXiv},
       eprint = {2502.04439},
 primaryClass = {cond-mat.supr-con},
       adsurl = {https://ui.adsabs.harvard.edu/abs/2025arXiv250204439L}
}

@article{jayaram2025probing,
  title = {Probing Vortex Dynamics in 2D Superconductors with Scanning Quantum Microscope},
  author = {Jayaram, Sreehari and Lenger, Malik and Zhao, Dong and Pupim, Lucas and Taniguchi, Takashi and Watanabe, Kenji and Peng, Ruoming and Scheffler, Marc and St\"ohr, Rainer and Scheurer, Mathias S. and Smet, Jurgen and Wrachtrup, J\"org},
  journal = {Phys. Rev. Lett.},
  volume = {135},
  issue = {12},
  pages = {126001},
  numpages = {7},
  year = {2025},
  month = {Sep},
  publisher = {American Physical Society},
  doi = {10.1103/1fzm-pb1d},
  url = {https://link.aps.org/doi/10.1103/1fzm-pb1d}
}

@article{han2025signatures,
  title={Signatures of chiral superconductivity in rhombohedral graphene},
  author={Han, Tonghang and Lu, Zhengguang and Hadjri, Zach and Shi, Lihan and Wu, Zhenghan and Xu, Wei and Yao, Yuxuan and Cotten, Armel A and Sharifi Sedeh, Omid and Weldeyesus, Henok and others},
  journal={Nature},
  volume={643},
  number={8072},
  pages={654--661},
  year={2025},
  publisher={Nature Publishing Group UK London},
  url = {https://doi.org/10.1038/s41586-025-09169-7}
}

@book{larkin2005theory,
  title={Theory of fluctuations in superconductors},
  author={Larkin, Anatoly and Varlamov, Andrei},
  volume={127},
  year={2005},
  publisher={OUP Oxford},
  url = {https://doi.org/10.1093/acprof:oso/9780198528159.001.0001}
}

@article{schmid1966time,
  title={A time dependent Ginzburg-Landau equation and its application to the problem of resistivity in the mixed state},
  author={Schmid, Albert},
  journal={Physik der kondensierten Materie},
  volume={5},
  number={4},
  pages={302--317},
  year={1966},
  publisher={Springer},
  url = {https://doi.org/10.1007/BF02422669}
}

@article{cyrot1973ginzburg,
doi = {10.1088/0034-4885/36/2/001},
url = {https://doi.org/10.1088/0034-4885/36/2/001},
year = {1973},
month = {feb},
publisher = {},
volume = {36},
number = {2},
pages = {103},
author = {M Cyrot},
title = {Ginzburg-Landau theory for superconductors},
journal = {Reports on Progress in Physics}
}

@article{schuller2006time,
  title={Time-dependent Ginzburg--Landau: from single particle to collective behavior},
  author={Schuller, Ivan K and Gray, KE},
  journal={Journal of superconductivity and novel magnetism},
  volume={19},
  pages={401--407},
  year={2006},
  publisher={Springer},
  url = {https://doi.org/10.1007/s10948-006-0179-2}
}

@article{kirtley:1999,
  title = {{{SCANNING SQUID MICROSCOPY}}},
  author = {Kirtley, John R. and Jr, John P. Wikswo},
  year = {1999},
  month = aug,
  journal = {Annual Review of Materials Research},
  volume = {29},
  number = {Volume 29, 1999},
  pages = {117--148},
  publisher = {Annual Reviews},
  issn = {1531-7331, 1545-4118},
  doi = {10.1146/annurev.matsci.29.1.117},
  langid = {english}
}

@article{kuboFDT1966,
  title = {The Fluctuation-Dissipation Theorem},
  author = {Kubo, R},
  year = {1966},
  month = jan,
  journal = {Reports on Progress in Physics},
  volume = {29},
  number = {1},
  pages = {255--284},
  issn = {00344885},
  doi = {10.1088/0034-4885/29/1/306},
  urldate = {2025-11-05},
}

@book{landau2013statistical,
  title={Statistical Physics: Volume 5},
  author={Landau, Lev Davidovich and Lifshitz, Evgenii Mikhailovich},
  volume={5},
  year={2013},
  publisher={Elsevier},
  url = {https://doi.org/10.1016/C2009-0-24487-4}
}

@book{landau1980statistical,
  title={Statistical physics: theory of the condensed state},
  author={Landau, Lev Davidovich and Lifshitz, Evgenii Mikhailovich and Pitaevskii, LP},
  volume={9},
  year={1980},
  publisher={Butterworth-Heinemann},
  url = {https://doi.org/10.1016/C2009-0-24308-X}
}

@article{machado2023quantum,
  title={Quantum noise spectroscopy of dynamical critical phenomena},
  author={Machado, Francisco and Demler, Eugene A and Yao, Norman Y and Chatterjee, Shubhayu},
  journal={Physical Review Letters},
  volume={131},
  number={7},
  pages={070801},
  year={2023},
  publisher={APS},
  url = {https://doi.org/10.1103/PhysRevLett.131.070801}
}

@article{MFisher89,
  title = {Vortex-glass superconductivity: A possible new phase in bulk high-${\mathrm{T}}_{\mathrm{c}}$ oxides},
  author = {Fisher, Matthew P. A.},
  journal = {Phys. Rev. Lett.},
  volume = {62},
  issue = {12},
  pages = {1415--1418},
  numpages = {0},
  year = {1989},
  month = {Mar},
  publisher = {American Physical Society},
  doi = {10.1103/PhysRevLett.62.1415},
  url = {https://link.aps.org/doi/10.1103/PhysRevLett.62.1415}
}

@article{fisherfisherhuse,
  title = {Thermal fluctuations, quenched disorder, phase transitions, and transport in type-II superconductors},
  author = {Fisher, Daniel S. and Fisher, Matthew P. A. and Huse, David A.},
  journal = {Phys. Rev. B},
  volume = {43},
  issue = {1},
  pages = {130--159},
  numpages = {0},
  year = {1991},
  month = {Jan},
  publisher = {American Physical Society},
  doi = {10.1103/PhysRevB.43.130},
  url = {https://link.aps.org/doi/10.1103/PhysRevB.43.130}
}

@article{DorseyHuse,
  title = {Nonlocal Conductivity in Type-{{II}} Superconductors},
  author = {Mou, Chung-Yu and Wortis, Rachel and Dorsey, Alan T. and Huse, David A.},
  year = {1995},
  month = mar,
  journal = {Phys. Rev. B},
  volume = {51},
  number = {10},
  pages = {6575--6587},
  issn = {0163-1829, 1095-3795},
  doi = {10.1103/PhysRevB.51.6575},
}

@article{Dorsey91,
  title = {Linear and nonlinear conductivity of a superconductor near ${\mathit{T}}_{\mathit{c}}$},
  author = {Dorsey, Alan T.},
  journal = {Phys. Rev. B},
  volume = {43},
  issue = {10},
  pages = {7575--7585},
  numpages = {0},
  year = {1991},
  month = {Apr},
  publisher = {American Physical Society},
  doi = {10.1103/PhysRevB.43.7575},
  url = {https://link.aps.org/doi/10.1103/PhysRevB.43.7575}
}

@article{halperin1979resistive,
  title={Resistive transition in superconducting films},
  author={Halperin, BI and Nelson, David R},
  journal={Journal of low temperature physics},
  volume={36},
  pages={599--616},
  year={1979},
  publisher={Springer},
  url = {https://doi.org/10.1007/BF00116988}
}

@article{Beasley,
  title = {Possibility of Vortex-Antivortex Pair Dissociation in Two-Dimensional Superconductors},
  author = {Beasley, M. R. and Mooij, J. E. and Orlando, T. P.},
  journal = {Phys. Rev. Lett.},
  volume = {42},
  issue = {17},
  pages = {1165--1168},
  numpages = {0},
  year = {1979},
  month = {Apr},
  publisher = {American Physical Society},
  doi = {10.1103/PhysRevLett.42.1165},
  url = {https://link.aps.org/doi/10.1103/PhysRevLett.42.1165}
}

@article{Nelson_VL,
  title = {Vortex Entanglement in High-${T}_{c}$ Superconductors},
  author = {Nelson, David R.},
  journal = {Phys. Rev. Lett.},
  volume = {60},
  issue = {19},
  pages = {1973--1976},
  numpages = {0},
  year = {1988},
  month = {May},
  publisher = {American Physical Society},
  doi = {10.1103/PhysRevLett.60.1973},
  url = {https://link.aps.org/doi/10.1103/PhysRevLett.60.1973}
}

@article{Sonin_RMP1987,
  title = {Vortex oscillations and hydrodynamics of rotating superfluids},
  author = {Sonin, E. B.},
  journal = {Rev. Mod. Phys.},
  volume = {59},
  issue = {1},
  pages = {87--155},
  numpages = {0},
  year = {1987},
  month = {Jan},
  publisher = {American Physical Society},
  doi = {10.1103/RevModPhys.59.87},
  url = {https://link.aps.org/doi/10.1103/RevModPhys.59.87}
}

@book{sonin2016dynamics,
  title={Dynamics of quantised vortices in superfluids},
  author={Sonin, Edouard B},
  year={2016},
  publisher={Cambridge University Press},
  url = {https://doi.org/10.1017/CBO9781139047616}
}

@article{CRD2019,
  title = {Diagnosing phases of magnetic insulators via noise magnetometry with spin qubits},
  author = {Chatterjee, Shubhayu and Rodriguez-Nieva, Joaquin F. and Demler, Eugene},
  journal = {Phys. Rev. B},
  volume = {99},
  issue = {10},
  pages = {104425},
  numpages = {23},
  year = {2019},
  month = {Mar},
  publisher = {American Physical Society},
  doi = {10.1103/PhysRevB.99.104425},
  url = {https://link.aps.org/doi/10.1103/PhysRevB.99.104425}
}

@article{berezinskii1971destruction,
  title={Destruction of long-range order in one-dimensional and two-dimensional systems having a continuous symmetry group I. Classical systems},
  author={Berezinskii, VL314399},
  journal={Sov. Phys. JETP},
  volume={32},
  number={3},
  pages={493--500},
  year={1971},
  url = {https://www.jetp.ras.ru//cgi-bin/e/index/e/32/3/p493?a=list}
}

@article{kosterlitz1973ordering,
  title={Ordering, metastability and phase transitions in two-dimensional systems},
  author={Kosterlitz, John Michael and Thouless, David James},
  journal={Journal of Physics C: Solid State Physics},
  volume={6},
  number={7},
  pages={1181},
  year={1973},
  publisher={IOP Publishing},
  doi = {10.1088/0022-3719/6/7/010}
}

@article{kosterlitz2016kosterlitz,
  title={Kosterlitz--Thouless physics: a review of key issues},
  author={Kosterlitz, J Michael},
  journal={Reports on Progress in Physics},
  volume={79},
  number={2},
  pages={026001},
  year={2016},
  publisher={IOP Publishing},
  doi = {10.1088/0034-4885/79/2/026001}
}

@article{casola2018probing,
  title={Probing condensed matter physics with magnetometry based on nitrogen-vacancy centres in diamond},
  author={Casola, Francesco and Van Der Sar, Toeno and Yacoby, Amir},
  journal={Nature Reviews Materials},
  volume={3},
  number={1},
  pages={1--13},
  year={2018},
  publisher={Nature Publishing Group},
  url = {https://doi.org/10.1038/natrevmats.2017.88}
}

@article{rovny2024nanoscale,
  title={Nanoscale diamond quantum sensors for many-body physics},
  author={Rovny, Jared and Gopalakrishnan, Sarang and Jayich, Ania C Bleszynski and Maletinsky, Patrick and Demler, Eugene and de Leon, Nathalie P},
  journal={Nature Reviews Physics},
  pages={1--16},
  year={2024},
  publisher={Nature Publishing Group UK London},
  url = {https://doi.org/10.1038/s42254-024-00775-4}
}

@misc{sahay2025noise,
      title={Noise Electrometry of Polar and Dielectric Materials}, 
      author={Rahul Sahay and Pavel A. Volkov and Satcher Hsieh and Eric Parsonnet and Lane W. Martin and Ramamoorthy Ramesh and Norman Y. Yao and Shubhayu Chatterjee},
      year={2025},
      eprint={2111.09315},
      archivePrefix={arXiv},
      primaryClass={cond-mat.mtrl-sci},
      url={https://arxiv.org/abs/2111.09315},journal={arXiv preprint}
}

@article{pearl1964current,
  title={Current distribution in superconducting films carrying quantized fluxoids},
  author={Pearl, J},
  journal={Applied Physics Letters},
  volume={5},
  number={4},
  pages={65},
  year={1964},
  url = {https://doi.org/10.1063/1.1754056}
}

@article{MinnhagenRMP,
  title = {The two-dimensional Coulomb gas, vortex unbinding, and superfluid-superconducting films},
  author = {Minnhagen, Petter},
  journal = {Rev. Mod. Phys.},
  volume = {59},
  issue = {4},
  pages = {1001--1066},
  numpages = {0},
  year = {1987},
  month = {Oct},
  publisher = {American Physical Society},
  doi = {10.1103/RevModPhys.59.1001},
  url = {https://link.aps.org/doi/10.1103/RevModPhys.59.1001}
}

@article{gottscholl2020BV,
  title={Initialization and read-out of intrinsic spin defects in a van der Waals crystal at room temperature},
  author={Gottscholl, Andreas and Kianinia, Mehran and Soltamov, Victor and Orlinskii, Sergei and Mamin, Georgy and Bradac, Carlo and Kasper, Christian and Krambrock, Klaus and Sperlich, Andreas and Toth, Milos and others},
  journal={Nature materials},
  volume={19},
  number={5},
  pages={540--545},
  year={2020},
  publisher={Nature Publishing Group UK London},
  url = {https://www.nature.com/articles/s41563-020-0619-6}
}

@article{saito2016naturereview,
  title={Highly crystalline 2D superconductors},
  author={Saito, Yu and Nojima, Tsutomu and Iwasa, Yoshihiro},
  journal={Nature Reviews Materials},
  volume={2},
  number={1},
  pages={1--18},
  year={2016},
  publisher={Nature Publishing Group},
  url = {https://doi.org/10.1038/natrevmats.2016.94}
}

@article{Wang2022Science,
author = {Hailong Wang  and Shu Zhang  and Nathan J. McLaughlin  and Benedetta Flebus  and Mengqi Huang  and Yuxuan Xiao  and Chuanpu Liu  and Mingzhong Wu  and Eric E. Fullerton  and Yaroslav Tserkovnyak  and Chunhui Rita Du },
title = {Noninvasive measurements of spin transport properties of an antiferromagnetic insulator},
journal = {Science Advances},
volume = {8},
number = {1},
pages = {eabg8562},
year = {2022},
doi = {10.1126/sciadv.abg8562},
URL = {https://www.science.org/doi/abs/10.1126/sciadv.abg8562},}

@article{Geim2013,
  title = {Van Der {{Waals}} Heterostructures},
  author = {Geim, A. K. and Grigorieva, I. V.},
  year = {2013},
  month = jul,
  journal = {Nature},
  volume = {499},
  number = {7459},
  pages = {419--425},
  issn = {1476-4687},
  doi = {10.1038/nature12385}
}

@article{kelly2025superconductivity,
      title={Superconductivity-enhanced magnetic field noise}, 
      author={Shane P. Kelly and Yaroslav Tserkovnyak},
      year={2025},
      eprint={2412.05465},
      archivePrefix={arXiv},
      primaryClass={cond-mat.supr-con},
      url={https://arxiv.org/abs/2412.05465},
      journal = {arXiv preprint}
}

@misc{numerical_integration,
  note = {The numerical integration is restricted to a small momentum range around the peak to avoid numerical instability.}
}

@article{maletinsky2012robust,
  title={A robust scanning diamond sensor for nanoscale imaging with single nitrogen-vacancy centres},
  author={Maletinsky, Patrick and Hong, Sungkun and Grinolds, Michael Sean and Hausmann, Birgit and Lukin, Mikhail D and Walsworth, Ronald L and Loncar, Marko and Yacoby, Amir},
  journal={Nature nanotechnology},
  volume={7},
  number={5},
  pages={320--324},
  year={2012},
  publisher={Nature Publishing Group UK London},
  url={https://www.nature.com/articles/nnano.2012.50}
}

@article{gong2023coherent,
  title={Coherent Dynamics of Strongly Interacting Electronic Spin Defects in Hexagonal Boron Nitride},
  author={Gong, Ruotian and He, Guanghui and Gao, Xingyu and Ju, Peng and Liu, Zhongyuan and Ye, Bingtian and Henriksen, Erik A and Li, Tongcang and Zu, Chong},
  journal={Nature Communications},
  volume={14},
  number={1},
  pages={3299},
  year={2023},
  publisher={Nature Publishing Group},
  url = {https://doi.org/10.1038/s41467-023-39115-y}
}

@article{stern2022room,
  title={Room-temperature optically detected magnetic resonance of single defects in hexagonal boron nitride},
  author={Stern, Hannah L and Gu, Qiushi and Jarman, John and Eizagirre Barker, Simone and Mendelson, Noah and Chugh, Dipankar and Schott, Sam and Tan, Hoe H and Sirringhaus, Henning and Aharonovich, Igor and others},
  journal={Nature Communications},
  volume={13},
  number={1},
  pages={618},
  year={2022},
  publisher={Nature Publishing Group},
  url = {https://doi.org/10.1038/s41563-024-01887-z}
}

@article{gong2023isotope,
  title={Isotope engineering for spin defects in van der Waals materials},
  author={Gong, Ruotian and Du, Xinyi and Janzen, Eli and Liu, Vincent and Liu, Zhongyuan and He, Guanghui and Ye, Bingtian and Li, Tongcang and Yao, Norman Y and Edgar, James H and others},
  journal={Nature Communications},
  volume={15},
  number={1},
  pages={104},
  year={2024},
  publisher={Nature Publishing Group UK London},
  url = {https://doi.org/10.1038/s41467-023-44494-3}
}

@article{pelliccione2016scanned,
  title={Scanned probe imaging of nanoscale magnetism at cryogenic temperatures with a single-spin quantum sensor},
  author={Pelliccione, Matthew and Jenkins, Alec and Ovartchaiyapong, Preeti and Reetz, Christopher and Emmanouilidou, Eve and Ni, Ni and Bleszynski Jayich, Ania C},
  journal={Nature nanotechnology},
  volume={11},
  number={8},
  pages={700--705},
  year={2016},
  publisher={Nature Publishing Group UK London},
  url = {https://doi.org/10.1038/nnano.2016.68}
}

@article{abrikosov1957magnetic,
  title={On the magnetic properties of superconductors of the second group},
  author={Abrikosov, Alexei A},
  journal={Soviet Physics-JETP},
  volume={5},
  pages={1174--1182},
  year={1957},
  url = {https://www.jetp.ras.ru//cgi-bin/e/index/e/5/6/p1174?a=list}
}

@article{abrikosovRMP,
  title = {Nobel Lecture: Type-II superconductors and the vortex lattice},
  author = {Abrikosov, A. A.},
  journal = {Rev. Mod. Phys.},
  volume = {76},
  issue = {3},
  pages = {975--979},
  numpages = {0},
  year = {2004},
  month = {Dec},
  publisher = {American Physical Society},
  doi = {10.1103/RevModPhys.76.975},
  url = {https://link.aps.org/doi/10.1103/RevModPhys.76.975}
}

@article{NMRqc_RMP,
  title = {NMR techniques for quantum control and computation},
  author = {Vandersypen, L. M. K. and Chuang, I. L.},
  journal = {Rev. Mod. Phys.},
  volume = {76},
  issue = {4},
  pages = {1037--1069},
  numpages = {0},
  year = {2005},
  month = {Jan},
  publisher = {American Physical Society},
  doi = {10.1103/RevModPhys.76.1037},
  url = {https://link.aps.org/doi/10.1103/RevModPhys.76.1037}
}

@article{VladPRX,
  title = {Superconductivity near a Ferroelectric Quantum Critical Point in Ultralow-Density Dirac Materials},
  author = {Kozii, Vladyslav and Bi, Zhen and Ruhman, Jonathan},
  journal = {Phys. Rev. X},
  volume = {9},
  issue = {3},
  pages = {031046},
  numpages = {23},
  year = {2019},
  month = {Sep},
  publisher = {American Physical Society},
  doi = {10.1103/PhysRevX.9.031046},
  url = {https://link.aps.org/doi/10.1103/PhysRevX.9.031046}
}

@article{VladPRL,
  title = {Synergetic Ferroelectricity and Superconductivity in Zero-Density Dirac Semimetals near Quantum Criticality},
  author = {Kozii, Vladyslav and Klein, Avraham and Fernandes, Rafael M. and Ruhman, Jonathan},
  journal = {Phys. Rev. Lett.},
  volume = {129},
  issue = {23},
  pages = {237001},
  numpages = {8},
  year = {2022},
  month = {Nov},
  publisher = {American Physical Society},
  doi = {10.1103/PhysRevLett.129.237001},
  url = {https://link.aps.org/doi/10.1103/PhysRevLett.129.237001}
}

@article{volkov2022superconductivity,
  title={Superconductivity from energy fluctuations in dilute quantum critical polar metals},
  author={Volkov, Pavel A and Chandra, Premala and Coleman, Piers},
  journal={Nature communications},
  volume={13},
  number={1},
  pages={4599},
  year={2022},
  publisher={Nature Publishing Group UK London},
  url = {https://doi.org/10.1038/s41467-022-32303-2}
}

@article{gozar2008high,
  title={High-temperature interface superconductivity between metallic and insulating copper oxides},
  author={Gozar, A and Logvenov, G and Kourkoutis, L Fitting and Bollinger, AT and Giannuzzi, LA and Muller, DA and Bozovic, I},
  journal={Nature},
  volume={455},
  number={7214},
  pages={782--785},
  year={2008},
  publisher={Nature Publishing Group UK London},
  url = {https://doi.org/10.1038/nature07293}
}

@article{gozar2016high,
  title={High temperature interface superconductivity},
  author={Gozar, A and Bozovic, I},
  journal={Physica C: Superconductivity and its Applications},
  volume={521},
  pages={38--49},
  year={2016},
  publisher={Elsevier},
  url = {https://doi.org/10.1016/j.physc.2016.01.003}
}

@article{song2021high,
  title={High temperature superconductivity at FeSe/LaFeO3 interface},
  author={Song, Yuanhe and Chen, Zheng and Zhang, Qinghua and Xu, Haichao and Lou, Xia and Chen, Xiaoyang and Xu, Xiaofeng and Zhu, Xuetao and Tao, Ran and Yu, Tianlun and others},
  journal={Nature Communications},
  volume={12},
  number={1},
  pages={5926},
  year={2021},
  publisher={Nature Publishing Group UK London},
  url = {https://doi.org/10.1038/s41467-021-26201-2}
}

@article{brun2016review,
  title={Review of 2D superconductivity: the ultimate case of epitaxial monolayers},
  author={Brun, Christophe and Cren, Tristan and Roditchev, Dimitri},
  journal={Superconductor Science and Technology},
  volume={30},
  number={1},
  pages={013003},
  year={2016},
  publisher={IOP Publishing},
  url = {10.1088/0953-2048/30/1/013003}
}

@book{Kardar_2007, place={Cambridge}, title={Statistical Physics of Particles}, publisher={Cambridge University Press}, author={Kardar, Mehran}, year={2007}, url={
https://doi.org/10.1017/CBO9780511815898}}

@article{Boson_amp_operator,
  title = {Ordered Expansions in Boson Amplitude Operators},
  author = {Cahill, K. E. and Glauber, R. J.},
  journal = {Phys. Rev.},
  volume = {177},
  issue = {5},
  pages = {1857--1881},
  numpages = {0},
  year = {1969},
  month = {Jan},
  publisher = {American Physical Society},
  doi = {10.1103/PhysRev.177.1857},
  url = {https://link.aps.org/doi/10.1103/PhysRev.177.1857}
}

@article{lopez2000matrix,
  title={Matrix elements for the one-dimensional harmonic oscillator},
  author={L{\'o}pez-Bonilla, J and Ovando, G},
  journal={Bull. Irish Math. Soc.(44)},
  volume={61},
  year={2000},
  url = {https://doi.org/10.33232/BIMS.0044.61.65}
}

@article{minowa2025direct,
  title={Direct excitation of Kelvin waves on quantized vortices},
  author={Minowa, Yosuke and Yasui, Yuki and Nakagawa, Tomo and Inui, Sosuke and Tsubota, Makoto and Ashida, Masaaki},
  journal={Nature Physics},
  volume={21},
  number={2},
  pages={233--238},
  year={2025},
  publisher={Nature Publishing Group UK London},
  url = {https://doi.org/10.1038/s41567-024-02720-9}
}

@article{potts2025spin,
  title={Spin-Qubit Noise Spectroscopy of Magnetic Berezinskii--Kosterlitz--Thouless Physics},
  author={Potts, Mark and Zhang, Shu},
  journal={Nano Letters},
  volume={25},
  number={51},
  pages={17677--17684},
  year={2025},
  publisher={ACS Publications},
  url = {https://doi.org/10.1021/acs.nanolett.5c04627}
}

@article{Mehler1866,
author = {Mehler, F.G.},
journal = {Journal für die reine und angewandte Mathematik},
language = {ger},
pages = {161-176},
title = {Ueber die Entwicklung einer Function von beliebig vielen Variablen nach Laplaceschen Functionen höherer Ordnung.},
url = {http://eudml.org/doc/147987},
volume = {66},
year = {1866},
}

@article{London1935,
    author = {London, F. and London, H.},
    title = {The electromagnetic equations of the supraconductor},
    journal = {Proceedings of the Royal Society of London. A. Mathematical and Physical Sciences},
    volume = {149},
    number = {866},
    pages = {71-88},
    year = {1935},
    month = {03},
    issn = {0080-4630},
    doi = {10.1098/rspa.1935.0048},
    url = {https://doi.org/10.1098/rspa.1935.0048},
}

@misc{2026detectinghalfquantum,
      title={Detecting half-quantum superconducting vortices by spin-qubit relaxometry}, 
      author={Gábor B. Halász and Nirjhar Sarkar and Yueh-Chun Wu and Joshua T. Damron and Chengyun Hua and Benjamin Lawrie},
      year={2026},
      eprint={2601.19975},
      archivePrefix={arXiv},
      primaryClass={cond-mat.supr-con},
      url={https://arxiv.org/abs/2601.19975}, 
}

@article{Labusch1969,
author = {Labusch, R.},
title = {Elastic Constants of the Fluxoid Lattice Near the Upper Critical Field},
journal = {physica status solidi (b)},
volume = {32},
number = {1},
pages = {439-442},
doi = {https://doi.org/10.1002/pssb.19690320145},
year = {1969}
}

@article{Brandt1977,
  title={Elastic energy of the vortex state in type II superconductors. II. Low inductions},
  author={Ernst Helmut Brandt},
  journal={Journal of Low Temperature Physics},
  year={1977},
  volume={26},
  pages={735-753},
  url={https://doi.org/10.1007/BF00654877}
}

@article{Fetter1966,
  title = {Stability of a Lattice of Superfluid Vortices},
  author = {Fetter, A. L. and Hohenberg, P. C. and Pincus, P.},
  journal = {Phys. Rev.},
  volume = {147},
  issue = {1},
  pages = {140--152},
  numpages = {0},
  year = {1966},
  month = {Jul},
  publisher = {American Physical Society},
  doi = {10.1103/PhysRev.147.140},
  url = {https://link.aps.org/doi/10.1103/PhysRev.147.140}
}

\begin{widetext}

\clearpage
\appendix
\section{Derivation of the transverse conductivity at zero-field}
\label{app:ZeroB}
In this appendix, we provide details of the calculation of the transverse conductivity arising from supercurrent fluctuations, which was omitted in the main text.
In particular, we derive the scaling functions $F_{\rm m, sc}(q\xi)$ that determine the transverse conductivity.

\subsection{Metallic side}
We first begin with the metallic side.
Recall that from Eq.~\eqref{main:sigmaTm}, the transverse conductivity is given by,
\begin{equation}
\label{appeq:sigmaTm}
\sigma_{\T}^{\rm m}(q) = \frac{1}{2 k_B T}
 \left( \frac{\hbar e^*}{ m^*}\right)^2
 \int_{\k,t} [\k - \hat{\q}(\hat{\q}\cdot\k)]^2 C(\k-\q,t)C(\k,t) \,,
\end{equation}
where the order-parameter correlation function $C(\q,t)$ is obtained by solving the linearized TDGL (Eq.~\eqref{eq:Langevin_TGzL}) in momentum and frequency space, using $\nabla\to i\q$, $\partial_t \to -i\omega$,
\begin{equation}
    \psi_{\q}(\omega)=\frac{\zeta_{\q}(\omega)}{-i \hbar \omega+\Gamma\left(r+K_m \q^2\right)} \implies  \left\langle\psi^*_{\q}(\omega) \psi_{\q}\left(\omega^{\prime}\right)\right\rangle= \frac{2 \Gamma \hbar k_B T(2 \pi) \delta\left(\omega-\omega^{\prime}\right)}{\hbar^2 \omega^2+\Gamma^2 \left(r+K_m \q^2\right)^2},\quad K_m = \hbar^2/(2m^*)
\end{equation}
and we have used $\expect{\zeta^*_\q(\omega)\zeta_\q(\omega')}=2 \Gamma \hbar k_B T(2 \pi) \delta\left(\omega-\omega^{\prime}\right)$ from FDT Eq.~\eqref{eq:Gamma_defintion},
\begin{equation}
\label{appeq:Cqtm}
\begin{aligned}
    C(\q, t)&=\left\langle\psi^*_\q(t) \psi_\q(0)\right\rangle=\int_{-\infty}^{+\infty} \frac{d \omega}{2 \pi}  \, e^{-i \omega t} \frac{2 \Gamma \hbar k_B T}{\hbar^2 \omega^2+\Gamma^2 \left(r+K_m \q^2\right)^2} = \frac{k_BT}{K_m\q^2+r}e^{-\Gamma(K_m\q^2 + r)|t|/\hbar}.
    \end{aligned}
\end{equation}

Applying Eq.~\eqref{appeq:Cqtm} on Eq.~\eqref{appeq:sigmaTm}, we find,
\begin{equation}\label{zero_scaling}
\begin{split}
\sigma_{\T}^{\rm m}(q) &= \frac{1}{k_B T}
\left( \frac{\hbar e^*}{ m^*}\right)^2
\int_\k [\k - \hat{\q}(\hat{\q}\cdot\k)]^2 \frac{(k_B T)^2 \hbar}{\Big(K_m(\k-\q)^2+r\Big) \Big(K_m \k^2+r \Big) \Gamma \Big[ K_m \big(\k^2 + (\k-\q)^2\big) + 2r \Big]} \\
&= \frac{k_B T \hbar}{\Gamma}
\left( \frac{\hbar e^*}{ m^* }\right)^2
\int \frac{k\,dk\,d\phi}{(2\pi)^2} \frac{k^2 \sin^2\phi}{\Big(K_m(k^2 - 2 k q \cos\phi + q^2) +r\Big) \Big(K_m k^2+r \Big) \Big(K_m (2k^2 - 2 k q \cos\phi + q^2) + 2r \Big)} \\
&= \frac{k_B T \hbar}{\Gamma r K_m^2}
\left( \frac{\hbar e^*}{ m^* }\right)^2
\int \frac{\tilde k\, d\tilde k\, d\phi}{(2\pi)^2} \frac{\tilde k^2 \sin^2\phi}{\Big(\tilde k^2 - 2 \tilde k \tilde q \cos\phi + \tilde q^2 + 1 \Big) \Big(\tilde k^2 + 1 \Big) \Big(2 \tilde k^2 - 2 \tilde k \tilde q \cos\phi + \tilde q^2 + 2 \Big)} \\
&= \frac{k_B T \hbar}{\Gamma r K_m^2}
\left( \frac{\hbar e^*}{ m^* }\right)^2
\frac{1}{4\tilde{q}^2} \int \frac{\tilde{k}\, d \tilde{k}}{2 \pi(1+\tilde{k}^2)^2} \cdot \left[\sqrt{(2+2\tilde{k}^2+\tilde{q}^2)^2-4\tilde{k}^2\tilde{q}^2}-\sqrt{(1+\tilde{k}^2+\tilde{q}^2)^2-4\tilde{k}^2\tilde{q}^2}-(1+\tilde{k}^2)\right] \\
&= \sigma_{\rm m}(0) F_{\rm m}(\tilde q) \textrm{ where } \\
& \sigma_{\T}^{\rm m}(0) = \frac{1}{8\pi}\left(\frac{e^{*}}{\hbar}\right)^2 k_BT\, \frac{\hbar}{r \Gamma } \,, \ \ 
F_{\rm m}(\tilde{q}) = \frac{2}{ \tilde{q}^2}\left[ \ln \frac{16}{(4+\tilde{q}^2)^2}+\frac{4\tilde{q}}{\sqrt{4+\tilde{q}^2}}\operatorname{arctanh}\frac{\tilde{q}}{\sqrt{4+\tilde{q}^2}}\right].    
\end{split} 
\end{equation}
In the first line, we have performed the  integration over time $t$.
In the third line, we reparametrized the variables, introducing $\tilde k =  k \xi_{\rm m}, \ \tilde q = q \xi_{\rm m}$, where $\xi_{\rm m} $ denotes the coherence length, given by $ \xi_{\rm m} = \sqrt{K_m/r}$.
In the fourth line, we carried out the angular integration over azimuthal angle $\phi$, which denotes the angle between the momenta $\q$ and $\k$.

\subsection{Superconducting side}
We now turn to the superconducting side of the transition.
Recall that from Eq.~\eqref{Eq:CurrentCorreltor_sc}, the transverse conductivity is given by,
\begin{equation}
\label{appeq:sigmaTsc}
\sigma_{\T}^{\rm sc}(q) = \frac{1}{2 k_B T}\left( \frac{\hbar e^*}{ m^*}\right)^2\int_{\k,t} [\k - \hat{\q}(\hat{\q}\cdot\k)]^2 C_{\chi\chi}(\k-\q,\nu)C_{\theta\theta}(\k,\nu)    ,
\end{equation}
where $C_{\chi\chi}$ and $C_{\theta\theta}$ denote the Higgs (amplitude) and Goldstone (phase) correlation functions, derived from the linearized Langevin equations Eq.~\eqref{eq:LangevinSc} subject to the FDT in Eq.~\eqref{ampphase_noise},

\begin{subequations}
\label{appeq:Cqtsc}
\begin{align}
     C_{\chi\chi} (\q,t) &= \frac{2 k_B T M_0}{K_m \q^2 + 2u M_0} e^{-\Gamma(K_m \q^2 + 2uM_0)|t|/\hbar}   ,\\
     C_{\theta\theta} (\q,t) &= \frac{k_B T}{2M_0 K_m \q^2} e^{-\Gamma K_m \q^2|t|/\hbar}  . 
\end{align}
\end{subequations}

Applying Eq.~\eqref{appeq:Cqtsc} on Eq.~\eqref{appeq:sigmaTsc}, we find,
\begin{equation}
\begin{split}
    \sigma_{\T}^{\rm sc}(q) &= \frac{k_B T \hbar}{\Gamma} \left(\frac{\hbar e^*}{ m^*}\right)^2 \int_{\k} \frac{[\k - \hat{\q}(\hat{\q}\cdot\k)]^2}{\Big(K_m (\k-\q)^2 + 2u M_0\Big) K_m \k^2 \Big[K_m\big((\k-\q)^2 + \k^2\big) + 2uM_0\Big]} \\
    &= \frac{k_B T \hbar}{\Gamma K_m} \left(\frac{\hbar e^*}{ m^*}\right)^2  \int \frac{k\,dk\,d\phi}{(2\pi)^2} \frac{\sin^2 \phi}{\Big(K_m (k^2 - 2kq\cos\phi +q^2) + 2u M_0\Big) \Big(K_m(2k^2 - 2kq \cos\phi + q^2) + 2uM_0 \Big)} \\
    &= \frac{k_B T \hbar}{\Gamma K_m^2 (2 uM_0)} \left(\frac{\hbar e^*}{ m^*}\right)^2  \int \frac{\tilde k\,d\tilde k\,d\phi}{(2\pi)^2} \frac{\sin^2 \phi}{ (\tilde k^2 - 2\tilde k \tilde q\cos\phi + \tilde q^2 + 1) (2\tilde k^2 - 2\tilde k \tilde q \cos\phi + \tilde q^2 + 1)} \\
    &= \frac{k_B T \hbar}{\Gamma K_m^2 (2 uM_0)} \left(\frac{\hbar e^*}{ m^*}\right)^2  \int \frac{\tilde k\, d\tilde k}{2\pi} \frac{\sqrt{(1+2\tilde{k}^2+\tilde{q}^2)^2-4\tilde{k}^2\tilde{q}^2}-\sqrt{(1+\tilde{k}^2+\tilde{q}^2)^2-4\tilde{k}^2\tilde{q}^2}-\tilde{k}^2}{4 \tilde k^4 \tilde q^2} \\
    &= \sigma_{\rm sc}(0) F_{\rm sc}(\tilde q) \textrm{ where } \\
    & \sigma_{\rm sc}(0) = \frac{\ln 2}{\pi} \left(\frac{e^*}{\hbar}\right)^2 k_B T \frac{\hbar}{2 u M_0 \Gamma}
    \,, \ \ F_{\rm sc}(\tilde{q}) = \frac{1}{\tilde{q}^2\ln2}\left[\ln\left(4\frac{1+\tilde{q}^2}{2+\tilde{q}^2}\right) - \frac{\ln(2+\tilde{q}^2)}{1+\tilde{q}^2}\right]   .
\end{split}
\end{equation}
Here, we have followed the same procedure as on the metallic side and introduced the dimensionless momentum variable $\tilde k = k \xi_{\rm sc}$, where $\xi_{\rm sc} = \sqrt{K_m / (2 |r|)}$ denotes the coherence length.

\section{Derivation of the transverse conductivity at finite-field}
\label{app:B_TDGL}
In this appendix, we present the details of the calculation of the transverse conductivity in the presence of a background magnetic field $\H$.
We first derive the order-parameter correlation function and effective correlation length $\xi_H$, followed by the transverse conductivity and its associated scaling function $F_H (q\xi_H, \ell / \xi_H)$.
Finally, we analyze the magnetic noise and elucidate how it differs from the zero–magnetic-field case.

\subsection{Order-parameter correlation function}
To begin with, under a finite magnetic field, the supercurrent \cite{London1935} is rewritten as
\begin{equation}
\begin{aligned}
\boldsymbol{J}_{s}&=\frac{\hbar e^*}{2 m^* i}\left(\psi^* \nabla \psi-\psi \nabla \psi^*\right)-\frac{\left(e^*\right)^2}{m^*}|\psi|^2 \A
=\frac{\hbar e^*}{2m^* i} \Big( \psi^* D \psi - \psi (D \psi)^* \Big)   ,
\end{aligned}
\end{equation}
where the differential operator $D_\mu=\partial_\mu-i e^* A_\mu / \hbar$. We may use Wick's theorem to compute the current-current correlation function $\left\langle J_{\mu 1} J_{\nu 2}\right\rangle=\left(\frac{\hbar e^*}{2 m^* i}\right)^2\left[\left(D_{\mu 1}-D_{\mu 3}^*\right)\left(D_{\nu 2}-D_{\nu 4}^*\right)\left\langle\psi_1^* \psi_3 \psi_2^* \psi_4\right\rangle\right]_{3=1,4=2}$. We can then apply the classical fluctuation–dissipation theorem (FDT) Eq.~\eqref{eq:sigma_current_correlator} to compute the nonlocal conductivity \cite{DorseyHuse}, which gives
\begin{equation}\label{app:sigma_H}
\sigma_{\mu \nu}(\q)= \frac{-1}{2 k_B T}\left(\frac{\hbar e^*}{2 m^*}\right)^2 \int \frac{d \Omega}{2 \pi} \int d^2\left(\bm{\rho}_1\!-\!\bm{\rho}_2\right) e^{i \q \cdot\left(\bm{\rho}_1-\bm{\rho}_2\right)} \left(D_{\mu 1}\!-\!D_{\mu 3}^*\right)\left(D_{\nu 2}\!-\!D_{\nu 4}^*\right) C\left(\bm{\rho}_2, \bm{\rho}_3, \Omega\right)  C\left(\bm{\rho}_1, \bm{\rho}_4, \Omega\right)\left.\right|_{3=1,4=2}  .
\end{equation}
The order-parameter correlation function $C$ is related to the response function $\chi$ by the classical FDT. In the classical (high-temperature or low-frequency) limit, $\hbar\Omega \ll k_B T$, the Bose factor reduces to $(1-e^{-\hbar \Omega / k_B T})^{-1} \approx k_B T / (\hbar \Omega)$ \cite{kuboFDT1966}:
\begin{equation}\label{psi_FDT}
C\left(\bm{\rho}, \bm{\rho}^{\prime} ; \Omega\right) =\int d\left(t\!-\!t^{\prime}\right) e^{i \Omega\left(t-t^{\prime}\right)}\left\langle\psi(\bm{\rho}, t) \psi^*\left(\bm{\rho}^{\prime}, t^{\prime}\right)\right\rangle=\frac{2 k_B T}{\hbar\Omega} \mathrm{Im}\, \chi\left(\bm{\rho}, \bm{\rho}^{\prime} ; \Omega\right)   ,
\end{equation}
where $\chi\left(\bm{\rho}, \bm{\rho}^{\prime} ; \Omega\right)$ is the order-parameter response function, defined via $\psi(\bm\rho, \Omega)=\int d^2 \bm\rho^{\prime} \, \chi\left(\bm\rho, \bm\rho^{\prime} ; \Omega\right) \zeta\left(\bm\rho^{\prime}, \Omega\right)$, which is the solution of the linearized TDGL equations,
\begin{subequations}\begin{align}
\hbar\Gamma^{-1}\partial_t \psi - K_m D^2 \psi + r \psi  & = \zeta   , \\
\left(-i \hbar\Gamma^{-1} \Omega- K_m D^2 + r\right) \chi\left(\bm{\rho} , \bm{\rho}^{\prime} ; \Omega\right) &=\delta^{(2)}\left(\bm{\rho}-\bm{\rho}^{\prime}\right)   .
\end{align}\end{subequations}
In the Landau gauge $\mathbf{A}=\mu_0 H(0, x, 0)$, the eigenfunctions of the operator $-K_m D^2$ are the standard Landau level eigenfunctions $u_n\left(x-x_0\right)$ with guiding centers $x_0= q_y\ell^2$; the parameter $r(T)$ simply shifts the energy eigenvalues $\epsilon_n$. The n-th energy level is $\epsilon_n = (n+1/2)\hbar\omega_c + r$, with the cyclotron frequency defined as $\omega_c = \mu_0 e^* H/m^*$. Using the completeness relation $\delta\left(x_1-x_2\right)=\sum_{n=0}^{\infty} u_n\left(x_1-x_0\right) u_n^*\left(x_2-x_0\right)$ and Eq.~\eqref{psi_FDT}, we obtain the correlation function
\begin{equation}
C\left(x_1, x_2 ; q_y, \Omega\right)=\frac{2 \hbar k_B T}{\Gamma} \sum_{n=0}^{\infty} \frac{u_n\left(x_1-x_0\right) u_n\left(x_2-x_0\right)}{(\hbar\Omega / \Gamma)^2+\epsilon_n^2}.
\label{appeq:CHxw}
\end{equation}
After integration over $q_y$, we arrive at the correlation function in real space, as shown in Eq.~\eqref{eq:C_H}. Here we have applied the eigenfunctions of Landau levels, given by,
\begin{equation}
u_n(x) = \left( \frac{1}{2^n n!\sqrt{\pi} \ell} \right)^{\frac{1}{2}} e^{-\frac{x^2}{2\ell^2}}H_n\left(\frac{x}{\ell}\right)   ,
\end{equation}
with magnetic length $\ell = \sqrt{\hbar/(\mu_0 e^* H)}$ and the Hermite polynomials $H_n(x) = (-1)^n e^{x^2} \frac{d^n}{d x^n} e^{-x^2}$. We can then define dimensionless eigenfunctions $\tilde{u}_n(X) = \sqrt{\ell}\, u_n(X\ell) = e^{-X^2/2} H_n(X) / (2^n n! \sqrt{\pi} )^{\frac12}$.

\subsection{Characteristic Lifetime \& Correlation Length}
We start by identifying the characteristic lifetime $\tau_H$ and the correlation length $\xi_H$ in the presence of a background magnetic field $\H$.
Fourier transforming Eq.~\eqref{appeq:CHxw}, we arrive at the correlation function in the time domian 
\begin{equation}
C\left(x_1, x_2; q_y, t\right)=\frac{2\hbar k_B T}{\Gamma} \int \frac{d \Omega}{2 \pi} e^{i \Omega t} \sum_{n=0}^{\infty} \frac{u_n\left(x_1-x_0\right) u_n\left(x_2-x_0\right)}{(\hbar\Omega / \Gamma)^2+\epsilon_n^2} =\sum_{n=0}^{\infty}\frac{k_B T}{\epsilon_n} e^{-\Gamma \epsilon_n|t|/\hbar} u_n\left(x_1-x_0\right) u_n\left(x_2-x_0\right),
\label{appeq:CHxt}
\end{equation}
From Eq.~\eqref{appeq:CHxt}, we can read off the lifetime $\tau_H= \max(\tau_n) = \hbar/(\Gamma \epsilon_0)$, which reduces to Eq.~\eqref{eq:tHxiH},
\begin{equation}
    \tau_{H} = \frac{\hbar}{\Gamma \left(r + \hbar \omega_c / 2\right)} \, .
\end{equation}
We note that in the limit of $\H \rightarrow 0$, we recover the lifetime at zero field, Eq.~\eqref{eq:tmxim} of $\tau_{\rm m} = \hbar / (\Gamma r)$.

To obtain the correlation length, we start from the equal-time correlator in real space by setting $t=0$, and integrating Eq.~\eqref{appeq:CHxt} over transverse momenta $q_y$ (equivalently guiding centers $x_0 = q_y\ell^2$), using $\int d q_y\, (2\pi)^{-1}\, e^{i q_y (y_1 - y_2)}$:
\begin{equation}\small
\begin{aligned}
C(\bm{\rho}_1,\bm{\rho}_2) &=  k_B T \sum_{n=0}^{\infty} \frac{1}{\epsilon_n}  \int \frac{d \bar{x}_0}{2 \pi \ell} \exp\left({i q_y (y_1 - y_2)}-\frac{(x_1-x_0)^2+(x_2-x_0)^2}{2\ell^2}\right) \frac{H_n\left(\frac{x_1-x_0}{\ell}\right) H_n\left(\frac{x_2-x_0}{\ell}\right)}{2^n n!\sqrt{\pi}\ell}   .  \\
&=\frac{k_B T}{\sqrt{\pi}\ell^2}  e^{-(\bar{x}_1^2+\bar{x}_2^2)/2} \int_{0}^{\infty} dp\,\sum_{n=0}^{\infty} e^{-\epsilon_n p}  \int \frac{d \bar{x}_0}{2\pi} \exp\Big(-\bar{x}_0^2 + \bar{x}_0 \big( \bar{x}_1 + \bar{x}_2 +i(\bar{y}_1 - \bar{y}_2) \big) \Big) \frac{H_n\left(\bar{x}_1-\bar{x}_0\right) H_n\left(\bar{x}_2-\bar{x}_0\right)}{2^n n!} \\
&= \frac{k_B T}{\sqrt{\pi}\ell^2}  e^{-(\bar{x}_1^2+\bar{x}_2^2)/2} \int_{0}^{\infty} dp\, e^{-\epsilon_0 p} \int_{-\infty}^{\infty} \frac{d \bar{x}_0}{2\pi} \exp\Big(-\bar{x}_0^2 + \bar{x}_0 \big( \bar{x}_1 + \bar{x}_2 +i(\bar{y}_1 - \bar{y}_2) \big) \Big) \frac{1}{\sqrt{1-\varsigma^2}} \exp \left(\frac{2 X_1 X_2 \varsigma-\left(X_1^2+X_2^2\right) \varsigma^2}{1-\varsigma^2}\right) \\
&= \frac{k_B T}{2\pi\ell^2} \int_{0}^{\infty} dp\, e^{-\epsilon_0 p} \frac{1}{1-\varsigma} \exp \Big( i \Delta_y \frac{\bar{x}_1+\bar{x}_2}{2}-\frac{1+\varsigma}{4(1-\varsigma)}\left(\Delta_y^2 + \Delta_x^2\right)\Big) \\
&\approx \frac{k_B T}{2\pi\ell^2}  \frac{1}{\hbar \omega_c} \exp \left(i \Delta_y \frac{\bar{x}_1+\bar{x}_2}{2}\right) \int_0^{\infty} \frac{d p}{p} e^{-\epsilon_0 p} e^{-\frac{\Delta^2}{2 \hbar \omega_c p}} = \frac{k_B T}{\pi\ell^2} \frac{1}{\hbar \omega_c} \exp \left(i \Delta_y \frac{\bar{x}_1+\bar{x}_2}{2}\right) K_0\left(\sqrt{\frac{2 \epsilon_0}{\hbar \omega_c}} \Delta\right)  .
\end{aligned}\label{corr_rho}
\end{equation}
In the second line, we introduced dimensionless coordinates $\bar{x}_0=x_0/\ell$, $\bar{x}_i = x_i /\ell$, $\bar{y}_i = y_i/\ell$, and rewrite $1/\epsilon_n = \int_0^\infty dp\, e^{-\epsilon_n p}$. In the third line, we extend the limits of integration to $(-\infty,+\infty)$, and apply the Mehler's Hermite polynomial formula \cite{Mehler1866}, 
\begin{equation}
\sum_{n=0}^{\infty} \frac{H_n\left(w_1\right) H_n\left(w_2\right)}{n!}\left(\frac{\varsigma}{2}\right)^n=\frac{1}{\sqrt{1-\varsigma^2}} \exp \left(\frac{2 w_1 w_2 \varsigma-\left(w_1^2+w_2^2\right) \varsigma^2}{1-\varsigma^2}\right), 
\end{equation}
where $\varsigma=e^{-\hbar \omega_c p}$, $X_i=\bar{x}_i - \bar{x}$. In the fourth line of Eq.~\eqref{corr_rho}, the separation is defined as $\Delta = \sqrt{\Delta_x^2 + \Delta_y^2}$, $\Delta_x=\bar{x}_1-\bar{x}_2$, and $\Delta_y=\bar{y}_1-\bar{y}_2$, the integral is dominated by small $p$ ($p\ll1$, $\varsigma \simeq 1-\hbar \omega_c p$), where we have used the identity $\int_0^{\infty} d p\, p^{-1} e^{-a p-b/p} =2 K_0(2 \sqrt{a b})$. For large separation $\Delta \gg 1$, modified Bessel function behaves like $K_0(x) \approx  e^{-x} \sqrt{\pi/(2x)}$. The asymptotic form reveals the exponential decay $C(\bm{\rho}_1,\bm{\rho}_2) \sim e^{-\Delta\rho/\xi_H}$, which allows us to find the correlation length, as defined in Eq.~\eqref{eq:tHxiH},
\begin{equation}\label{app:s}
\xi_H = \frac{\ell}{\sqrt{\frac{2 \epsilon_0}{\hbar \omega_c}}} = \frac{\hbar}{\sqrt{2 m^* (r + \frac{1}{2} \hbar \omega_c)}}, \quad s=\frac{\ell}{\xi_H}=\sqrt{\frac{2 r}{\hbar \omega_c}+1}   .
\end{equation}
The phase factor $\exp \left(i \Delta_y \frac{\bar{x}_1+\bar{x}_2}{2}\right)$ arises from the gauge choice. Since a physical correlator must be gauge invariant, we restore invariance by inserting the Peierls phase acquired along a straight‑line path between the two points:
\begin{equation}
C_{\mathrm{inv}}(\rr_1,\rr_2,\Omega) = \frac{2\hbar k_BT}{\Gamma} \exp\left(-i\frac{e^*}{\hbar} \int_{\bm{\rho}_1}^{\bm{\rho}_2} \mathbf{A} \cdot d \bm{\rho} \right) \int_{q_y}e^{iq_y(y_1-y_2)}\sum_{n=0}^{\infty}\frac{u_n(x_1-x_0)u_n(x_2-x_0)}{(\hbar\Omega/\Gamma)^2 + \epsilon_n^2}   .
\end{equation}

\subsection{Scaling function}
The dimensionless energy levels $\tilde\epsilon_n$ are obtained by scaling the energy levels with the cyclotron energy,
\begin{equation}\label{rescaled_energy}
\tilde\epsilon_n = \frac{\epsilon_n}{\hbar \omega_c} = \frac{\hbar \omega_c (n + \frac{1}{2}) + r}{\hbar \omega_c} = n + \frac{\ell^2}{2\xi_H^2} = n + \frac{s^2}{2}  .
\end{equation}
where the dimensionless ratio $s = \ell /\xi_H$, defined in Eq.~\eqref{app:s}, characterizes the magnetic field strength relative to the energy scale $r$. When the field is weak, $s\to+\infty$, and when the field is strong, $s\to 0^+$.

Similar to zero-field case, the transverse conductivity can be  factorized into a zero‑momentum part $\sigma_H(0, s)$ and a universal scaling function $F_H(q\xi_H, s)$,
\begin{equation}
\sigma^H_{\T}(q, s) = \sigma_H(0, s) F_H(q\xi_H, s)  .
\end{equation}
Based on \cref{Eq:CurrentCorreltor_H,app:sigma_H,corr_rho}, the scaling function is proportional to 
\begin{equation}\small\label{scaling_str}
\begin{aligned}
F_H(\tilde{q}, s) &\propto \int_{-\infty}^{\infty} \frac{d \Omega}{2\pi}\int_{\mathbb{R}^2} d^2(\rr_1 \!-\! \rr_2) \,e^{i q(x_1-x_2)} (D_{y_1} \!-\! D^*_{y_3})(D_{y_2} \!-\! D^*_{y_4})C(\rr_2,\rr_3,\Omega)C(\rr_1,\rr_4,\Omega) \Bigg|_{\substack{3=1\\4=2}} \left(\frac{2 \hbar k_B T}{\Gamma}\right)^{-2} \\
&= \int_{-\infty}^{\infty} \frac{d \Omega}{2\pi} \int_{-\infty}^{\infty} d(x_1 \!-\! x_2) \int_{-\infty}^{\infty} d(y_1 \!-\! y_2) \,e^{i q(x_1-x_2)} \left\{\left[ \partial_{y_1} - \partial_{y_3} - i \frac{e^* }{\hbar}\mu_0 H(x_1 \!+\! x_3)\right]\left[\partial_{y_2} - \partial_{y_4} - i \frac{e^* }{\hbar}\mu_0 H(x_2 \!+\! x_4)\right] \right.\\
&\qquad\qquad\times \left.\left.\int_{q_y^{\prime}}e^{iq_y^{\prime}(y_2-y_3)}\sum_{m=0}^{\infty}\frac{u_m(x_2-x_0)u_m(x_3-x_0)}{(\hbar\Omega/\Gamma)^2 + \epsilon_m^2}\int_{q_y}e^{iq_y(y_1-y_4)}\sum_{n=0}^{\infty}\frac{u_n(x_1-x_0)u_n(x_4-x_0)}{(\hbar\Omega/\Gamma)^2 + \epsilon_n^2} \right\}\right|_{3=1,4=2} \\
&= \frac{1}{(2\pi)^2}\int_{-\infty}^{\infty} d \Omega \int_{-\infty}^{\infty} d(x_1 \!-\! x_2) \,e^{i q(x_1-x_2)} \int_{-\infty}^{\infty} d q_y\left(2 q_y-\frac{2 e^*}{\hbar}  \mu_0 H x_1\right)\left(2 q_y-\frac{2 e^*}{\hbar}  \mu_0 H x_2\right) \\ 
&\qquad\qquad\qquad\qquad\qquad\qquad\qquad\qquad\quad\times\sum_{m=0}^{+\infty}\sum_{n=0}^{+\infty} \frac{u_m\left(x_2-x_0\right) u_m\left(x_1-x_0\right) u_n\left(x_1-x_0\right) u_n\left(x_2-x_0\right)}{[(\hbar\Omega/\Gamma)^2 + \epsilon_m^2][(\hbar\Omega/\Gamma)^2 + \epsilon_n^2]}\\
&= \left(\frac{2\ell^{-2}}{2\pi}\right)^2\int_{-\infty}^{\infty} d \Omega \int \frac{d x_0}{\ell^2} \int d\left(x_1\!-\!x_2\right) e^{i q\left(x_1-x_2\right)}\left(x_1\!-\!x_0\right)\left(x_2\!-\!x_0\right)\sum_{n,m=0}^{+\infty} \frac{u_m\left(x_2\!-\!x_0\right) u_m\left(x_1\!-\!x_0\right) u_n\left(x_1\!-\!x_0\right) u_n\left(x_2\!-\!x_0\right)}{[(\hbar\Omega/\Gamma)^2 + \epsilon_m^2][(\hbar\Omega/\Gamma)^2 + \epsilon_n^2]} \\
&= \left(\frac{1}{\pi\ell^3}\right)^2\int_{-\infty}^{\infty} d \Omega \sum_{n,m=0}^{+\infty} \frac{\left|\int_{-\infty}^{+\infty} d (x_i\!-\!x_0)\, u_m^*(x_i\!-\!x_0) [(x_i\!-\!x_0) e^{- i q (x_i-x_0)}] u_n(x_i\!-\!x_0)\right|^2 }{\left[(\hbar\Omega/\Gamma)^2+\epsilon_n^2\right]\left[(\hbar\Omega/\Gamma)^2+\epsilon_m^2\right]} 
\end{aligned}
\end{equation}
The evaluation proceeds as follows. In the second line, we act with the derivatives before imposing the identification $3=1$, $4=2$. The derivatives with respect to $y_1$, $y_2$, $y_3$, and $y_4$ act exclusively on the exponential factors, yielding the substitutions $\partial_{y_1} \to iq_y$, $\partial_{y_2} \to iq_y'$, $\partial_{y_3} \to -iq_y'$, and $\partial_{y_4} \to -iq_y$. 
After setting $3=1$, $4=2$, the product of the two exponentials becomes $e^{i(q_y-q_y')(y_1-y_2)}$. Integration over the relative coordinate $(y_1\!-\!y_2)$ generates a delta function $\int_{-\infty}^{\infty} d (y_1\!-\!y_2) e^{i(q_y-q_y^{\prime}) (y_1-y_2)}=2 \pi \delta\left(q_y\!-\!q_y^{\prime}\right)$, which collapses the double integral to a single integral over $q_y$.
In the penultimate line, we exploit the relation between the transverse momenta $q_y$ and the guiding centers, $q_y = x_0/\ell^2 = x_0 \hbar / (e^* \mu_0 H)$. This substitution converts the combination $(2q_y - 2e^*\mu_0 H x_i/\hbar) = -2\ell^{-2}(x_i - x_0)$. The final step involves recognizing that the resulting expression factorizes into the modulus squared of the dipole matrix element $\langle m | (x_i\!-\!x_0) e^{-iq(x_i-x_0)} | n \rangle$, weighted by the appropriate Landau level wavefunctions.

For the numerator of the scaling function $F_H(\tilde{q},s)$ in Eq.~\eqref{scaling_str}, one can define a matrix function $A_{mn}(x)$, with displacement in the $x$-direction relative to the guiding centers $x_i\!-\!x_0=X\ell$, rescaled momenta $\tilde{q} = q\xi_H$, satisfying $\int_{-\infty}^{+\infty} d (x_i-x_0)\, u_m^*(x_i-x_0) [(x_i-x_0) e^{- i q (x_i-x_0)}] u_n(x_i-x_0)  = \ell \int_{-\infty}^{+\infty} dX\, \tilde{u}_m^*(X) \left( X e^{-i \tilde{q} s X} \right) \tilde{u}_n(X) =\ell A_{mn}(\tilde{q}s) $. Note that the matrix function $A_{mn}(x)$ can be written in its first quantization form:
\begin{equation}\label{A_mn_first_quantized}
\begin{aligned}
    A_{mn}(x) &= i\frac{d}{dx}\int_{-\infty}^{+\infty} dX\, \tilde{u}_m^*(X) e^{-ixX} \tilde{u}_n(X)  = i \frac{d}{dx} \langle m | e^{-i x \hat{X}} | n \rangle .
\end{aligned}
\end{equation} 
Here, $|n\rangle$ is the $n$-th Landau level eigenstate with wavefunction $\langle X|n\rangle = \tilde{u}_n(X)$. In the Landau gauge, the position operator $\hat{X}$ can be expressed in terms of the harmonic oscillator ladder operators as $\hat{X} = (\hat{a} + \hat{a}^\dagger)/\sqrt{2}$.
The matrix element $\langle m | e^{-ix\hat{X}}| n \rangle$ can be evaluated following Refs.~\onlinecite{lopez2000matrix,Boson_amp_operator}, with associated Laguerre polynomial $L_n^{(\alpha)}$,
\begin{equation}
\begin{aligned}
\langle m| e^{-i x \hat{X}}|n\rangle & =\int_{-\infty}^{+\infty} d X\, \tilde{u}_m^*(X)\left(e^{-i x X}\right) \tilde{u}_n(X)= \left(\frac{2^n n!}{2^m m!}\right)^{\frac{1}{2}}  e^{-\frac{x^2}{4}} (-ix)^{m-n} L_{n}^{(m-n)} \left(\frac{x^2}{2}\right), \quad m\geqslant n  .
\end{aligned}
\end{equation}
For $m < n$, the result follows from the Hermiticity of $\hat{X}$, which implies the symmetry $\langle m | e^{-ix\hat{X}} | n \rangle = (\langle n | e^{ix\hat{X}} | m \rangle)^*$.
Plug above expression back to Eq.~\eqref{A_mn_first_quantized}, we find the explicit form of the matrix function,
\begin{subequations}
    \begin{align}
        &A_{mn}(x) =  i\left(\frac{2^n n!}{2^m m!}\right)^{\frac{1}{2}} e^{-\frac{x^2}{4}}(-i x)^{m-n}\left[\left(\frac{m-n}{x}-\frac{x}{2}\right) L_n^{(m-n)}\left(\frac{x^2}{2}\right)-x L_{n-1}^{(m-n+1)}\left(\frac{x^2}{2}\right)\right], \quad m\geqslant n\\
        &|A_{mn}(x)|^2 = \frac{\min (m, n)!}{\max (m, n)!}\left(\frac{x^2}{2}\right)^{|m-n|} e^{-\frac{x^2}{2}}\left[\left(\frac{|m-n|}{x}-\frac{x}{2}\right) L_{\min (m, n)}^{|m-n|}\left(\frac{x^2}{2}\right)-x L_{\min (m, n)-1}^{(|m-n|+1)}\left(\frac{x^2}{2}\right)\right]^2    ,\label{A_mn_num}
    \end{align}
\end{subequations}
where the modulus square of $A_{mn}(x)$ shown in Eq.~\eqref{A_mn_num} is part of the numerator.  

For the denominator of the scaling function Eq.~\eqref{scaling_str}, the integral is performed via the residue theorem,
\begin{equation}\label{scaling_deno}
\int_{-\infty}^{\infty} \frac{d \Omega}{\left[(\hbar\Omega/\Gamma)^2+\epsilon_n^2\right]\left[(\hbar\Omega/\Gamma)^2+\epsilon_m^2\right]}  = \frac{\pi \Gamma/\hbar}{(\hbar\omega_c)^3}\frac{1}{\tilde\epsilon_n \tilde\epsilon_m\left(\tilde\epsilon_n+\tilde\epsilon_m\right)}   .
\end{equation}
where we have applied the rescaled energy levels $\tilde\epsilon_n$ defined in Eq.~\eqref{rescaled_energy}.
Combine the numerator Eq.~\eqref{A_mn_num} and denominator Eq.~\eqref{scaling_deno}, the normalized scaling function takes the form:
\begin{equation}\label{general_scaling}
F_H(\tilde{q},s) = \frac{s^2}{g(s)} \sum_{n,m=0}^{+\infty} \frac{|A_{mn}(\tilde{q}s)|^2}{\tilde{\epsilon}_n(s) \tilde{\epsilon}_m(s) [\tilde{\epsilon}_n(s) + \tilde{\epsilon}_m(s)]}   ,
\end{equation}
where a normalization function $g(s)$ is introduced to ensure $F_H(0,s)=1$, which can be evaluated by setting $\tilde{q}=0$, 
\begin{equation}\label{norm_factor}
\begin{aligned}
\frac{g(s)}{s^2}&=\sum_{n, m=0}^{+\infty} \frac{\left|A_{m n}(0)\right|^2}{\left(\frac{s^2}{2}+n\right)\left(\frac{s^2}{2}+m\right)\left(s^2+n+m\right)} = \sum_{n=0}^{+\infty} \frac{n+1}{(\frac{s^2}{2}+n)(\frac{s^2}{2}+n+1)(s^2+2 n+1)}  \\
&= \sum_{n=0}^{+\infty}\left(-\frac{\frac{s^2}{2}-1}{n+\frac{s^2}{2}}-\frac{\frac{s^2}{2}}{n+\frac{s^2}{2}+1}+\frac{4 \frac{s^2}{2}-2}{2 n+2 \frac{s^2}{2}+1}\right)
=1+\left(s^2-1\right)\left[\psi\left(\frac{s^2}{2}\right)-\psi\left(\frac{s^2+1}{2}\right)\right]    .
\end{aligned}
\end{equation}
The normalization follows from the ladder-operator identity $\langle m|\hat{X}|n\rangle = \sqrt{\frac{n}{2}} \delta_{m,n-1} + \sqrt{\frac{n+1}{2}} \delta_{m,n+1}$, which gives $\left|A_{m n}(0)\right|^2=|\langle m| \hat{X} | n\rangle|^ 2=\frac{1}{2}\big((n+1) \delta_{m, n+1}+n \delta_{m, n-1}\big)$, which is equivalent to $(n+1) \delta_{m, n+1}$ in the double sum. The sum is performed using the digamma‑function relations $\sum_{n=0}^N \frac{1}{n+C}=\psi(N+1+C)-\psi(C)$ and $\psi(C+1)=\psi(C)+\frac{1}{C}$.
The function $g(s)$ decreases monotonically from $\lim_{s \to 0} g(s) = 2$ to $\lim_{s \to +\infty} g(s) = \frac{1}{2}$ as $s$ increases.

Overall, the scaling function $F_H(\tilde{q},s)$ thus obtained in plotted in Fig.~\ref{fig:FHApp}. 

\begin{figure}[!t]
    \centering
    \includegraphics[width=0.6\linewidth]{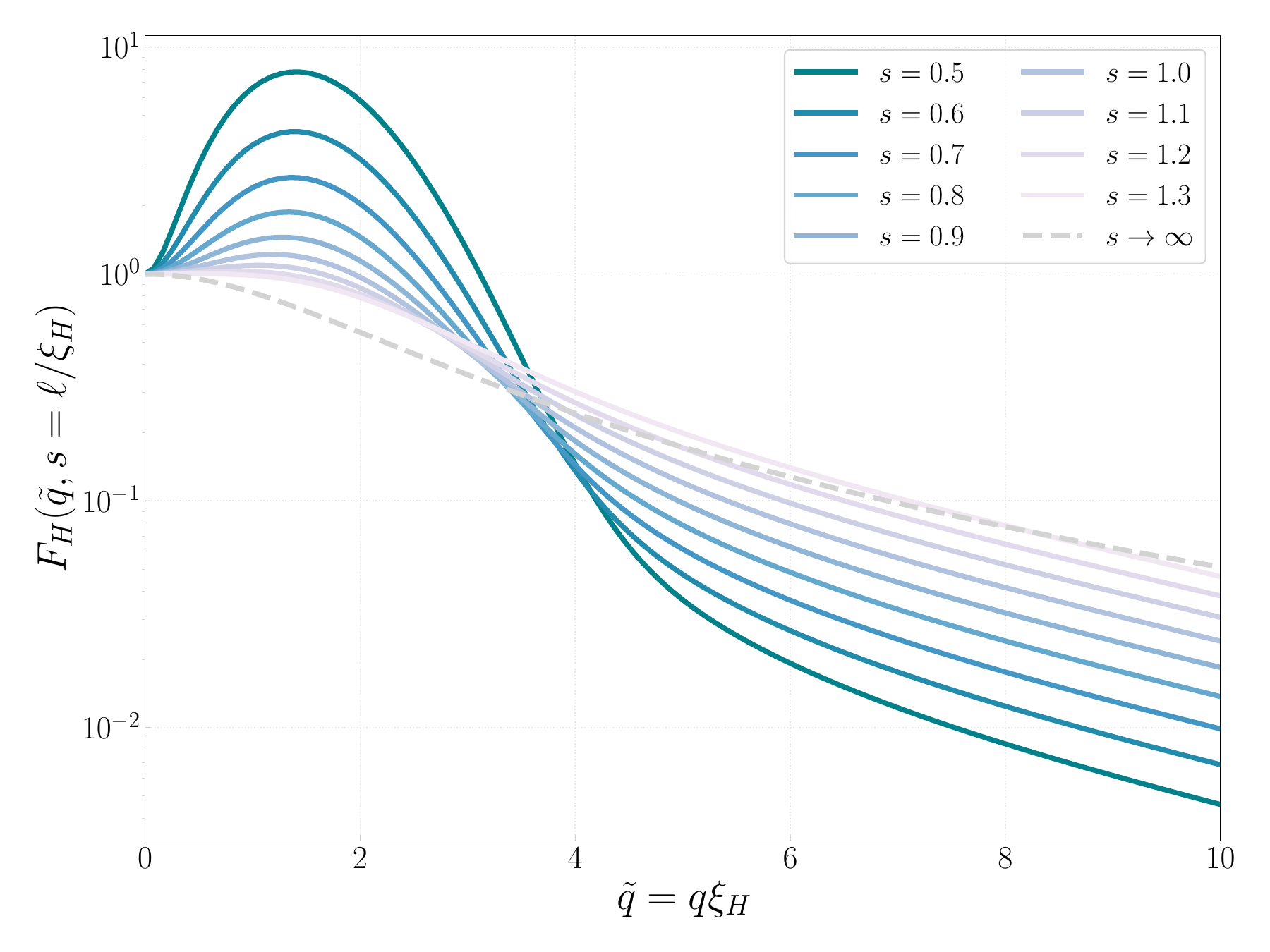}
    \caption{Scaling function $F_H(\tilde{q},s)$.}
    \label{fig:FHApp}
\end{figure}

Next, we derive the limiting cases of the scaling function $F_H$.
In the \textit{zero-field limit}, $B=0$, $s \to +\infty$: the scaling function should reduce to the known zero-field scaling function in Eq.~\eqref{zero_scaling}
\begin{equation}
F_H(\tilde{q}, s \to +\infty) = \frac{2}{\tilde{q}^2}\left[ \ln \frac{16}{(4+\tilde{q}^2)^2}+\frac{4\tilde{q}}{\sqrt{4+\tilde{q}^2}}\operatorname{arctanh}\frac{\tilde{q}}{\sqrt{4+\tilde{q}^2}}\right]  .
\end{equation}
As a \textit{weak magnetic field} is turned on, $B\to0^+$, $s\to+\infty$: we consider expansion of the scaling function near $s=+\infty$. 
A small-$x$ expansion of the matrix $A_{mn}(x)$ using Taylor series, $A_{m n}(x)=\langle m| \hat{X}|n\rangle-i x\langle m| \hat{X}^2|n\rangle-\frac{1}{2} x^2\langle m| \hat{X}^3|n\rangle+\mathcal{O}\left(x^3\right), ~ x=\tilde{q}s$. Note that odd powers vanish after taking the modulus squared, therefore $|A_{mn}(x)|^2$ is,
\begin{equation}\label{matrix_small_q}\small
\begin{aligned}
|A_{mn}(x)|^2 &=\frac{1}{2} \left[\left(n+1\right) \delta_{m, n+1} + n \delta_{m, n-1}\right] \\  &\quad  + x^2\left[\left(n+\frac{1}{2}\right)^2 \delta_{m, n}+\frac{(n+1)(n+2)}{4} \delta_{m, n+2}+\frac{n(n-1)}{4} \delta_{m, n-2}-\frac{3 n^2}{4} \delta_{m, n-1}-\frac{3(n+1)^2}{4} \delta_{m, n+1}\right] + \mathcal{O}\left(x^4\right)  ,
\end{aligned}
\end{equation}
applying Eq.~\eqref{matrix_small_q} on Eq.~\eqref{general_scaling} yields the small-momentum expansion of scaling function in the large-$s$ limit,
\begin{equation}\small
\begin{aligned}
    F_H(\tilde{q}, s\to+\infty) &= 1 + (\tilde{q}s)^2 \frac{s^2}{g(s)} \left\{ \frac{3\left(s^2-1\right)^2}{4} \left[\psi\left(\frac{s^2}{2}\right)-\psi\left(\frac{s^2+1}{2}\right)\right] - \frac{s^2-1}{2} \psi^{(1)}\left(\frac{s^2}{2}\right)-\frac{\left(s^2-1\right)^2}{16} \psi^{(2)}\left(\frac{s^2}{2}\right) +\frac{3(2s^2-1)}{8} \right\} +\cdots\\
    &= 1 + \tilde{q}^2 \left[-\frac{5}{24}-\frac{5}{24 s^2}+\mathcal{O}\left(\frac{1}{s^4}\right)\right] + \tilde{q}^4 \left[\frac{11}{240}+\frac{11}{120 s^2}+\mathcal{O}\left(\frac{1}{s^4}\right)\right] + \mathcal{O}\left(\tilde{q}^6\right)   .
\end{aligned}
\end{equation}
In the \textit{strong-field limit}, $B\to+\infty$, $s\to0^+$: the dominating term in the double sum Eq.~\eqref{general_scaling} is the lowest landau level $n=m=0$, 
\begin{equation}\label{scaling_lowest_laudau}
\begin{aligned}
F_H(\tilde{q},s\to 0)
& \simeq \frac{s^2}{g(0)} \frac{\left|A_{00}(\tilde{q}s)\right|^2}{\left(\frac{s^2}{2}+0\right)\left(\frac{s^2}{2}+0\right)\left(s^2+0+0\right)} = \frac{2 s^2}{s^6}  e^{-\frac{\tilde{q}^2s^2}{2}}\left[\left(-\frac{\tilde{q}s}{2}\right) L_0^{(0)}\left(\frac{(\tilde{q}s)^2}{2}\right)-\tilde{q}s L_{-1}^{(1)}\left(\frac{(\tilde{q}s)^2}{2}\right)\right]^2 \\
& = \frac{2}{s^4} e^{-\frac{\tilde{q}^2s^2}{2}}\left(-\frac{\tilde{q}s}{2}\right)^2=\frac{1}{2 s^2} \tilde{q}^2 e^{-\frac{\tilde{q}^2s^2}{2}}  .
\end{aligned}
\end{equation}
Finally, collecting the prefactors from \cref{app:sigma_H,eq:C_H,scaling_str,scaling_deno,norm_factor}, the zero-momentum conductivity $\sigma_H(0, s)$ is determined to be
\begin{equation}\label{sigma_T_H}
\begin{aligned}
\sigma_H(0, s) & = \frac{1}{2 k_B T } \left( \frac{\hbar e^*}{2 m^*}\right)^2 \left(\frac{2 \hbar k_B T}{\Gamma}\right)^2 \left(\frac{\mu_0 e^* H}{\pi\hbar}\right)^2 \frac{\pi\Gamma/\hbar}{(\hbar \omega_c)^3}\sum_{n, m=0}^{+\infty} \frac{\left|A_{m n}(0)\right|^2}{\tilde{\epsilon}_n(s) \tilde{\epsilon}_m(s)\left[\tilde{\epsilon}_n(s)+\tilde{\epsilon}_m(s)\right]}\\
& = \frac{1}{2\pi} k_B T \left(\frac{e^*}{\hbar}\right)^2 \frac{m^* \xi_H^2}{\hbar\Gamma} g(s)= \frac{1}{4\pi} k_B T \left(\frac{e^*}{\hbar}\right)^2 \frac{\hbar}{\Gamma(r + \frac{1}{2} \hbar \omega_c)} g(s)   \,,
\end{aligned}
\end{equation}
thus arriving at Eq.~\eqref{eq:sigmaTB} in the main text.
In the limit $s = 0$, corresponding to zero magnetic field, this expression reduces to the zero-field conductivity, $\sigma_{H}(0,s=0) = \sigma_{\rm m}(0)$ in Eq.~\eqref{eq:sigma_Tm}, providing a useful consistency check.

\subsection{Asymptotic behavior of noise from critical fluctuations}
As discussed in Refs.~\onlinecite{Agarwal2016,DC2022}, the noise tensor of magnetic fluctuations can be related to the correlations of sources within the sample by solving Maxwell's equations. 
Neglecting retardation effects, most of the noise arises due to evanescent electromagnetic (EM) modes. 
The transverse noise $\mathcal{N}_{\T}$ can be related to the non-local conductivity $\sigma_{\T}$ as,
\begin{equation}\label{noise_original}
\mathcal{N}_{\T}(\Omega)=\frac{\mu_0 k_B T}{16 \pi \Omega z_0^3} \int_0^{\infty} d x\, x^2 e^{-x} \operatorname{Im}\left[r_s\left(\frac{x}{2 z_0}, \Omega\right)\right],
\end{equation}
where $r_s(\q, \Omega)$ is the reflection coefficient for $s$-polarized EM waves which couple to transverse currents, $\q \cdot \j^{\T}(\q)=0$. Here $\q$ is the in-plane momentum and the out-of-plane momentum is substituted with $q_z \approx i q$ ($q\gg\Omega/c$ so that we consider only evanescent waves). One can relate the reflection coefficient $r_s$ to the transverse conductivity \cite{Agarwal2016},
\begin{equation}\label{r_s}
r_s(\q, \Omega) = -\left(1+\frac{2 i q}{\mu_0 \Omega \sigma_{\T}(\q, \Omega)}\right)^{-1}   .
\end{equation}
Further restricting to the experimentally relevant limit of $\hbar\Omega \ll k_B T$, in small-$\Omega$ limit $\operatorname{Im}\left[r_s\right] \simeq \mu_0 \Omega \operatorname{Re}[\sigma_{\T}] /(2 q)$, plug Eq.~\eqref{r_s} back to Eq.~\eqref{noise_original} with momentum $q = x /(2 z_0)$, the transverse magnetic noise $\mathcal{N}_{\T}$ is given by:
\begin{equation}
\mathcal{N}_{\T}=\frac{\mu_0^2 k_B T}{4 \pi} \int_0^\infty dq \, q \, e^{-2 q z_0} \, \text{Re}\left[ \sigma_{\T}(q,\Omega) \right]  .
\end{equation}
Then we can apply Eq.~\eqref{sigma_T_H} and express the noise $\mathcal{N}_{\T}^H$ with zero-momentum conductivity and scaling functions, 
\begin{equation}
    \begin{aligned}
        \mathcal{N}_{\T}^H&=\frac{\mu_0^2 k_B T}{4 \pi} \sigma_H(0, s) \int_0^{\infty} d q\, q\, e^{- 2 q z_0}\, F_H\left( q\xi_H, s\right)= \frac{\mu_0^2 k_B T}{4 \pi} \frac{\sigma_H(0, s)}{\xi_H^2} \int_0^{\infty}  d \tilde{q}\, \tilde{q}\, e^{-2\tilde{q}\tilde{z}}\, F_H\left(\tilde{q}, s\right)   ,
    \end{aligned}
\end{equation}
as shown in Eq.~\eqref{eq:Noise} and Eq.~\eqref{eq:NB_int}, with $\tilde{q}=q\xi_H$, $\tilde{z}=z_0/\xi$.

In the \textit{zero-field limit}, if the correlation length $\xi_H \gg z_0$ $(\tilde{z}\ll1)$, which corresponds to 
$\lim_{\tilde{z}\to0^{+}} F_H\left(\tilde{q}, +\infty\right)=F_m\left(\tilde{q}\to+\infty\right)\sim \tilde{q}^{-2}$, the noise diverges logarithmically, $\int_0^{\infty}  d \tilde{q}\, \tilde{q} e^{-2\tilde{q}\tilde{z}} F_H\left(\tilde{q}, +\infty\right)\sim \int_0^{\infty}   d \tilde{q}\, \tilde{q}^{-1} e^{-2\tilde{q}\tilde{z}} \sim -\ln \tilde{z}$, meaning that the noise saturates when the probe height is much shorter than the correlation length, as shown in Fig.~\ref{Fig:Metal_B=0},
\begin{equation}
    \mathcal{N}_{\T} \sim  (\mu_0 k_B T)^2\left(\frac{e^*}{\hbar}\right)^2 \frac{m^*}{\hbar\Gamma} \ln \left(\frac{\xi_H}{z_0} \right)   ,
\end{equation}
and if instead $\xi_H \ll z_0$ $(\tilde{z}\gg1)$, which corresponds to 
$\lim_{\tilde{z}\to+\infty} F_H\left(\tilde{q}, +\infty\right)=F_m\left(\tilde{q}\to0^+, +\infty\right)\simeq 1-5 \tilde{q}^2/24 \to 1$, the integral is finite, $\int_0^\infty x e^{-x} \cdot 1 = 1$, so the noise is inversely proportional to $\tilde{z}^2$, as captured in Fig.~\ref{Fig:Metal_B=0},
\begin{equation}\label{similar}
    \mathcal{N}_{\T} \sim  (\mu_0 k_B T)^2\left(\frac{e^*}{\hbar}\right)^2 \frac{m^*}{\hbar\Gamma}\left(\frac{\xi_H}{z_0}\right)^2   ,
\end{equation}

In the \textit{strong-field limit}, if the correlation length $\xi_H \gg z_0$ $(\tilde{z}\ll1)$, the scaling function asymptotically approaches   
$F_H\left(\tilde{q}, s\to0\right) \simeq \tilde{q}^2(2 s^2)^{-1} e^{-\tilde{q}^2s^2/2}$ according to Eq.~\eqref{scaling_lowest_laudau}, and $\lim_{\tilde{z}\to0^+}\int_0^{\infty}  d \tilde{q}\, \tilde{q}\, e^{-2\tilde{q}\tilde{z}}\, F_H\left(\tilde{q}, s\to0^{+}\right) \sim s^{-6}$,
\begin{equation}
    \mathcal{N}_{\T} \sim (\mu_0 k_B T)^2\left(\frac{e^*}{\hbar}\right)^2 \frac{m^*}{\hbar\Gamma}  \frac{1}{s^6}   ,
\end{equation}
and if the correlation length $\xi_H \ll z_0$ $(\tilde{z}\gg1)$, $\lim_{\tilde{z}\to+\infty} F_H(0, s\to 0^+)=1$, the noise behaves similar to Eq.~\eqref{similar}, i.e. inversely proportional to $\tilde{z}^2$ as the probe height is much larger than the correlation length.

\section{Magnetic field from a single vortex}
\label{app:C_mag}
In this appendix, we determine the magnetic field generated by a single isolated vortex in several distinct settings.
We begin with the field produced by an Abrikosov vortex (three-dimensional line vortex), which is applicable when the film thickness $d$ exceeds the penetration depth $\lambda_L$.
We first derive the magnetic field for a static line vortex and then generalize the result to the case of an undulating vortex line.
Finally, we derive the magnetic field generated by a Pearl vortex in two dimensions, which applies when the film thickness $d$ is much smaller than the bulk penetration depth $\lambda_L$.

\subsection{Abrikosov vortex}
We first consider an isolated static vortex along the $z$-axis in a superconductor which occupies the lower half space $z < 0$.
We will further assume a type-II superconductor with $\kappa = \lambda_L/\xi \gg 1$, so that we can neglect the variation of the superconducting order parameter $\psi$ restricted to the core of size $\xi$. 
We also neglect surface variation of the order parameter as $z \to 0^-$, and solve the following Maxwell-London equations
\begin{equation}
\nabla^2 \B_0=\left\{\begin{array}{l}
\dfrac{1}{\lambda_L^2}\left[\B_0-\hat{z}\, \Phi_0 \,\delta^{(2)}(\boldsymbol{\rho})\right], \quad z<0 \\
0, \quad z>0  
\end{array}\right.  ,
\end{equation}
where $\Phi_0=h/(2e)$ is the magnetic flux quantum \cite{London1935}. 
Solving the Laplace equations, we find that the magnetic field deep inside the superconductor is given by,
\begin{equation}\label{field_z<0_3D}
\B_0\big(\r = (\bm{\rho},z)\big) = \hat{z} \left( \frac{\Phi_0}{2\pi \lambda_L^2} \right) K_0\left( \frac{\rho}{\lambda_L}\right), ~~~ z < 0   .
\end{equation}

We now consider the magnetic field above the superconductor.
For the magnetic field outside the superconductor ($z>0$), we can find it through the scalar potential $\phi_M$, satisfying $\B_0 = -\nabla \phi_M$ and $\nabla \times \B_0 = \nabla^2 \phi_M = 0$. 
In momentum space, Laplace’s equation becomes $\partial_z^2 \phi_M-k^2 \phi_M(\boldsymbol{k}, z)=0 \implies \phi_M\sim e^{-kz}$ is the physical solution that decays as one goes farther from the source.
Therefore, we can express $\phi_M$ and $\B_0$ in 2D Fourier transform form, with $k=|\k|$,
\begin{equation}\label{field_z>0_3D}
\phi_M(\bm{\rho},z) = \int \frac{d^2\k}{(2\pi)^2} e^{i \k \cdot \bm{\rho}} e^{-k z} \phi_M(\k) \implies \B_0(\bm{\rho},z) =  -\int \frac{d^2\k}{(2\pi)^2} e^{i \k \cdot \bm{\rho}} e^{-k z} (i k_x, i k_y, -k) \phi_M(\k)  ,\quad z>0  .   
\end{equation}
Now, we impose the boundary condition that the normal component of $\B_0$ (along $z$) has to be continuous across the superconductor vacuum boundary at $z =0$, apply Eq.~\eqref{field_z<0_3D} and Eq.~\eqref{field_z>0_3D} to solve $B_0^z(\bm{\rho}, 0^-) = B_0^z(\bm{\rho}, 0^+)$,
\begin{align}\label{eq:mag_pot}
\phi_M(\k) &= \frac{\Phi_0}{2\pi k \lambda_L^2} \int d^2\bm{\rho} \, e^{-i \k \cdot \bm{\rho}}K_0\left( \frac{\rho}{\lambda_L}\right) = \frac{\Phi_0}{k \lambda_L^2} \int_0^\infty d\rho \, \rho \, J_0(k \rho) K_0\left( \frac{\rho}{\lambda_L}\right) \nonumber \\ 
&= \frac{\Phi_0}{k} \int_0^\infty d\bar{\rho} \, \bar{\rho} \, J_0(k \lambda_L \bar{\rho}) K_0(\bar{\rho}) = \frac{\Phi_0}{k(1 + k^2 \lambda_L^2)}   ,
\end{align}
where we have used the identity $\int_0^{2 \pi} d \theta\, e^{ \pm i k \rho \cos \theta}=2 \pi J_0(k \rho),~\int_0^{\infty} u J_0(a u) K_0(b u) d u=(a^2+b^2)^{-1}$. 
Plugging this back in Eq.~\eqref{field_z>0_3D}, with a dimensionless wavevector $\tilde{k}=k \lambda$, we find the scalar potential in real space, 
\begin{equation}\label{eq:mag_pot_full}
\phi_M(\bm{\rho},z) = \frac{\Phi_0}{2\pi} \int_0^\infty dk \, \frac{J_0(k \rho) e^{-k z}}{1 + k^2 \lambda_L^2} = \frac{\Phi_0}{2\pi \lambda_L} \int_0^{\infty} d\tilde{k} 
 \frac{ J_0\big( \rho \tilde{k} /\lambda_L \big) e^{- z \tilde{k}/\lambda_L}}{1 + \tilde{k}^2}  .
\end{equation}
Accordingly, applying $\B_0 = -\nabla\phi_M=-(\partial_\rho \phi_M)\hat{\rho}-(\partial_z \phi_M)\hat{z}$ and $-\partial_\rho J_0(k \rho)=k J_1(k \rho)$, the magnetic field $\B_0(\r)$ in the space above the superconductor is given by
\begin{equation}\label{field_3D}
\B_0(\r) = \frac{\Phi_0}{2\pi \lambda_L^2}\left[  \left( \int_0^{\infty} d\tilde{k} \, \frac{\tilde{k}  J_1\left( \rho \tilde{k} /\lambda_L \right) e^{- z \tilde{k}/\lambda_L}}{1 + \tilde{k}^2} \right)\hat{\rho}  + \left( \int_0^{\infty} d\tilde{k} \, \frac{\tilde{k} \, J_0\left( \rho \tilde{k} /\lambda_L \right) e^{- z \tilde{k}/\lambda_L}}{1 + \tilde{k}^2} \right) \hat{z} \right],\quad z>0  .
\end{equation}

We now investigate asymptotic limits of the potential in Eq.~\eqref{eq:mag_pot_full}, 
\begin{equation}
\phi_M(\,\rho \gg \lambda_L \text{~or~} z \gg \lambda_L) \approx \frac{\Phi_0}{2\pi \lambda_L} \int_0^{\infty} d\tilde{k} \, J_0\left( \frac{\rho \tilde{k} }{\lambda_L} \right) e^{- z \tilde{k}/\lambda_L} = \frac{\Phi_0}{2\pi \lambda_L} \frac{\lambda_L}{\sqrt{\rho^2 + z^2}} = \frac{\Phi_0}{2\pi r}
\end{equation}
For $\rho \gg \lambda_L$ or $z \gg \lambda_L$, the integral is dominated by $\tilde{k} \ll 1$.
This is the potential of a point \textit{magnetic charge} at the origin, and therefore the magnetic field at large distances is radial, i.e., $\B_0(\r) =  \hat{r}~\Phi_0/(2\pi r^2), \text{ for }  r = \sqrt{\rho^2 + z^2} \gg \lambda_L$.

We can also investigate the limits $\rho \ll \lambda_L$, using Taylor series $J_0(x) = 1 - (x/2)^2+\cdots$ in Eq.~\eqref{eq:mag_pot_full}, the potential and magnetic field components are summarized in Table.~\ref{tab:asymptotic}.
The extra factor $\rho/\lambda_L$ of $B_\rho$ indicates strong suppression of the radial field near the $z$-axis from $\rho \ll \lambda_L$, as expected from cylindrical symmetry the radial field on the $z$-axis must go to zero since $\nabla \cdot \B_0(\r) = 0$.
\begin{table}[!t]
\centering
\renewcommand{\arraystretch}{2.3}
\begin{tabular}{ccc}
\hline
\hline
 & \textbf{Case 1}: \(r \gg \lambda_L\) & \textbf{Case 2}: \(\rho \ll \lambda_L\)  \\
\hline
\(\phi_M\) & 
\(\dfrac{\Phi_0}{2\pi r}\) & 
\(\dfrac{\Phi_0}{2\pi\lambda_L} \left[ \displaystyle\int_0^\infty \dfrac{e^{-z\tilde{k}/\lambda_L}}{1+\tilde{k}^2} d\tilde{k} - \left( \dfrac{\rho}{2\lambda_L} \right)^2 \displaystyle\int_0^\infty \dfrac{\tilde{k}^2 e^{-z\tilde{k}/\lambda_L}}{1+\tilde{k}^2} d\tilde{k} \right]\)  \\
%\hline
\(B_0^\rho\) & 
\(\dfrac{\Phi_0}{2\pi r^2} \dfrac{\rho}{r}\) & 
\(\dfrac{\Phi_0}{4\pi\lambda_L^2} \left( \dfrac{\rho}{\lambda_L} \right) \displaystyle\int_0^\infty \dfrac{\tilde{k}^2 e^{-z\tilde{k}/\lambda_L}}{1+\tilde{k}^2} d\tilde{k}\)  \\
%\hline
\(B_0^z\) & 
\(\dfrac{\Phi_0}{2\pi r^2} \dfrac{z}{r}\) & 
\(\dfrac{\Phi_0}{2\pi\lambda_L^2} \displaystyle\int_0^\infty \dfrac{\tilde{k} e^{-z\tilde{k}/\lambda_L}}{1+\tilde{k}^2} d\tilde{k}\) \\
%\hline
\textbf{Behavior:} & 
Monopole (upper half-space) & 
Cylindrical symmetry, \(B_0^\rho \to 0\) on axis  \\
\hline
\hline
\end{tabular}
\caption{Asymptotic limits of the scalar potential and magnetic field of a Abrikosov vortex in a type-II superconductor. \textbf{Case 1}: far-field limit; \textbf{Case 2}: near-axis limit.}
\label{tab:asymptotic}
\end{table}

\subsection{Undulating Abrikosov vortex}
Now, we extend our results to include undulations of the vortex along its axis, i.e., we assume that the displacement $\u_t(z)$ is a function of both time $t$ and the vertical coordinate $z$. 
In this case, we need to solve the following generalization of the Maxwell-London equation:
\begin{equation}
\nabla^2 \B=\left\{\begin{array}{l}
\dfrac{1}{\lambda_L^2}\left[\B-\hat{z}\, \Phi_0 \,\delta\big(\boldsymbol{\rho}-\u_t(z)\big)\right], \quad z<0 \\
0, \quad z>0  
\end{array}\right.    .
\label{eq:UVEOM}
\end{equation}
Similarly, we consider a Fourier representation of the field, that generalizes our calculation for the rigid vortex:
\begin{equation}
\B(\bm{\rho}, z, t) = \sum_{n=0}^{N_\ell} \int \frac{d^2k} {(2\pi)^2} \bm{b}_n(\bm{k},t) e^{i (\bm{k} \cdot \bm{\rho} + k_n z)}, ~~~~~ z \leqslant 0   ,
\label{eq:UVA}
\end{equation}
where $k_n$ denotes momenta along the $z$-direction, which is fixed by periodic boundary conditions to be $k_n = n (2 \pi/d)$ for a material of thickness $d$ and inter-layer spacing $a_c$, i.e., number of layers $N_\ell = d/a_c$.
Plugging the ansatz in Eq.~\eqref{eq:UVA} into Eq.~\eqref{eq:UVEOM} and using the linear independence of the distinct Fourier modes, we find that $b_n(\bm{k},t) \equiv \hat{z} \cdot \bm{b}_n(\bm{k},t)$ satisfies the  Maxwell-London equation in momentum space:
\begin{equation}
\begin{aligned}
\left[ \lambda_L^2 (\k^2 + k_n^2) + 1 \right]b_n(\bm{k},t) & = \frac{\Phi_0}{d} \int_0^d dz \, e^{-i k_n z} e^{- i \k \cdot \u_t(z)}\\
&= \Phi_0\sum_{m=0}^{\infty} \frac{[-i \k \cdot \u_t(0)]^m}{m!} \underbrace{\left( \frac{1}{d}  \int_0^d dz \, e^{i (m k_M - k_n) z} \right)}_{ \delta_{k_n, m k_M}}
  .    
\end{aligned}
\label{eq:bFourier}
\end{equation}
where we have assumed a fixed $k_z = k_M$ for analytical progress, i.e., $\u_t(z) = \u_t(0) e^{i k_M z}$. 
We note that only the Fourier modes $k_n = m k_M$ which are the harmonics of the vortex undulation mode contribute to the magnetic field. 
Accordingly, we can set $b_n = 0$ in Eq.~\eqref{eq:bFourier} unless $k_n = m k_M$, and write the non-zero $b_{n}$ as
\begin{equation}
b_{k_n = m k_M}(\k,t) = \frac{\Phi_0}{1 + \lambda_L^2 (\k^2 + (m k_M)^2)} \frac{[- i \k \cdot \u_t(0)]^m}{m!}.
\end{equation}
Therefore, for an undulating vortex with $\u_t(z) = \u_t(0) e^{i k_M z}$, the $z$-component of the field $\B$ inside the superconductor is given by
\begin{equation}
B^z(\bm{\rho}, z \leqslant 0, t) =  \sum_{m} \int \frac{d^2k} {(2\pi)^2} \frac{\Phi_0}{1 + \lambda_L^2 (\k^2 + (m k_M)^2)} \frac{[- i \k \cdot \u_t(0)]^m}{m!}  e^{i (\bm{k} \cdot \bm{\rho} + m k_M z)}.
\label{eq:Bz_und}
\end{equation}
At the same time, for $z > 0$, we can simply use that $B^z$ satisfies the Laplace equation outside the superconductor with boundary conditions set by $ B^z(0^-)$, to write 
\begin{equation}
B^z(\bm{\rho}, z \geqslant 0, t) = \int \frac{d^2k} {(2\pi)^2} B^z(\k, z=0, t) e^{i \bm{k} \cdot \bm{\rho}} e^{- k z} = \sum_m \int \frac{d^2k} {(2\pi)^2} \left[ \frac{\Phi_0}{1 + \lambda_L^2 (\k^2 + (m k_z)^2)} \frac{[- i \k \cdot \u_t(0)]^m}{m!} \right] e^{i \bm{k} \cdot \bm{\rho}} e^{- k z}   ,
\label{eq:BzUV}
\end{equation}
where we have made the replacement $k_M \to k_z$. 
Since the magnetic field can be written as a gradient of a scalar potential in the upper half space ($z > 0$), Eq.~\eqref{eq:BzUV} implies that all components of the field may be obtained by generalizing the scalar potential $\phi_M(\k)$ for the rigid vortex to 
\begin{equation}
\phi(\k, k_z) = 
\sum_{m} \frac{\Phi_0}{k(1 + \lambda_L^2 (\k^2 + (m k_z)^2))} \frac{[- i \k \cdot \u_t(0)]^m}{m!}, \implies \B(\bm{\rho}, z \geqslant 0, t) = \int \frac{d^2k} {(2\pi)^2} (-i \k, k) \phi(\k, k_z) e^{i \bm{k} \cdot \bm{\rho}} e^{- k z},
\end{equation}
as claimed in the Eq.~\eqref{mag_k_z_modes}. 

\subsection{Pearl vortex}
Now, we consider the two-dimensional limit. 
The Maxwell-London equation now only holds over the film thickness $d$ which is assumed to be finite and much less than the bulk penetration depth $\lambda_L$ ($d\ll\lambda_L$) :
\begin{equation}\label{london_pearl}
\nabla^2 \B_0=\left\{\begin{array}{l}
\dfrac{1}{\lambda_L^2}\left[\B_0-\hat{z}\, \Phi_0 \,\delta^{(2)}(\boldsymbol{\rho})\right], \quad-\dfrac{d}{2}<z<\dfrac{d}{2} \\
0, \qquad\qquad \qquad\qquad \qquad\qquad |z|>\dfrac{d}{2}  
\end{array}\right. .
\end{equation}
In the thin-film ($d \to 0$) limit, it is convenient to introduce a sheet-current density which depends only on the in-plane coordinate $\bm{\rho} = (x,y)$, by the assumption that the current density is uniformly distributed in the $z$-direction, i.e., 
\begin{equation}
\j(\bm{\rho},z) = \begin{cases}
    \K(\bm{\rho})/d, \quad-\dfrac{d}{2} \leqslant z \leqslant \dfrac{d}{2} \\ 
    0, \qquad \qquad \quad |z|> \dfrac{d}{2}  
\end{cases} .
\end{equation}
This will allow us to relate the magnetic field above and below the superconducting sheet. 
We consider the $z$-component of the above equation, using vector identity $\nabla \times(\nabla \times \B_0)=\nabla(\nabla \cdot \B_0)-\nabla^2 \B_0$, Maxwell equations $\nabla \times \B_0= \mu_0\j$ and London equation Eq.~\eqref{london_pearl}, and integrate both sides over $z$ from $-d/2$ to $d/2$,
\begin{align}
\frac{d}{\lambda_L^2} B_0^z - \frac{\Phi_0 d}{\lambda_L^2} \delta^{(2)}(\bm{\rho})  = -  \mu_0 \underbrace{\int_{-\frac{d}{2}}^{\frac{d}{2}} dz \, (\nabla \times \j)_z}_{ (\nabla \times \K(\bm{\rho}))_z} &  = - \mu_0 \left(\partial_x \int_{-\frac{d}{2}}^{\frac{d}{2}} dz \, j_x - \partial_y \int_{-\frac{d}{2}}^{\frac{d}{2}} dz \, j_x \right)   ,
\end{align}
the boundary conditions for the in-plane field $\B_0^{\parallel}$ and sheet-current density $\K$ (taking the limit of small thickness, i.e., $d \to 0$) are: (suppressing $\bm{\rho}$ for clarity)
\begin{subequations}
    \begin{align}
        &\B_0^{\parallel}(z \to 0^+) - \B_0^{\parallel}(z \to 0^-) = \mu_0 \left( \K \times \hat{z} \right)   ,\\
&\B_0^{\parallel}(z \to 0^+) = -\B_0^{\parallel}(z \to 0^-) = \frac{\mu_0 (\K \times \hat{z})}{2}  ,
\label{eq:sheetBC}
    \end{align}
\end{subequations}
where Eq.~\eqref{eq:sheetBC} is from the symmetry of the problem. Using the above result, we can express the curl of the sheet current density in terms of $B_0^z$, $\mu_0 (\nabla \times \K(\bm{\rho}))_z = 2 [\nabla \times (\hat{z} \times  \B_0^{\parallel}(0^+))]_z  = - 2 \partial_z B_0^z(0^+)$.
Finally, we can plug this result back into the $z$-integrated London equation to express everything in terms of $B_0^z$:
\begin{equation}\label{Pearl_boundary}
2\partial_z B_0^z = \frac{d}{\lambda_L^2} B_0^z - \frac{\Phi_0 d}{\lambda_L^2} \delta^{(2)}(\bm{\rho}) \implies B_0^z - \Lambda \partial_z B_0^z = \Phi_0 \delta^{(2)}(\bm{\rho}), ~~~ z \to 0^+   ,
\end{equation}
where we have defined the Pearl length $\Lambda = 2\lambda_L^2/d$. 
Now, we are finally in a position to solve for the magnetic field. 
To do so, we again use the scalar potential $\phi_M(\r)$ for $z > 0$, such that $\B_0(\r) = - \nabla\phi_M(\r)$, and impose $\nabla\cdot \B_0 = 0$ to get $\nabla^2 \phi_M =0$. 
The general solution for a decaying solution $\sim e^{-k z}$ for $z > 0$ can be adapted from Eq.~\eqref{field_z>0_3D},
\begin{equation}\label{field_z>0_2D}
\phi_M(\bm{\rho},z) = \int \frac{d^2\k}{(2\pi)^2} e^{i \k \cdot \bm{\rho}} e^{-k z} \phi_M(\k) \implies \B_0(\bm{\rho},z) =  -\int \frac{d^2\k}{(2\pi)^2} e^{i \k \cdot \bm{\rho}} e^{-k z} (i k_x, i k_y, -k) \phi_M(\k)  .   
\end{equation}
Applying Eq.~\eqref{field_z>0_2D} on Eq.~\eqref{Pearl_boundary}, we can solve the scalar potential $\phi_M$,
\begin{equation}\label{eq:mag_pot_Pearl}
\begin{aligned}
B_0^z(\bm{\rho},z) =  \int \frac{d^2\k}{(2\pi)^2} k e^{i \k \cdot \bm{\rho}} e^{-k z} \phi_M(\k) &\implies B_0^z - \Lambda \partial_z B_0^z = \int \frac{d^2\k}{(2\pi)^2} k (1 + \Lambda k) e^{i \k \cdot \bm{\rho}} e^{-k z} \phi_M(\k)  = \Phi_0 \delta^{(2)}(\bm{\rho})\\
&\implies \phi_M(\k) = \frac{\Phi_0}{k(1 + \Lambda k)}  ,
\end{aligned}
\end{equation}
and find the scalar potential in real space,
\begin{equation}\label{eq:mag_pot_realspace_2D}
\phi_M(\bm{\rho},z) = \Phi_0 \int \frac{d^2\k}{(2\pi)^2} \frac{e^{i \k \cdot \bm{\rho}} e^{-k z}}{k(1 + \Lambda k)} = \frac{\Phi_0}{2\pi \Lambda} \int_0^\infty d\tilde{k} \frac{J_0\big( \tilde{k}z /\Lambda \big) e^{- \tilde{k}z /\Lambda} }{1 + \tilde{k}}  , ~z>0  ,
\end{equation}
where we have taken advantage of radial symmetry to evaluate the angular integral, following a similar way of deriving Eq.~\eqref{eq:mag_pot_full}, and rescaled by $\Lambda$ to define a dimensionless $\tilde{k} = k \Lambda$.
We note now that all physics is sensitive to the Pearl length $\Lambda = 2 \lambda_L^2/d$, which can be much larger than the penetration depth in the thin film limit $d \ll \lambda_L$. 
Accordingly, the magnetic fields are given by
\begin{equation}\label{field_2D}
\B_0(\r) = \frac{\Phi_0}{2\pi \Lambda^2}\left[  \left( \int_0^{\infty} d\tilde{k} \, \frac{\tilde{k}  J_1\left( \rho \tilde{k} /\Lambda \right) e^{- z \tilde{k}/\Lambda}}{1 + \tilde{k}} \right)\hat{\bm{\rho}}  + \left( \int_0^{\infty} d\tilde{k} \, \frac{\tilde{k} \, J_0\left( \rho \tilde{k}/\Lambda \right) e^{- z \tilde{k}/\Lambda}}{1 + \tilde{k}} \right) \hat{z} \right]   .
\end{equation}
We now investigate asymptotic limits of the potential in Eq.~\eqref{eq:mag_pot_realspace_2D}, 
\begin{equation}
\phi_M(\,\rho \gg \Lambda \text{~or~} z \gg \Lambda) \approx \frac{\Phi_0}{2\pi \Lambda} \int_0^\infty d\tilde{k} \, J_0\left( \tilde{k} \frac{\rho}{\Lambda}\right) e^{- \frac{\tilde{k}z}{\Lambda}} = \frac{\Phi_0}{2 \pi r} \implies \B_0(\r)= \frac{\Phi_0}{2\pi r^2}  \hat{r} , ~z>0    .
\end{equation}
For $\rho \gg \Lambda$ or $z \gg \Lambda$,  we have $r = \sqrt{\rho^2 + z^2} \gg \Lambda$, which is identical to what we found for an Abrikosov vortex in the far-field limit.
This is because far away from the vortex the internal structure becomes irrelevant and one essentially sees a \textit{point magnetic monopole}.

For $\rho \ll \Lambda$, we compute and summarize the potential and magnetic field in Table.~\ref{tab:pearl_asymptotic}.
As expected, the radial magnetic field almost vanishes for $\rho \ll \Lambda$ (except when $z \to 0$), while the axial component $B_0^z$ does not. Right at the surface of the superconductor ($z\to0^+$), both the axial and radial magnetic fields $B_0^z$ and $B_0^\rho$ diverge as $1 / \rho$. Contrast this with the previous Abrikosov case, when $B_0^z$ diverges only weakly as $\ln \rho $ and $B_0^\rho \rightarrow 0$ (cutoff momentum $1/\xi$) at the superconductor vacuum interface.
\begin{table}[!t]
\centering
\renewcommand{\arraystretch}{2.3}
\begin{tabular}{ccc}
\hline
\hline
 & \textbf{Case 1:} \(r \gg \Lambda\) & \textbf{Case 2:} \(\rho \ll \Lambda\)  \\
\hline
\(\phi_M\) & 
\(\dfrac{\Phi_0}{2\pi r}\) & 
\(\dfrac{\Phi_0}{2\pi\Lambda} \displaystyle\left[\int_0^{\infty} d \tilde{k} \dfrac{e^{-z \tilde{k}/\Lambda}}{1+\tilde{k}}-\left(\frac{\rho}{2 \Lambda}\right)^2 \int_0^{\infty} d \tilde{k} \dfrac{\tilde{k}^2 e^{-z \tilde{k}/\Lambda}}{1+\tilde{k}}\right] \)  \\
\(B_0^\rho\) & 
\(\dfrac{\Phi_0}{2\pi r^2} \dfrac{\rho}{r}\) & 
\(\dfrac{\Phi_0}{4 \pi \Lambda^2}\displaystyle\left(\frac{\rho}{\Lambda}\right) \int_0^{\infty} d \tilde{k} \frac{\tilde{k}^2 e^{-z \tilde{k}/\Lambda}}{1+\tilde{k}}\)  \\
\(B_0^z\) & 
\(\dfrac{\Phi_0}{2\pi r^2} \dfrac{z}{r}\) & 
\(\displaystyle\frac{\Phi_0}{2 \pi \Lambda^2} \int_0^{\infty} d \tilde{k} \frac{\tilde{k} e^{-z \tilde{k}/\Lambda}}{1+\tilde{k}}\)  \\
\textbf{Behavior:} & 
Monopole (upper half-space) & 
Cylindrical symmetry, \(B_0^\rho \to 0\) on axis  \\
\hline
\hline
\end{tabular}
\caption{Asymptotic limits of the scalar potential and magnetic field of a Pearl vortex in a type-II superconductor ($d \ll \lambda_L$). \textbf{Case 1}: far-field limit; \textbf{Case 2}: near-axis limit.}
\label{tab:pearl_asymptotic}
\end{table}

\subsection{Derivatives of magnetic field components}\label{derivative_z_dependence}
\subsubsection{Abrikosov vortex}
Take derivatives of Eq.~\eqref{field_3D}, using properties of Bessel functions $\partial_x J_0(x) = -J_1(x)$ and $\partial_x J_1(x)=J_0(x)-J_1(x)/x$,
\begin{subequations}
\begin{align}
\partial_\rho B_0^\rho & =\frac{\Phi_0}{2 \pi \lambda_L^2} \int_0^{\infty} d \tilde{k} \frac{\tilde{k} e^{-z \tilde{k}/\lambda_L}}{1+\tilde{k}^2}\left[\frac{\tilde{k}}{\lambda_L} J_0\left(\frac{\rho \tilde{k}}{\lambda_L}\right)-\frac{1}{\rho} J_1\left(\frac{\rho \tilde{k}}{\lambda_L}\right)\right]   , \\
\partial_\rho^2 B_0^z & =-\frac{\Phi_0}{2 \pi \lambda_L^3} \int_0^{\infty} d \tilde{k} \frac{\tilde{k}^2 e^{-z \tilde{k}/\lambda_L}}{1+\tilde{k}^2}\left[\frac{\tilde{k}}{\lambda_L} J_0\left(\frac{\rho \tilde{k}}{\lambda_L}\right)-\frac{1}{\rho} J_1\left(\frac{\rho \tilde{k}}{\lambda_L}\right)\right]  .
\end{align}
\end{subequations}
If we look at the center of the vortex, i.e., $\rho=0$, with $J_0(0)=1, \lim_{x\to0^+}J_1(x) / x = 1 / 2$,
\begin{equation}\label{field_derivative_3D}
\partial_\rho B_0^\rho(\rho=0)=\frac{\Phi_0}{4 \pi \lambda_L^3} \int_0^{\infty} d \tilde{k} \frac{\tilde{k}^2 e^{-z \tilde{k}/\lambda_L}}{1+\tilde{k}^2}, \quad \partial_\rho^2 B_0^z(\rho=0)=-\frac{\Phi_0}{4 \pi \lambda_L^4} \int_0^{\infty} d \tilde{k} \frac{\tilde{k}^3 e^{-z \tilde{k}/\lambda_L}}{1+\tilde{k}^2}   .
\end{equation}
For $z \gg \lambda_L$, the exponential strongly constrains the integral to $\tilde k \ll \lambda_L/z$ ($1+\tilde{k}^2 \approx 1$), and we find that the above derivatives become $
\partial_\rho B_0^\rho(\rho=0) = \Phi_0/(2\pi z^3),~
\partial_\rho^2 B_0^z(\rho=0) = -3\Phi_0/(2\pi z^4)$; On the other hand, for $z \ll \lambda_L$, the exponential is weak ($\tilde{k}\gg1$), and we find $\partial_\rho B_0^\rho(\rho=0) = \Phi_0/(4\pi \lambda_L^2 z) ,~\partial_\rho^2 B_0^z(\rho=0) = -\Phi_0/(4\pi \lambda_L^2 z^2)$.

More explicitly, evaluating these integrals in Eq.~\eqref{field_derivative_3D} at $\rho = 0$ give $z$-dependent expressions:
\begin{align}
(\partial_{\rho} B_{0}^{\rho})_{\rho = 0} &=  \frac{\Phi_0}{2} \int \frac{d^2\k}{(2\pi)^2} \,  \frac{k \, e^{- k z} }{1 + k^2\lambda_L^2} = \frac{\Phi_0}{4\pi \lambda_L^3} \bigg[ \frac{\lambda_L}{z}-\frac{\pi}{2} \cos \left(\frac{z}{\lambda_L}\right) -\sin \left(\frac{z}{\lambda_L}\right) \operatorname{Ci}\left(\frac{z}{\lambda_L}\right) +\cos \left(\frac{z}{\lambda_L}\right) \operatorname{Si}\left(\frac{z}{\lambda_L}\right) \bigg], \nonumber \\
(\partial_{\rho}^2 B_{0}^{z})_{\rho=0} & = \frac{\Phi_0}{2} \int \frac{d^2\k}{(2\pi)^2} \, \frac{k^2 e^{- k z}}{1 + k^2\lambda_L^2} = \frac{\Phi_0}{4\pi \lambda_L^4} \frac{G_{1,3}^{3,1}\left(\begin{array}{c}
-1 \\
-1,0, \frac{1}{2}
\end{array}\left\lvert\,\dfrac{z^2}{4\lambda_L^2}  \right.\right)}{2 \sqrt{\pi}}     .
\label{eq:derB2D}
\end{align}
where cosine/sine integrals are $\operatorname{Ci}(x)=-\displaystyle\int_x^{\infty} \dfrac{\cos \varsigma}{\varsigma} d \varsigma$, $\operatorname{Si}(x)=\displaystyle\int_0^x \dfrac{\sin \varsigma}{\varsigma} d s$, $G_{p, q}^{m, n}\left(\begin{array}{c}
\mathbf{a}_{\p}  \\
\mathbf{b}_{\q} 
\end{array} \bigg| x\right) $ is the Meijer G-function.

\subsubsection{Pearl vortex}
Take derivatives of Eq.~\eqref{field_2D}, using properties of Bessel functions $\partial_x J_0(x) = -J_1(x)$ and $\partial_x J_1(x)=J_0(x)-J_1(x)/x$,
\begin{subequations}
\begin{align}
\partial_\rho B_0^\rho & =\frac{\Phi_0}{2 \pi \Lambda^2} \int_0^{\infty} d \tilde{k} \frac{\tilde{k} e^{-z \tilde{k}/\Lambda}}{1+\tilde{k}}\left[\frac{\tilde{k}}{\Lambda} J_0\left(\frac{\rho \tilde{k}}{\Lambda}\right)-\frac{1}{\rho} J_1\left(\frac{\rho \tilde{k}}{\Lambda}\right)\right]   , \\
\partial_\rho^2 B_0^z & =-\frac{\Phi_0}{2 \pi \Lambda^2} \int_0^{\infty} d \tilde{k} \frac{\tilde{k}^2 e^{-z \tilde{k}/\Lambda}}{1+\tilde{k}}\left[\frac{\tilde{k}}{\Lambda} J_0\left(\frac{\rho \tilde{k}}{\Lambda}\right)-\frac{1}{\rho} J_1\left(\frac{\rho \tilde{k}}{\Lambda}\right)\right]  .
\end{align}
\end{subequations}
If we look at the center of the vortex, i.e., $\rho=0$, with $J_0(0)=1, \lim_{x\to0^+}J_1(x) / x = 1 / 2$,
\begin{equation}\label{field_derivative_2D}
\partial_\rho B_0^\rho(\rho=0)=\frac{\Phi_0}{4 \pi \Lambda^3} \int_0^{\infty} d \tilde{k} \frac{\tilde{k}^2 e^{-z \tilde{k}/\Lambda}}{1+\tilde{k}}, \quad \partial_\rho^2 B_0^z(\rho=0)=-\frac{\Phi_0}{4 \pi \Lambda^4} \int_0^{\infty} d \tilde{k} \frac{\tilde{k}^3 e^{-z \tilde{k}/\Lambda}}{1+\tilde{k}}.
\end{equation}
For $z \gg \Lambda$, the exponential strongly constrains the integral to $\tilde k \ll \Lambda/z$ ($1+\tilde{k}^2 \approx 1$), and we find that the above derivatives become $
\partial_\rho B_0^\rho(\rho=0) = \Phi_0/(2\pi z^3),~
\partial_\rho^2 B_0^z(\rho=0) = -3\Phi_0/(2\pi z^4)$, which matches $z$-asymptotics for Abrikosov case; On the other hand, for $z \ll \Lambda$, the exponential is weak ($\tilde{k}\gg1$), and we find $\partial_\rho B_0^\rho(\rho=0) = \Phi_0/(4\pi \Lambda z^2) ,~\partial_\rho^2 B_0^z(\rho=0) = -\Phi_0/(2\pi \Lambda z^3)$.

More explicitly, evaluating these integrals in Eq.~\eqref{field_derivative_2D} at $\rho = 0$ give $z$-dependent expressions:
\begin{align}
(\partial_{\rho} B_{0}^{\rho})_{\rho = 0} &=  \frac{\Phi_0}{2} \int \frac{d^2\k}{(2\pi)^2} \,  \frac{k \, e^{- k z_0} }{1 + \Lambda k} = \frac{\Phi_0}{4\pi \Lambda^3} \left[ -\frac{\Lambda}{z_0} + \left(\frac{\Lambda}{z_0}\right)^2 + e^{\frac{z_0}{\Lambda}} \textrm{E}_1\left(\frac{z_0}{\Lambda} \right) \right], \nonumber \\
(\partial_{\rho}^2 B_{0}^{z})_{\rho=0} & = \frac{\Phi_0}{2} \int \frac{d^2\k}{(2\pi)^2} \, \frac{k^2 e^{- k z_0}}{1 + \Lambda k}  = \frac{\Phi_0}{4\pi \Lambda^4} \bigg[ \frac{\Lambda}{z_0} - \left(\frac{\Lambda}{z_0}\right)^2 + 2\left(\frac{\Lambda}{z_0}\right)^3 - e^{\frac{z_0}{\Lambda}} \textrm{E}_1\left(\frac{z_0}{\Lambda} \right) \bigg]   .
\label{eq:derB2D}
\end{align}
where $\textrm{E}_1(x) = \displaystyle \int_{x}^{\infty} d\varsigma \, \frac{e^{-\varsigma}}{\varsigma}$ is the exponential integral function.\\

\section{Details on the pinned vortex / vortex dynamics}
\label{app:pinned}

In this appendix, we provide details of the calculation of the magnetic noise generated by a pinned vortex, which were omitted in the main text.
Foremost, while the main text focused on the magnetic noise produced by a vortex undulating at a single wavevector, here we consider the more general case in which the vortex supports fluctuations at multiple wavevectors.
As we show below, the transverse magnetic noise contributions from different modes add independently, leading to a simple series summation over the undulation modes.
Second, we provide a detailed derivation that connects the displacement correlation function to the expressions for both the transverse $\left(\mathcal{N}_{x x}\right)$ and longitudinal $\left(\mathcal{N}_{z z}\right)$ magnetic noise.

\subsection{Multiple undulating modes}
We first discuss the magnetic noise produced by a pinned vortex undulating with multiple wavevectors, and show that the resulting transverse magnetic noise is a simple summation of the noise at all the different wavevectors.

To this end, we first determine the magnetic field generated by a vortex whose transverse displacement at depth $z$ is given by, 
 \begin{equation}
     \u_t(z) = \sum_n \u_{t, n} e^{i k_n z}  ~~~ k_n = \frac{2\pi n}{d} \, .
     \label{appeq:u_t_gen}
 \end{equation}

Following Eq.~\eqref{eq:UVA}, we introduce a Fourier representation of the magnetic field, and Taylor expand to leading order in the vortex displacement $\u_t$ Eq.~\eqref{eq:bFourier}.
We find,
$$\left[ \lambda_L^2 (\k^2 + k_n^2) + 1 \right]b_n(\bm{k},t) \simeq \frac{\Phi_0}{d} \int_0^d dz \, e^{-i k_n z} (1 - i \k \cdot \u_t(z) ).$$
Inserting Eq.~\eqref{appeq:u_t_gen} and performing the $z$ integration, we find that to leading order in $\u_{t, n}$, $b_n(k,t)$ is given by,
\begin{equation}
    b_n(\k,t) = \Phi_0 \left[ \frac{-i \k \cdot \u_{t, n}}{\lambda_L^2 (\k^2 + k_n^2)  + 1} \right].
    \label{appeq:bn_gen}
\end{equation}
Finally, the real-space magnetic field is then obtained by performing the inverse Fourier transform of Eq.~\eqref{appeq:bn_gen}.
Putting the qubit directly on top of the vortex at $\r_q = (\rr_q, z_q) = (0,0,z_0)$, we find that the time-dependence of the in-plane component of magnetic field $\delta \B_\parallel$ is given by,
\begin{equation}
    \delta \B_{\parallel}(t) = - \sum_n \left[ \frac{\Phi_0}{2} \int \frac{d^2 \k}{(2\pi)^2} \frac{k \, e^{-k z_0}}{(1 + \lambda_L^2 (\k^2 + k_n^2))} \right] \u_{t, n},
\end{equation}
where we have used that the angular average $\langle \hat{k}_\alpha \hat{k}_\beta \rangle = \delta_{\alpha \beta}/2$ inside the $\k$ integral.

Finally, let us compute the transverse magnetic noise due to the vortex undulations.
\begin{equation}\label{app:noise_multi_modes}
\begin{split}
    \cN_{\rm T}(\Omega) = \cN_{xx}(\Omega) + \cN_{yy}(\Omega) &= \int dt\, e^{i \Omega t} \langle \delta \B_\parallel(t) \cdot \delta \B_\parallel(0) \rangle \\
    &= \int dt\, e^{i \Omega t} \sum_n \left[ \frac{\Phi_0}{2} \int \frac{d^2 k}{(2\pi)^2} \frac{k \, e^{-k z_0}}{(1 + \lambda_L^2 (\k^2 + k_n^2))} \right]^2 \langle \u_{t, n} \cdot  \u_{0, -n} \rangle.
\end{split} 
\end{equation}
Here in the second line, we have used the fact that in the Langevin dynamics of Eq.~\eqref{eq:eom_pinned} and \eqref{eq:WN3D}, the modes with different wavevectors decouple from one another, and $\langle u^\alpha_{t, n}  u^\beta_{t^\prime, m} \rangle \propto \delta^{\alpha\beta}\delta_{n, -m}$.
Consequently, the transverse magnetic noise in the presence of multiple undulation modes reduces to a simple sum over contributions from individual modes.
For small drag coefficient $\eta$, each mode remains weakly damped, and the magnetic noise spectrum exhibits well-resolved peaks associated with the discrete mode frequencies.
In contrast, for large $\eta$, strong damping broadens the spectral features, causing the contributions from different modes to overlap and merge into a smooth continuum.

\subsection{Displacement correlations from Langevin dynamics}

We now provide a detailed derivation for the transverse and longitudinal magnetic noise.
Our primary focus is to derive the correlation functions of the vortex displacement $\expect{\u_t \u_0}$ and $\expect{|\u_t|^2 |\u_0|^2}$, since these correlation functions determine the magnetic noise through Eq.~\eqref{eq:Npinned}.

Recall that from Eq.~\eqref{eq:eom_pinned}, the Langevin equation for the undulating vortex is given by,
\begin{equation}\label{app:langevin_eom}
\mu_v \ddot{\u}_t = n_s h (\dot{\u}_t \times \hat{z}) -K \u_t + \varepsilon_l \partial_z^2 \u_t - \eta \dot{\u}_t + \bm{\zeta}_t \,.
\end{equation}
Applying the Fourier-space ansatz for the displacement from Eq.~\eqref{eq:u_Fourier}, $\u_{t}(z) = \u_{\Omega,k_z} e^{i ( k_z z- \Omega t)}$, and restricting to a single longitudinal mode with wavevector $k_z$, the equation of motion simplify to Eq.~\eqref{eq:eom_frequency},
\begin{align}
& M_{\Omega, k_z}^{-1} ~\u_{\Omega,k_z} = \bm\zeta_{\Omega,k_z}, \text{ where }~ M_{\Omega,k_z}^{-1} =  \begin{pmatrix}
        - \mu_v \Omega^2 - i\eta\Omega + \overline{K}(k_z) & i n_s h \Omega \\
        -i n_s h \Omega & - \mu_v \Omega^2 - i\eta\Omega + \overline{K}(k_z)
    \end{pmatrix} \,.
\end{align}
Here, we have defined the effective spring constant $\overline{K}(k_z) = K + \varepsilon_l k_z^2$.

Let us first find $\expect{\u_t \u_0}$, which determines the transverse magnetic noise.
Applying $\u = M \bm\zeta$ on Eq.~\eqref{eq:WN3D},
\begin{equation}\label{app:uuCorr}
\langle u^\alpha_{\Omega,k_z} u^\beta_{-\Omega,-k_z} \rangle = 2 (\eta/a_c) k_B T  [M_{\Omega, k_z} M^\dagger_{\Omega, k_z}]^{\alpha \beta} ,~\implies ~\left\langle \u_{\Omega,k_z} \cdot \u_{-\Omega,-k_z} \right\rangle=2 (\eta/a_c) k_B T  \Tr\left[M_{\Omega, k_z} M^\dagger_{\Omega, k_z}\right]   ,
\end{equation}
where the trace of the matrix product $\Tr\left[ M M^{\dagger} \right]$ is determined from the inverse of the $2\times 2$ matrix $M_{\Omega,k_z}^{-1}$,
\begin{equation}\label{app:uu_derivation}
\begin{aligned}
&\Tr\left[ M M^{\dagger} \right] = \frac{\Tr\left[ M^{-1} (M^{-1})^{\dagger} \right]}{\left|\det (M^{-1})\right|^2} =2 \frac{K^{\prime 2} + (\eta\Omega)^2 + \left(n_s h \Omega\right)^2}{\left|(K^{\prime}-i \eta \Omega)^2-\left(n_s h \Omega\right)^2\right|^2}    \\
 &\implies \left\langle \u_{\Omega,k_z} \cdot \u_{-\Omega,-k_z} \right\rangle  = 4(\eta/a_c) k_B T \frac{K^{\prime 2}+\tilde\eta^{2}\Omega^2 }{(K^{\prime 2}-\Omega^2\tilde\eta^{2})^2+4\Omega^2 \eta^2  K^{\prime2}} \,.
\end{aligned}
\end{equation}
Here, we have defined $\tilde{\eta} = \sqrt{\eta^2 +(n_s h)^2}$, and $K^{\prime}  = K + \varepsilon_l k_z^2 - \mu_v \Omega^2$ for notational convenience.
Finally, applying Eq.~\eqref{app:uu_derivation} to Eq.~\eqref{eq:Npinned}, we determine the transverse magnetic noise to be,
\begin{equation}
    \mathcal{N}_{x x}(\Omega)=\mathcal{N}_{y y}(\Omega) \simeq \left(\partial_\rho B_{0}^{\rho}\right)_{\rho=0}^2 ~\left(2 \frac{\eta}{a_c}  k_B T\right)\frac{K^{\prime 2}+\Omega^2 \tilde{\eta}^2}{\left(K^{\prime 2}-\Omega^2 \tilde{\eta}^2\right)^2+4 \Omega^2 \eta^2 K^{\prime 2}} \,,
    \label{app:N_xx_general}
\end{equation}
thus arriving at Eq.~\eqref{eq:Nxx_pinned} of the main text.

We now turn to $\expect{|\u_t|^2 |\u_0|^2}$, which determines the longitudinal magnetic noise.
To this end, we define a displacement correlation matrix $\mathcal{U}_{\Omega}$, with elements $\mathcal{U}_{\Omega}^{\alpha\beta} = \expect{u_{\Omega}^{\alpha} u_{-\Omega}^\beta}$ obtained via Eq.~\eqref{app:uuCorr} (we temporarily suppress the $k_z$ index for clarity). 
We note that by symmetry $\mathcal{U}_{\Omega}^{xx} = \mathcal{U}_{\Omega}^{yy}$, and $\mathcal{U}_{\Omega}^{xy} = -\mathcal{U}_{\Omega}^{yx}$,
\begin{equation}
\begin{aligned}
\mathcal{U}_{\Omega}^{xx} = \mathcal{U}_{\Omega}^{yy} &=\frac{2 (\eta/a_c) k_B T\left(K^{\prime 2}+\Omega^2 \tilde\eta^{2}\right)}{\left(K^{\prime 2}-\Omega^2 \tilde\eta^{2}\right)^2+4 \Omega^2 \eta^2 K^{\prime 2}}, \\
\mathcal{U}_{\Omega}^{xy} = -\mathcal{U}_{\Omega}^{yx}&=-i \frac{4 (\eta/a_c) k_B T \cdot n_s h K^{\prime} \Omega}{(K^{\prime 2}-\Omega^2\tilde\eta^{2})^2+4\Omega^2 \eta^2  K^{\prime2}} .
\end{aligned}
\end{equation}
The four-point correlator can be decomposed into products of two-point correlators using Wick's theorem,
\begin{equation}
\begin{aligned}
\left\langle u^{\alpha}(t)^2 u^\beta(0)^2\right\rangle &=\left\langle u^{\alpha}(t)^2\right\rangle\left\langle u^\beta(0)^2\right\rangle+2\left\langle u^{\alpha}(t) u^\beta(0)\right\rangle^2   , \\
\langle|\u_t|^2|\u_0|^2\rangle &= \sum_{\alpha, \beta}\left[\left\langle u^{\alpha}(t)^2\right\rangle\left\langle u^\beta(0)^2\right\rangle+2~\mathcal{U}^{\alpha\beta}(t)^2\right]   .
\end{aligned}
\end{equation}
After integrating over time, 
\begin{equation}\label{app:u4_derivation}
\begin{aligned}
\int_{-\infty}^{\infty} d t\, e^{i \Omega t} \langle\left| \u_t\right|^2\left|\u_0\right|^2\rangle &=\int_{-\infty}^{\infty} d t\, e^{i \Omega t}\left[|2 ~\mathcal{U}^{xx} (0)|^2+2\left(2 |\mathcal{U}^{xx}(t)|^2+ 2 |\mathcal{U}^{xy}(t)|^2\right)\right]\\
&=2 \pi \delta(\Omega) |2~ \mathcal{U}^{xx} (0)|^2 + 4 \int_{-\infty}^{\infty} d t\, e^{i \Omega t}\left[|\mathcal{U}^{xx}(t)|^2+ |\mathcal{U}^{xy}(t)|^2\right]\\
&=\frac{2}{\pi}\delta(\Omega) \left(\int_{-\infty}^{\infty} d \omega\,~ \mathcal{U}_{\omega}^{xx}\right)^2 + \frac{2}{\pi}\left( \int_{-\infty}^{\infty} d \omega\,~ \mathcal{U}_{\omega}^{xx} ~\mathcal{U}_{\Omega-\omega}^{xx} - \int_{-\infty}^{\infty} d \omega\, ~\mathcal{U}_{\omega}^{xy} ~\mathcal{U}_{\Omega-\omega}^{xy} \right)\\
&= \frac{2}{\pi}\delta(\Omega) \left(\int_{-\infty}^{\infty} d \omega\,~ \mathcal{U}_{\omega}^{xx}\right)^2 + \frac{1}{\pi}\left(\int_{-\infty}^{\infty} d \omega~ \mathcal{U}_\omega^{+} \mathcal{U}_{\Omega-\omega}^{+}+ \int_{-\infty}^{\infty} d \omega~ \mathcal{U}_\omega^{-} \mathcal{U}_{\Omega-\omega}^{-} \right) .
\end{aligned}
\end{equation}
Here, in the last line, we have introduced the two-point correlators in circular basis $\mathcal{U}_{\Omega}^{ \pm}=\mathcal{U}_{\Omega}^{x x} \pm i \mathcal{U}_{\Omega}^{x y}$,
\begin{equation}
\begin{aligned}
\mathcal{U}_{\Omega}^{ \pm}=\frac{2(\eta/a_c) k_B T}{\left(K^{\prime}\mp\Omega n_s h\right)^2+\Omega^2 \eta^2} .
\end{aligned}    
\end{equation}

Applying Eq.~\eqref{app:u4_derivation} back into Eq.~\eqref{eq:Npinned}, we determine the longitudinal magnetic noise.
\begin{equation}
\mathcal{N}_{z z}(\Omega) \simeq \left(\partial_\rho^2 B^{z}_0\right)_{\rho=0}^2 ~ \frac{1}{4\pi}\left(\int_{-\infty}^{\infty} d \omega~ \mathcal{U}_\omega^{+} \mathcal{U}_{\Omega-\omega}^{+}+ \int_{-\infty}^{\infty} d \omega~ \mathcal{U}_\omega^{-} \mathcal{U}_{\Omega-\omega}^{-} \right) . 
\label{app:N_zz_general}
\end{equation}

While the integral above is algebraically cumbersome, it simplifies significantly in several physically relevant limits.
We begin by considering the massless limit ($\mu_v = 0$). In this limit, the two-point correlators simplify to,
\begin{equation}
\begin{aligned}
&\mathcal{U}_{\Omega}^{ \pm}=\frac{2(\eta/a_c) k_B T}{\left(\overline{K}(k_z)\mp\Omega n_s h\right)^2+\Omega^2 \eta^2} =  \frac{2(\eta/a_c) k_B T}{\tilde{\eta}^2\left(\Omega\mp\Omega_c\right)^2+\left(\eta\overline{K}(k_z)/\tilde\eta\right)^2}, \quad \Omega_c = \frac{n_s h \overline{K}(k_z) }{\tilde\eta^2}.
\end{aligned}
\end{equation}
Convoluting this simplified correlator, we determine the correlator of the squared displacements to be,
\begin{equation}\label{app:4_point_corr_massless}
\begin{aligned}
\int dt \, \ e^{i\Omega t}\expect{|\u_t|^2|\u_0|^2} &=  \frac{1}{\pi}\left(\int_{-\infty}^{\infty} d \omega~ \mathcal{U}_\omega^{+} \mathcal{U}_{\Omega-\omega}^{+}+ \int_{-\infty}^{\infty} d \omega~ \mathcal{U}_\omega^{-} \mathcal{U}_{\Omega-\omega}^{-} \right) \\
 &= \frac{1}{\pi\tilde\eta^4}\int_{-\infty}^{\infty} d \omega~ \frac{2(\eta/a_c) k_B T}{\left(\omega-\Omega_c\right)^2+\left(\eta\overline{K}(k_z)/\tilde\eta^2\right)^2} \frac{2(\eta/a_c) k_B T}{\left(\Omega-\omega-\Omega_c\right)^2+\left(\eta\overline{K}(k_z)/\tilde\eta^2\right)^2} + \left(\Omega_c \rightarrow -\Omega_c \right) \\
 &= \frac{8\eta}{\tilde\eta^2\overline{K}(k_z)}\left(\frac{k_B T}{a_c}\right)^2 \left[\frac{1}{(\Omega-2\Omega_c)^2+\left(2 \eta\overline{K}(k_z)/\tilde\eta^2\right)^2} + \frac{1}{(\Omega+2\Omega_c)^2+\left(2 \eta\overline{K}(k_z)/\tilde\eta^2\right)^2}  \right]\\
 &= \frac{16\eta}{\overline{K}(k_z)}\left(\frac{k_B T}{a_c}\right)^2 \frac{4\overline{K}(k_z)^2 + \Omega^2 \tilde\eta^2}{\left(4\overline{K}(k_z)^2 + \Omega^2 \tilde\eta^2\right)^2 - 16\Omega^2 \left(n_s h \overline{K}(k_z)\right)^2} .
\end{aligned}
\end{equation}
Here, in the second-to-last line, we have used the convolution identity,
\begin{equation}\label{app:convolution_thm}
\int_{-\infty}^{\infty} d \omega\,\frac{1}{(\omega-x)^2+\gamma^2} \frac{1}{(\Omega-\omega-y)^2+\gamma^2}=\frac{2 \pi}{\gamma} \frac{1}{(\Omega-x-y)^2+(2 \gamma)^2} .
\end{equation}

Next, in the massive or high frequency limit, we can neglect the Magnus forces set by $n_s h$. 
Here, the correlator $\mathcal{U}_{\Omega}^{ \pm}$ can be well-approximated by a sum of two Lorentzians centered at $\pm\Omega_l$, provided the damping is weak, i.e., $\mu_v \Omega_l \gg \eta$.
\begin{equation}
\begin{aligned}
\mathcal{U}_{\Omega}^{ \pm}&=\frac{2(\eta/a_c) k_B T}{\left(\overline{K}(k_z) - \mu_v \Omega^2\right)^2+\Omega^2 \eta^2} =  \frac{2(\eta/a_c) k_B T}{\mu_v^2\left(\Omega^2-\Omega_l^2\right)^2+\eta^2 \Omega^2}, \quad \Omega_l = \sqrt{\frac{\overline{K}(k_z)}{\mu_v}} \\
&\approx \frac{2(\eta/a_c) k_B T}{4 \mu_v^2 \Omega_l^2}\left[\frac{1}{\left(\Omega-\Omega_l\right)^2+(\eta/2\mu_v)^2}+\frac{1}{\left(\Omega+\Omega_l\right)^2+(\eta/2\mu_v)^2}\right], ~\text{ for }\Omega\to\pm\Omega_l.
\end{aligned}
\end{equation}
With this, the correlator of the squared displacements is given by,
\begin{equation}\label{app:4_point_corr_massive}
\begin{aligned}
\int dt \, \ e^{i\Omega t}\expect{|\u_t|^2|\u_0|^2} & =  \frac{1}{\pi}\left(\int_{-\infty}^{\infty} d \omega~ \mathcal{U}_\omega^{+} \mathcal{U}_{\Omega-\omega}^{+}+ \int_{-\infty}^{\infty} d \omega~ \mathcal{U}_\omega^{-} \mathcal{U}_{\Omega-\omega}^{-} \right) \\
 &= \frac{2}{\pi}\int_{-\infty}^{\infty} d \omega~\left(\frac{2(\eta/a_c) k_B T}{4 \mu_v^2 \Omega_l^2}\right)^2\left[\frac{1}{\left(\omega-\Omega_l\right)^2+(\eta/2\mu_v)^2}+\frac{1}{\left(\omega+\Omega_l\right)^2+(\eta/2\mu_v)^2}\right] \\
 & ~~~~~~~~~~~~~~~~~~~~~~~~~~\times \left[\frac{1}{\left((\Omega-\omega)-\Omega_l\right)^2+(\eta/2\mu_v)^2}+\frac{1}{\left((\Omega-\omega)+\Omega_l\right)^2+(\eta/2\mu_v)^2}\right]\\
 & = \frac{2\eta}{\mu_v}\left(\frac{k_B T}{a_c\overline{K}(k_z)}\right)^2\left[\frac{1}{\left(\Omega-2 \Omega_l\right)^2+(\eta/\mu_v)^2}+\frac{1}{\left(\Omega+2 \Omega_l\right)^2+(\eta/\mu_v)^2}+\frac{2}{\Omega^2+(\eta/\mu_v)^2}\right] ,
\end{aligned}
\end{equation}
where we have used the convolution identity  Eq.~\eqref{app:convolution_thm}.\\

Last, in the limit of large viscosity $\eta$ ($\eta\gg n_sh,\mu_v\Omega$), we can neglect $n_s h\Omega$ and $\mu_v\Omega^2$ in the two-point correlators, upon which the correlator of the squared displacements simplifies to,
\begin{equation}\label{app:4_point_corr_damping}
\begin{aligned}
&\mathcal{U}_{\Omega}^{ \pm}=\frac{2(\eta/a_c) k_B T}{\left(\overline{K}(k_z)\right)^2+\Omega^2 \eta^2}
\implies& \int dt \, \ e^{i\Omega t}\expect{|\u_t|^2|\u_0|^2} =\frac{16}{\eta\overline{K}(k_z)}\left(\frac{k_B T}{a_c}\right)^2 \frac{1}{\Omega^2+(2 \overline{K}(k_z)/\eta)^2}  .
\end{aligned}
\end{equation}
Applying \cref{app:4_point_corr_massless,app:4_point_corr_massive,app:4_point_corr_damping} on Eq.~\eqref{app:N_zz_general}, we arrive at \cref{eq:Nzz_pinned_massless,eq:Nzz_pinned_massive,eq:Nzz_pinned_damping} in the main text.

\section{Energetics of vortices}

\subsection{Line tension of a single vortex}
\label{app:E_self_energy}
In this appendix, we will calculate the self-energy of a single static Abrikosov (line) vortex. 
We will assume that the vortex axis is along the $z$-direction, and that the vortex is translation invariant in the $z$-direction, so we will only care about self-energy per unit length. 
This quantity can also be interpreted as the line-tension $\varepsilon_l$ of the vortex, as the energy required to stretch the vortex by unit length in the $z$-direction is given by $\varepsilon_l$.
Analytical results can be obtained in the extreme type-II limit, which is $\kappa = \lambda/\xi \gg 1$, i.e., the core of the vortex, where the order parameter varies in amplitude, is assumed to be very small.

We start with the full Ginzburg-Landau free energy for a charged superfluid, which is given by
\begin{equation}
\mathcal{F} = \frac{1}{2m^*} \int d^3\r \, |(-i \hbar \nabla - e^* \bm{A})\psi |^2 + \int d^3\r\, \left[ \frac{|\nabla \times \bm{A}|^2}{2\mu_0} + \frac{1}{2 \varepsilon_0} \bm{E}^2_0 \right]  .
\label{eq:GLF}
\end{equation}
We will further assume that there is local charge density such that $\nabla \cdot \bm{E}_0 = 0$ (one may think this is not true for the vortex core where the order-parameter amplitude is indeed changing, but that region has unpaired electrons that balance the positive charge background so that the system is electrically neutral everywhere), and that the magnetic field is static, such that $\nabla \times \bm{E}_0 = 0$, together implying that  $\bm{E}_0 = 0$. 
Finally, we note that the gauge-invariant superfluid velocity (where the order-parameter amplitude variation can be neglected) is given by
\begin{equation}
    \j_s = n_s e^* \v_s = n_s e^* \left(\frac{\hbar \nabla \theta - e^* \bm{A}}{m^*}\right)  .
\end{equation}
where we used $\psi = |\psi| e^{i \theta}$, and $|\psi|^2 = n_s$ is the superfluid density.
Hence, the first term in the GL free energy density can be interpreted as the kinetic energy density of the superfluid. 
\begin{equation}
\begin{aligned}
    \frac{1}{2m^*} |(-i \hbar \nabla - e^* \bm{A})\psi |^2 &= \frac{|\psi|^2 m^*}{2} \left|\frac{\hbar \nabla \theta - e^* \bm{A}}{m^*} \right|^2 =  \frac{1}{2} n_s m^* \v_s^2 \\
    &  = \frac{m^*}{2 n_s (e^*)^2} \j_s^2 = \frac{\mu_0}{2 } \times \frac{m^*}{\mu_0
n_s (e^*)^2} \j_s^2 = \frac{\mu_0 \lambda_L^2}{2} \j_s^2  ,
\end{aligned}
\end{equation}
where we have used the definition of the London penetration depth $\lambda_L^{-2} = \mu_0 n_s (e^*)^2/m^*$.
Using this, and the Maxwell equation $\nabla \times \B_0 = \mu_0 \j_s$,  the free energy can be written entirely in terms of the magnetic field $\B_0$ as 
\begin{equation}
\mathcal{F}= \frac{1}{2\mu_0} \left[ \int d^3\r \, (\lambda_L \mu_0 \j_s)^2 +  \int d^3\r \, \B^2_0 \right] =  \frac{1}{2\mu_0} \left[ \int d^3\r \, \left( \lambda_L^2 |\nabla \times \B_0|^2 + |\B_0|^2 \right) \right]  .
\label{eq:GLF2}
\end{equation}
Eq.~\eqref{eq:GLF2} is the most convenient form to evaluate free energy.

The calculation of the line tension ($\varepsilon_l = \mathcal{F}/ d$) can be done in (2d) momentum space, using our expression for $\phi_M(\k)$ from Eq.~\eqref{eq:mag_pot}, that gives the correct expression for the $\B_0$ field as $z \to 0^+$ in the half-space sample, and therefore by translation invariance at all $z$ in the 3D sample with no boundaries. 
In this case 
\begin{equation}
\B_0 = \B_0(\bm{\rho}) \hat{z}= - \partial_z \phi_M(\bm{\rho},z)|_{z \to 0^+} = \int \frac{d^2\k}{(2\pi)^2} \, e^{i \k \cdot \bm{\rho}}\, k\, \phi_M(\k), \text{ such that we can define } \B_0(\k) = k\, \phi_M(\k)\, \hat{z}
\end{equation}
has a component only in the $z$-direction, so the current in purely azimuthal. That gives line tension, 
\begin{equation}
\begin{aligned}
    \varepsilon_l &= \frac{\lambda_L^2}{2\mu_0}\int d^2 \rr\, |\nabla \, \times \B_0|^2  + \frac{1}{2\mu_0}\int d^2 \rr\, |\B_0|^2  \\
    & = \frac{\lambda_L^2}{2\mu_0}\int \frac{d^2\k}{(2\pi)^2} \, |i \k \times \B_0(\k)|^2 + \frac{1}{2\mu_0} \int \frac{d^2\k}{(2\pi)^2} |\B_0(\k)|^2   \\
    &= \frac{\lambda_L^2}{2\mu_0}\int \frac{d^2\k}{(2\pi)^2} \, k^4 |\phi_M(\k)|^2 + \frac{1}{2\mu_0}\int \frac{d^2\k}{(2\pi)^2} \,  k^2|\phi_M(\k)|^2 \\
    & =  \frac{\Phi_0^2}{2\mu_0} \int \frac{d^2\k}{(2\pi)^2} \frac{\lambda_L^2 k^2}{(1 + \lambda_L^2 k^2)^2}  +  \frac{\Phi_0^2}{2\mu_0} \int \frac{d^2\k}{(2\pi)^2} \frac{1}{(1 + \lambda_L^2 k^2)^2} \\
 &= \frac{\Phi_0^2}{2\mu_0} \frac{1}{2\pi \lambda_L^2} \int_0^\infty d\tilde{k} \, \frac{\tilde{k}^3}{(1 + \tilde{k}^2)^2} +   \frac{1}{2\mu_0} \frac{\Phi_0^2}{2 \pi \lambda_L^2} \int_0^\infty d\tilde{k} \frac{\tilde{k}}{(1 + \tilde{k}^2)^2}   .
\end{aligned}
\end{equation}
The first term of RHS is UV-divergent, so we need to regulate it by remembering that our vortex solution stops being valid for $\rho \lesssim \xi$, implying that we need to cut-off the integral at $k \approx \xi^{-1}$, i.e., $\tilde{k} \approx \lambda_L/\xi = \kappa \gg 1$~\cite{tinkham}.
This yields
\begin{equation}
 \frac{\Phi_0^2}{2\mu_0} \frac{1}{2\pi \lambda_L^2} \int_0^{\kappa} d\tilde{k} \, \frac{\tilde{k}^3}{(1 + \tilde{k}^2)^2} = \frac{1}{2\mu_0} \frac{\Phi_0^2}{2\pi \lambda_L^2} \left[ \frac{1}{2}\ln\left(1 + \frac{\lambda_L^2}{\xi^2} \right) - \frac{1}{2(1 + (\xi/\lambda_L)^2)} \right] \approx \frac{1}{2\mu_0}  \frac{\Phi_0^2}{2\pi \lambda_L^2} \left[ \ln\left(\frac{\lambda_L}{\xi}\right) - \frac{1}{2} \right]   .
\end{equation}
For the second term, we find that there is no issue with UV divergence
\begin{equation}
\frac{1}{2\mu_0} \frac{\Phi_0^2}{2 \pi \lambda_L^2} \int_0^\infty d\tilde{k} \frac{\tilde{k}}{(1 + \tilde{k}^2)^2} =  \frac{1}{2\mu_0} \frac{\Phi_0^2}{2 \pi \lambda_L^2} \frac{1}{2}  .
\end{equation}
The magnetic field contribution exactly cancels off the subleading contribution due to current density, leading to a net line tension, i.e., free energy per unit length in the $z$-direction, given by
\begin{equation}
\varepsilon_l = \frac{1}{2\mu_0}  \frac{\Phi_0^2}{2\pi \lambda_L^2} \ln\left( \frac{\lambda_L}{\xi} \right) = \frac{1}{2\mu_0} \underbrace{\left( \frac{\Phi_0}{\pi \lambda_L^2} \right)^2}_{\text{average magnetic field}} \times \underbrace{(\pi \lambda_L^2)}_{\text{area where field is significant}} \times \underbrace{\frac{1}{2} \ln\left( \frac{\lambda_L}{\xi} \right)}_{\text{U.V. physics}}.
\end{equation}

\subsection{Interaction energy of two vortices}
\label{app:E_int_energy}

We assume that the two vortices (labeled 1 and 2) are placed at a distance $\rho_0 ~(\gg \xi)$ from each other, and that the total super-current density and the magnetic fields are the superpositions of those of the individual vortices. 
Therefore, using Eq.~\eqref{eq:GLF2}, the total energy is given by 
\begin{equation}
\mathcal{F}(\rho_0) = \frac{1}{2\mu_0} \left[ \int d^3\r \, (\lambda_L \mu_0)^2( \j_{s,1} + \j_{s,2})^2 +  \int d^3\r \, (\B_{0, 1} + \B_{0, 2})^2 \right] =  \mathcal{F}_1 + \mathcal{F}_2 + V(\rho_0)  ,
\end{equation}
where $\mathcal{F}_1$ and $\mathcal{F}_2$ are the self-energies of the two vortices calculated in the previous section, and $V(\rho_0)$ is interaction energy between two vortices, given by
\begin{equation}
V(\rho_0) = \frac{1}{\mu_0} \left[ \int d^3\r \, (\lambda_L \mu_0)^2( \j_{s,1} \cdot \j_{s,2}) +  \int d^3\r \, \B_{0, 1} \cdot \B_{0, 2} \right]   .
\label{eq:F2v}
\end{equation}
More generally, in case of multiple vortices, we may write the most general interaction energy respecting translation invariance as 
\begin{equation}
\mathcal{H}_{\rm interaction} = \frac{1}{2} \int d^2 \rr \int d^2 \rr^\prime\, n_v(\bm{\rho}) V(\bm{\rho} - \bm{\rho}^\prime) n_v(\bm{\rho}^\prime)  .
\end{equation}
This reduces to the expression for two vortices in Eq.~\eqref{eq:F2v} when we consider two point vortices, such $n_v(\bm{\rho}) = \delta^{(2)}(\bm{\rho}) + \delta^{(2)}(\bm{\rho} - \bm{\rho_0})$.
Our goal is to find $V(\bm{\rho}) = V(|\bm{\rho}|)$, and subsequently its Fourier transform $\tilde{V}(\k)$.

\subsubsection{Abrikosov vortex}
In this case, we again assume translation invariance in the $z$-direction, and focus on the energy per unit length. 
\begin{align}
\frac{V(\rho_0)}{d} &= \frac{1}{\mu_0} \left[ \int d^2 \rr\, (\lambda_L \mu_0)^2( \j_{s,1} \cdot \j_{s,2}) +  \int d^2 \rr\, \B_{0, 1} \cdot \B_{0, 2} \right] \nonumber \\
&= \frac{1}{\mu_0} \left[ \int \frac{d^2k}{(2\pi)^2} (\lambda_L \mu_0)^2 \, \j_{s}(\k) \cdot \j_s(-\k) e^{i \k \cdot \bm{\rho_0}} + \int \frac{d^2k}{(2\pi)^2} \B_0(\k) \cdot \B_0(-\k) e^{i \k \cdot \bm{\rho}_0} \right] \nonumber \\
&= \frac{1}{\mu_0} \int \frac{d^2k}{(2\pi)^2} (k^2 \lambda_L^2 + 1) |B_0^z(\k)|^2 e^{i \k \cdot \bm{\rho}_0}  \nonumber  \\
&=  \frac{\Phi_0^2}{\mu_0} \int \frac{d^2k}{(2\pi)^2} 
\frac{1}{1 + k^2 \lambda_L^2} e^{i \k \cdot \bm{\rho}_0} 
\label{eq:2Vintk} \\
& = \frac{\Phi_0^2}{2\pi \mu_0 \lambda_L^2} \int_0^\infty d\tilde{k} \, \frac{\tilde{k} J_0(\tilde{k}\rho_0/\lambda_L) }{1 + \tilde{k}^2} =  \frac{\Phi_0^2}{2\pi \mu_0 \lambda_L^2} K_0\left( 
\frac{\rho_0}{\lambda_L} \right)   ,
\end{align}
where we used $B_0^z(\k) = k \phi_M(k) =  \Phi_0/(1 + k^2 \lambda_L^2)$.
Remarkably, this has a similar analytic form as the single vortex self-energy, except that the modified Bessel function $K_0$ has an argument that scales with the distance $\rho_0$ between the vortices relative to the penetration depth $\lambda_L$. 
Thus, we may use the asymptotic form of $K_0(x)$ to determine the behavior is the small and large distance limits. 
For small $x$ we have $K_0(x) \sim -\ln x$, so the interaction scales as $\ln(\lambda_L/\rho_0)$ for small $\rho_0/\lambda_L$, i.e, we recover the expected logarithmic interaction of unscreened vortices at lengthscales below the penetration depth $\lambda_L$.
For large $x$ we have $K_0(x) \sim e^{-x}/\sqrt{x}$, so the interaction between vortices drop exponentially beyond the penetration depth $\lambda_L$.

Finally, from Eq.~\eqref{eq:2Vintk}, we can also read off the Fourier space interaction (per unit length) for pairwise interaction between vortices:
\begin{equation}
\frac{\tilde{V}(\k)}{d} = \int d^2 \rr\,  e^{- i \k \cdot \bm{\rho}} \frac{V(\rho)}{d} = \frac{\Phi_0^2}{\mu_0(1 + k^2 \lambda_L^2)}  .
\label{eq:Vintk3d}
\end{equation}

\subsubsection{Pearl vortex}
The result follows directly from the self-energy calculation for the Pearl vortex and the interaction energy calculation for 3D vortices (except that we now calculate the total energy rather than the energy per unit length). 
\begin{align}
V(\rho_0) &= \frac{1}{\mu_0} \left[ \int d^2 \rr\, (\lambda_L \mu_0)^2( \j_{s,1} \cdot \j_{s,2}) +  \int d^2 \rr\, \B_{0, 1} \cdot \B_{0, 2} \right] \nonumber \\
& = \frac{2\Lambda}{\mu_0} \int \frac{d^2k}{(2\pi)^2} |\B_\parallel(\k)|^2 e^{i \k \cdot \bm{\rho}_0} + \frac{1}{\mu_0}\int \frac{d^2k}{(2\pi)^2} \frac{ (|\B_{\parallel}(\k)|^2 + |\B_\perp(\k)|^2) e^{i \k \cdot \bm{\rho}_0}}{k} \nonumber \\
 & = \frac{2}{\mu_0} \int \frac{d^2 k}{(2\pi)^2} \, k(\Lambda k  + 1) |\phi_M(\k)|^2 e^{i \k \cdot \bm{\rho}_0} \nonumber \\
 & = \frac{2 \Phi_0^2}{\mu_0}  \int \frac{d^2 k}{(2\pi)^2} \, \frac{1}{k(1 + \Lambda k)} e^{i \k \cdot \bm{\rho}_0} 
\label{eq:2PVintk} \\
 & = \frac{\Phi_0^2}{\pi \mu_0 \Lambda} \int_0^\infty d\tilde{k} \frac{J_0(\tilde{k}\rho_0/\lambda_L)}{1 + \tilde{k}} = \frac{\Phi_0^2}{2 \mu_0 \Lambda} \left[ H_0\left( \frac{\rho_0}{\Lambda} \right) - Y_0\left(\frac{\rho_0}{\Lambda} \right)  \right]  ,
\end{align}
where we have used that $\phi_M(\k) = \Phi_0/[k(1 + \Lambda k)]$ for Pearl vortices. 
Once again, we may study the asymptotic limits --- we have $H_0(x) - Y_0(x) \approx - \ln x$ if $x \ll 1$, which leads to the expected unscreened logarithmic interaction between two Pearl vortices at short distances $\rho_0$ compared to the Pearl length $\Lambda$. 
For large $\rho_0/\Lambda$, we need the asymptotic form $H_0(x) - Y_0(x) \approx 1/x$ for $x \gg 1$, which leads to $V(\rho_0) \approx \Phi_0^2/(2 \mu_0 \rho_0)$. 
This is remarkably different from 3D vortices, and indicates a long range Coulomb-like $1/\rho$ interaction between two pointlike Pearl vortices. 
We will later show that this long-range interaction will have important consequences for the longitudinal elastic modulus of a lattice of Pearl vortices.

Finally, from Eq.~\eqref{eq:2PVintk}, we can again read off the Fourier transform of the pair-interaction potential for Pearl vortices:
\begin{equation}\label{eq:Vintk2d}
\tilde{V}(\k) = \frac{2 \Phi_0^2}{\mu_0 k(1 + k \Lambda)}  .
\end{equation}

\section{Vortex lattice phonon dispersion}
\label{app:solid}

In this appendix, we will first focus on the derivation of the propagators $G(\k)$ that connect the in-plane displacements to the linear order expansion of the magnetic fields. From there, we can compute the vortex phonon noise, integrated from longitudinal correlation function $C_{\L}(\k,\Omega) = \langle \k \cdot \u_{\k, \Omega} \,~ \k \cdot \u_{-\k,-\Omega} \rangle$. 
We then determine the elastic moduli of the vortex lattice by vortex interactions, which together with $G(\k)$ lead to the full expression of $C_{\L}$.

\subsection{Propagators $G(\boldsymbol{k})$}
For $z \gg \lambda_L$, the magnetic field of a vortex behaves like that of a magnetic monopole, as discussed in Appendix~\ref{app:C_mag},
\begin{equation}
\phi_M(\bm{\rho}, z \gg \lambda_L) = \frac{\Phi_0}{2\pi r}, \quad \B_0(\r) = \frac{\Phi_0}{2\pi r^2} \hat{r} ~~~\text{ for }~ r = \sqrt{\rho^2 + z^2} \gg \lambda_L  .
\end{equation}
The total static magentic field at position $\r_q = (\boldsymbol{0}, z_0)$ generated by all vortices in the lattice is
\begin{equation}
\B_{\rm lat, 0}(\r_q, t) = \frac{\Phi_0}{2\pi}\sum_{i} \frac{\r_q - \r_i(t)}{\left|\r_q - \r_i(t)\right|^3}  .
\end{equation}
where $\r_i(t) = (\R_i + \u_i(t), 0)$, $\R_i$ is the equilibrium position of the $i$-th vortex in the lattice, and $\u_i(t)$ is its displacement from equilibrium. For small displacements $|\u_i(t)| \ll a_\Delta$, we can expand the magnetic field to linear order in $\u_i(t)$, by letting $\boldsymbol{\varrho}_i(t)=\r_q-\r_i(t)=\r_q-\R_i-\u_i(t)$, $\boldsymbol{\varrho}_{0 i}=\r_q-\R_i$, 
\begin{equation}
\left|\boldsymbol{\varrho}_i(t)\right|^{-3} \approx \varrho_{0 i}^{-3}\left[1+\frac{3 \boldsymbol{\varrho}_{0 i} \cdot \u_i}{\varrho_{0 i}^2}\right]  \implies \frac{\boldsymbol{\varrho}_{0 i}-\u_i}{\left|\boldsymbol{\varrho}_{0 i}-\u_i\right|^3} =(\boldsymbol{\varrho}_{0 i}-\u_i)\left|\boldsymbol{\varrho}_i(t)\right|^{-3} \approx \frac{\boldsymbol{\varrho}_{0 i}}{\varrho_{0 i}^3}+\frac{3 \boldsymbol{\varrho}_{0 i}\left(\boldsymbol{\varrho}_{0 i} \cdot \u_i\right)}{\varrho_{0 i}^5}-\frac{\u_i}{\varrho_{0 i}^3}  ,
\end{equation}
that help us find the linear order expansion of the magnetic field,
\begin{equation}
\begin{aligned}
\B\left(\r_q, t\right)    & =  \frac{\Phi_0}{2 \pi} \sum_i\left[\frac{\r_q-\R_i}{\left|\r_q-\R_i\right|^3}+\frac{3\left(\r_q-\R_i\right)\left[\left(\r_q-\R_i\right) \cdot \u_i(t)\right]-\u_i(t)\left|\r_q-\R_i\right|^2}{\left|\r_q-\R_i\right|^5}\right]  .  \\
\end{aligned}
\end{equation}
Applying $\boldsymbol\varrho_{0 i} = \r_q - \R_i = (\boldsymbol{0} - \R_i, z_0)$, $\varrho_{0 i}^2 = R_i^2 + z_0^2$, $\boldsymbol{\varrho}_{0 i} \cdot \u_i = -\R_i \cdot \u_i$ and the lattice symmetry $\sum_i\frac{\R_i}{\left|\r_q-\R_i\right|^3}=\sum_i\frac{\R_i}{\left(z_0^2+R_i^2\right)^{3 / 2}}=0$, we arrive at Eq.~\eqref{eq:Blat},
\begin{equation}
\begin{aligned}
\B\left(\r_q, t\right) & =\B_{\rm lat, 0}\left(\r_q\right)+\delta B_z\left(\r_q, t\right) \hat{z}+\delta \B_{\|}\left(\r_q, t\right), \text { where } \\
\B_{\rm lat, 0}\left(\r_q\right) & =\frac{\Phi_0}{2 \pi}\left(\sum_i \frac{z_0}{\left(z_0^2+R_i^2\right)^{3 / 2}}\right) \hat{z}   ,\\
\delta B_z\left(\r_q, t\right) & =-\frac{\Phi_0}{2 \pi} \sum_i \frac{3 z_0 \R_i \cdot \u_i(t)}{\left(z_0^2+R_i^2\right)^{5 / 2}}, \text { and } \\
\delta \B_{\|}\left(\r_q, t\right) & =\frac{\Phi_0}{2 \pi} \sum_i \frac{3 \R_i\left(\R_i \cdot \u_i(t)\right)-\u_i(t) \left(z_0^2+R_i^2\right)}{\left(z_0^2+R_i^2\right)^{5 / 2}}   .
\end{aligned}
\end{equation}
The first term is the static field from the vortex lattice, and the second and third terms are the fluctuating fields due to vortex motion (out-of-plane and in-plane components, respectively).

For $z_0 \gg a_{\Delta}$, the sum over the vortex lattice can be approximated by the continuum integral $\sum_i \rightarrow n_v \int d^2 R$,
\begin{equation}
\begin{aligned}
\B_{\rm lat, 0}\left(\r_q\right) & = \frac{\Phi_0 n_v}{2 \pi} \int_{\mathbb{R}^2} d^2 R \frac{z_0}{\left(R^2+z_0^2\right)^{3 / 2}} \hat{z} =\frac{\Phi_0 n_v z_0}{2 \pi} \int_0^{2 \pi} d \phi \int_0^{\infty} \frac{R\, d R}{\left(R^2+z_0^2\right)^{3 / 2}} \hat{z} = \Phi_0 n_v \hat{z}\\
\delta B_z(t)&=-\frac{\Phi_0 n_v}{2 \pi} \int d^2 R \frac{3 z_0 \R \cdot \u(\R, t)}{\left(R^2+z_0^2\right)^{5 / 2}}, \\
\delta \boldsymbol{B}_{\|}(t)&=\frac{\Phi_0 n_v}{2 \pi} \int d^2 R \frac{3 \R(\R \cdot \u(\R, t))-\u(\R, t)\left(R^2+z_0^2\right)}{\left(R^2+z_0^2\right)^{5 / 2}} .
\end{aligned}
\end{equation}
We could consider to write fluctuating fields in Fourier space for convenience, using the displacement field $\u_i(t) = N_v^{-1/2} \sum_{\k} \u(\k, t) e^{i \k \cdot \R_i}$, where $N_v$ is the total number of vortices in the lattice, the out-of-plane component becomes
\begin{equation}
\begin{aligned}
\delta B_z(t)&=\frac{1}{\sqrt{N_v}} \sum_{\k} G_z{ }^\mu(\k) u_\mu(\k, t),\\
G_z{ }^\mu(\k)&=-\frac{\Phi_0 n_v}{2 \pi} \int d^2 R \frac{3 z_0 R^\mu}{\left(R^2+z_0^2\right)^{5 / 2}} e^{i \k \cdot \R} = i \frac{\Phi_0 n_v}{2 \pi}\cdot 3 z_0\cdot \frac{\partial}{\partial k_\mu} \int d^2 R \frac{e^{i \k \cdot \R}}{\left(R^2+z_0^2\right)^{5 / 2}}  \\
& = i \frac{\Phi_0 n_v}{2\pi} \cdot 3 z_0 \cdot \frac{\partial}{\partial k_\mu} \frac{2 \pi}{3 z_0^3}\left(1+k_\mu z_0\right) e^{-k z_0} = \Phi_0 n_v (-i k^\mu) e^{-k z_0} =-n_v \Phi_0 \frac{k_\mu k_\nu}{k} e^{-k z_0}  ,
\end{aligned}
\end{equation}
for in-plane component,
\begin{equation}
\begin{aligned}
\delta B_{\|}{ }^\mu(t)&=\frac{1}{\sqrt{N_v}} \sum_{\k} G_{\|}{ }^{\mu \nu}(\k) u_\nu(\k, t),\\
G_{\|}{ }^{\mu \nu}(\k)&=\frac{\Phi_0 n_v}{2 \pi} \int d^2 R \frac{3 R^\mu R^\nu-\delta^{\mu \nu}\left(R^2+z_0^2\right)}{\left(R^2+z_0^2\right)^{5 / 2}} e^{i \k \cdot \R} \\
&= \frac{\Phi_0 n_v}{2 \pi}\left[3 \frac{\partial^2}{\partial k_\mu \partial k_\nu} I_5(\k)-\delta^{\mu \nu} I_3(\k)\right] = -n_v \Phi_0 \frac{k_\mu k_\nu}{k} e^{-k z_0}  .
\end{aligned}
\end{equation}
These expressions yield Eq.~\eqref{eq:BlatFourier}. In the derivations, we have used the following integral formulae (for $z_0>0$):
\begin{subequations}\begin{align}
&I_1(\k)=\int_{\mathbb{R}^2} \frac{e^{i \k \cdot \R}}{\left(R^2+z_0^2\right)^{1 / 2}} d^2 R=2 \pi \int_0^{\infty} \frac{R J_0(k R)}{\sqrt{R^2+z_0^2}} d R =\frac{2 \pi}{k} e^{-k z_0},\\
&I_3(\k)=\int_{\mathbb{R}^2} \frac{e^{i \k \cdot \R}}{\left(R^2+z_0^2\right)^{3 / 2}} d^2 R=-\frac{1}{z_0} \frac{\partial}{\partial z_0} I_1(\k)=-\frac{1}{z_0} \frac{\partial}{\partial z_0}\left(\frac{2 \pi}{k} e^{-k z_0}\right)=\frac{2 \pi}{z_0} e^{-k z_0},\\
&I_5(\k)=\int_{\mathbb{R}^2} \frac{e^{i \k \cdot \R}}{\left(R^2+z_0^2\right)^{5 / 2}} d^2 R=-\frac{1}{3 z_0} \frac{\partial}{\partial z_0} I_3(\k)=-\frac{1}{3 z_0} \frac{\partial}{\partial z_0}\left(\frac{2 \pi}{z_0} e^{-k z_0}\right)=\frac{2 \pi}{3 z_0^3}\left(1+k z_0\right) e^{-k z_0}.
\end{align}\end{subequations}

\subsection{Elastic moduli of vortex lattices}
In this section, we provide a brief derivation of the longitudinal elastic moduli $c_{11}(k)$ of the vortex lattice for rigid Abrikosov vortices and Pearl vortices, and indicate how to compute the transverse modulus $c_{66}(k)$.
To this end, we will take a continuum approach, and assume that small lattice displacements can be quantified by a continuum displacement field $\u(\bm{\rho})$, where $\bm{\rho}= (x,y)$ denotes the coordinate in the 2D plane. 
Accordingly, the vortex number density will change as 
\begin{equation}
    \delta n_v(\bm{\rho}) = - \bar{n}_v \bm{\nabla} \cdot \u(\bm{\rho})  ,
\end{equation}
where we have used $\bar{n}_v$ to denote the average density of vortices (and avoid confusion with the spatially fluctuating local density $n_v(\bm{\rho}$): in the main text this average density has been denoted simply by $n_v$). 
Accordingly, in Fourier space, we can write the vortex density as $\delta n_v(\k) = - i \bar{n}_v \k \cdot \u(\k) = - i \bar{n}_v k u_\L(\k)$, which indicates that only longitudinal displacements $u_\L(\k) = \hat{\k} \cdot \u(\k)$ change the vortex density. 
Given the pairwise interaction potential $\tilde{V}(\k)$ between vortices, the change in energy due to such a longitudinal displacement of the vortex lattice is given by
\begin{equation}
\delta E = \frac{1}{2} \int \frac{d^2k}{(2\pi)^2} \tilde{V}(\k) \delta n_v(\k) \delta n_v(-\k) = \frac{1}{2}  \int \frac{d^2k}{(2\pi)^2} \bar{n}_v^2 \tilde{V}(\k) k^2 |u_\L(\k)|^2   .
\end{equation}
Comparing with the continuum elastic Hamiltonian, where $\delta E = \frac{1}{2}\int_{\k} \cone(\k) k^2 |u_\L(\k)|^2$ for a longitudinal deformation (e.g., a compression), we find that:
\begin{equation}
    \cone(k) = \bar{n}_v^2 \tilde{V}(k)   .
\end{equation}
Thus, the compression modulus is directly determined by the pair-wise interaction potential $\tilde{V}(\k)$ between vortices, as quoted in the main text. 
The relevant interaction potentials for Abrikosov vortices (per unit length) and Pearl vortices were calculated in Appendix~\ref{app:E_int_energy}.

Further, we note that our results for the compression modulus can also be derived more formally by expanding the interaction Hamiltonian for the vortex lattice without appealing to the continuum limit. 
We sketch the derivation here as it also leads naturally the shear modulus (which is zero in the continuum approximation).
The energy per unit area (per unit volume for Abrikosov vortices) due to a lattice displacement field $\u(\k)$ can generally be written as 
\begin{equation}
\frac{\mathcal{H}_{\rm int}}{A} = \frac{1}{2} \int \frac{d^2k}{(2\pi)^2} u^\alpha(\k) \Phi_{\alpha \beta}(\k) u^\beta(-\k), \text{ with }  \Phi_{\alpha \beta}(\k) = \sum_\G \tilde{V}_{\alpha \beta}^{\prime \prime}(\G) - \tilde{V}_{\alpha \beta}^{\prime \prime}(\k + \G),
\end{equation}
where $\G$ denote the reciprocal vectors of the vortex lattice, and we have defined $\tilde{V}^{\prime \prime}_{\alpha \beta}(\k)$ as the Fourier transform of $\partial_{\alpha} \partial_\beta V(\rho)|_{\bm{\rho} = \R_{i}}$, i.e., 
\begin{equation}
\tilde{V}^{\prime \prime}_{\alpha \beta}(\k) = \sum_i e^{- i \k \cdot \R_i} \partial_{\alpha} \partial_\beta V(\rho)|_{\bm{\rho} = \R_{i}} \xrightarrow[]{\rm{continuum ~ limit}} n_v \int d^2 \rr\,  e^{- i \k \cdot \bm{\rho}} \partial_{\alpha} \partial_\beta V(\rho) .
\end{equation}
If the sum on $\G$ is convergent, we may further integrate by parts to write 
\begin{equation}
\tilde{V}_{\alpha \beta}^{\prime \prime}(\G) - \tilde{V}_{\alpha \beta}^{\prime \prime}(\k + \G) = \left[ (\k + \G)_\alpha (\k + \G)_\beta - \G_\alpha \G_\beta \right] \tilde{V}(\k + \G)   .
\end{equation}
In general, for a triangular lattice, $\Phi_{\alpha \beta}(\k)  = k_\alpha k_\beta \cone(k) + (k^2 \delta_{\alpha \beta} - k_\alpha k_\beta) \csix(k)$
where $\cone(k)$ is the compression modulus and $\csix(k)$ is the shear modulus. 
We see that if we keep only the $\G = 0$ term to determine the elastic free energy, we get consistent results with the continuum theory.
Specifically, within this approximation, the compressional modes with $\nabla \cdot \u \neq 0$ cost finite energy, while the shear modes cost zero energy, which is the characteristic of a vortex liquid phase. 
Therefore, we turn to an explicit computation of the shear modulus $\csix$ for the vortex lattice, noting that the shear modulus is not expected to have strong momentum dependence~\cite{BlatterRMP, Brandt95}.

Instead of evaluating sum over reciprocal lattice vectors, we consider an alternate strategy and directly compute the change in energy density $\delta E=\frac{1}{2} c_{66} \gamma^2$ due to a shear deformation, where $\gamma=\partial_x u^y+\partial_y u^x$ is the shear strain. 
We choose a specific a shear deformation on the vortex lattice points: $\R_i = (x_i, y_i) \to \R_i^{\prime} = (x_i + \epsilon y_i, y_i)$, where $\epsilon\ll1$ parametrizes the shear.
To preserve the periodicity of $e^{i \G \cdot \R}$ in the deformed lattice, the reciprocal vectors transform as $\G = (G_x, G_y) \to \G^{\prime} = (G_x, G_y - \epsilon G_x)$. 
This implies that the reciprocal space is sheared with a same angle as the real space, but in the perpendicular direction, with momenta transforming as, $\k = (k_x, k_y) \to \k^{\prime} = (k_x, k_y - \epsilon k_x)$.
Therefore the shear elastic energy density $\delta E=\frac{1}{2} c_{66} \epsilon^2$ can also be expressed in terms of the change in the vortex interaction potential due to deformation (with $\R_i\neq 0$ since a vortex does not interact with itself \cite{Brandt1977}):
\begin{equation}
\begin{aligned}
\delta E&=\frac{1}{2}\bar{n}_v\sum_{\R_i \neq 0}\left[V(\R_i')-V(\R_i)\right] =\frac{1}{2}\bar{n}_v\int \frac{d^2 k}{(2 \pi)^2}\left[\tilde{V}(\k^{\prime})-\tilde{V}(\k)\right] \sum_{\R_i \neq 0} e^{i \k \cdot \R_i}\\
&=\frac{1}{2} \bar{n}_v \Bigg\{\bar{n}_v\sum_{\G \neq 0} \left[ \tilde{V}(\G^{\prime}) - \tilde{V}(\G) \right] -\int \frac{d^2 k}{(2\pi)^2} \left[ \tilde{V}(\k^{\prime}) - \tilde{V}(\k) \right]\Bigg\}\\
\end{aligned}
\end{equation}
Here, we have performed a Fourier transform by using the identity $\sum_{\R_i \neq 0} e^{i \k \cdot \R_i}=\left[(2 \pi)^2 \bar{n}_v \sum_{\G} \delta^{(2)}(\k-\G)\right]-1$. 
Then, $\tilde{V}\left(\k^{\prime}\right)$ is expanded to second order in $\Delta \k=\k^{\prime}-\k$, where the argument $(\k)$ suppressed in $\tilde{V}^{\prime}, \tilde{V}^{\prime\prime}$ for clarity,
\begin{equation}
\begin{aligned}
& \tilde{V}\left(\k^{\prime}\right) \approx \tilde{V}(\k)+\left(\tilde{V}^{\prime}_x \Delta k_x+\tilde{V}^{\prime}_y \Delta k_y\right)+\frac{1}{2}\left[\tilde{V}^{\prime\prime}_{xx} \left(\Delta k_x\right)^2+2 \tilde{V}^{\prime\prime}_{xy} \Delta k_x \Delta k_y+\tilde{V}^{\prime\prime}_{yy}\left(\Delta k_y\right)^2\right],\\
& \tilde{V}^{\prime}_x=\tilde{V}^{\prime} \frac{k_x}{k}, \quad \tilde{V}^{\prime}_y=\tilde{V}^{\prime} \frac{k_y}{k},\quad \tilde{V}^{\prime\prime}_{xx}=\tilde{V}^{\prime \prime} \frac{k_x^2}{k^2}+\tilde{V}^{\prime}\frac{k_y^2}{k^3},\quad  \tilde{V}^{\prime \prime}_{yy}=\tilde{V}^{\prime \prime} \frac{k_y^2}{k^2}+\tilde{V}^{\prime} \frac{k_x^2}{k^3} ,\quad \tilde{V}^{\prime \prime}_{xy}=\tilde{V}^{\prime \prime} \frac{k_x k_y}{k^2}-\tilde{V}^{\prime}\frac{k_x k_y}{k^3},
\end{aligned}
\end{equation}
we find the shear moduli from the second-order expansion of the elastic energy,
\begin{equation}
\begin{aligned}
&\tilde{V}(\k^{\prime}) - \tilde{V}(\k) = -\epsilon \tilde{V}^{\prime}(k) \frac{k_x k_y}{k}  + \frac{\epsilon^2}{2}\left(\tilde{V}^{\prime \prime} \frac{k_x^2 k_y^2}{k^2}+\tilde{V}^{\prime}\frac{k_x^4}{k^3}\right) =\frac{\epsilon^2}{16}\left(k^2 \tilde{V}^{\prime \prime} +3k \tilde{V}^{\prime} \right) \\
& \implies c_{66} = \frac{2\cdot\delta E}{\epsilon^2} = \frac{\bar{n}_v^2}{16}\sum_{\G \neq 0}\left(G^2 \tilde{V}^{\prime \prime} +3G \tilde{V}^{\prime} \right) - \frac{\bar{n}_v}{16} \int \frac{d^2 k}{(2\pi)^2} \left( k^2 \tilde{V}^{\prime \prime} +3k \tilde{V}^{\prime} \right).
\end{aligned}
\end{equation}
Here, the first-order term vanishes after summing over $\k$ due to lattice symmetry. Also, we have used the angular average of $\expect{ k_x^2} =k^2\expect{\cos^2\phi}  ={k^2}/2, \quad\expect{ k_x^2 k_y^2}=k^4\expect{\cos^2\phi\sin^2\phi}  =k^4/8, \quad \expect{ k_x^4}=k^4\expect{\cos^4\phi}  =3 k^4/8$ to simplify the expression. As noted in Ref.~\onlinecite{Brandt1977}, the sum over reciprocal lattice vectors contributes a negligibly small negative amount, so the integral term dominates. 

We now evaluate the shear modulus for specific vortex lattice types. For a lattice of Pearl vortices, using the interaction potential in momentum space found in Eq.~\eqref{eq:Vintk2d} and its derivatives, with an ultraviolet cutoff $k_{\max} \sim \xi^{-1}$ set by the coherence length, we obtain the shear modulus,
\begin{equation}
\begin{aligned}
c_{66}&\approx \frac{\bar{n}_v}{16}\int_{0}^{k_{\max}} \frac{d^2 k}{{(2 \pi)^2}} \frac{2\Phi_0^2}{\mu_0} \frac{1+3 k \Lambda}{k(1+k \Lambda)^3} \\
& = \frac{\Phi_0 H}{16 \pi} \int_{0}^{\xi^{-1}} d k\,\frac{1+3 k \Lambda}{(1+k \Lambda)^3}   = \frac{\Phi_0 H}{16\pi\Lambda}\int_{0}^{\Lambda/\xi} dx\,\left(\frac{3}{(1+x)^2}-\frac{2}{(1+x)^3}\right) \\
& = \frac{\Phi_0 H}{16 \pi \Lambda} \frac{\Lambda(\xi+2 \Lambda)}{(\xi+\Lambda)^2} \approx \frac{\Phi_0 (\mu_0 H)}{8 \pi \mu_0 \Lambda}   \quad \text{for}~\Lambda/\xi \gg 1\quad.
\end{aligned}
\end{equation}
For a vortex lattce of Abrikosov vortices, applying the interaction potential per unit length founded in Eq.~\eqref{eq:Vintk3d} and its derivatives, the shear modulus is found to be
\begin{equation}
\begin{aligned}
c_{66}&\approx\frac{\bar{n}_v}{16}\int_{0}^{k_{\max}} \frac{d^2 k}{(2 \pi)^2} \frac{8 \Phi_0^2}{\mu_0} \frac{k^2 \lambda_L^2}{\left(1+k^2 \lambda_L^2\right)^3} \\
& =\frac{\bar{n}_v}{16} \frac{2 \Phi_0^2 \lambda_L^2}{\pi \mu_0} \int_{0}^{\xi^{-2}} \frac{k^2}{\left(1+k^2 \lambda_L^2\right)^3} d\left(k^2\right) = \frac{\bar{n}_v}{16} \frac{2 \Phi_0^2 \lambda_L^2}{\pi \mu_0} \frac{1}{\lambda_L^4} \int_{1}^{1+\kappa^2} \frac{u-1}{u^3} d u \\
& =\frac{\Phi_0 H}{8 \pi \lambda_L^2}\frac{\kappa^4}{2\left(1+\kappa^2\right)^2} \approx \frac{\Phi_0 (\mu_0 H)}{16 \pi \mu_0 \lambda_L^2}  \quad \text{for}~\kappa = \lambda_L/\xi \gg 1~ \text{(extreme type-II limit)}.
\end{aligned}
\end{equation}
Note that the shear moduli for the two vortex types are related by $c_{66}^{\rm Pearl}/d \approx c_{66}^{\rm Abrikosov}$.
These expressions correspond to the elastic moduli presented in \cref{eq:elastic_moduli_2D,eq:elastic_moduli_3D} of the main text.

Finally, we carefully note again that our calculations above are performed in the London limit. 
For magnetic fields near the upper critical field ($H\sim H_{c2}$), a linearized Ginzburg-Landau approach can be used to obtain modified nonlocal elastic moduli \cite{Labusch1969, Brandt1977, Fetter1966}.

\subsection{Vortex phonon correlation function}
In this section, we derive the longitudinal vortex phonon correlation function $C_\L(\k,\Omega)$ specified in Eq.~\eqref{eq:uuLongCorrVL}.
We start with the equation of motion in momentum space (Eq.~\eqref{eq:MinvVL} in the main text): 
\begin{align}
[M^{-1}_{\k,\Omega}]_{\mu \nu} u^{\nu}_{\k,\Omega} = \zeta^\mu_{\k,\Omega},  \text{where }  M^{-1}_{\k,\Omega} =   \, ( c_{11}(k)k^2/n_v -\mu_v \Omega^2 - i \eta \Omega )P_\L(\k)  
+ (c_{66}(k) k^2/n_v - \mu_v \Omega^2 - i \eta \Omega) P_{\T}(\k) + i \Omega n_s h \varepsilon  .
\label{eq:MinvVLapp}
\end{align}
Next, we note that the inverse of a matrix $M^{-1}_{\k,\Omega} = a_{\k,\Omega} \, P_\L(\k) + b_{\k,\Omega} \, P_\T(\k) + c_{\k,\Omega}\, \varepsilon$ takes the form
\begin{equation}
M_{\k,\Omega} = \frac{1}{(a_{\k,\Omega}\, b_{\k,\Omega} + c^2_{\k,\Omega})}  \left[  b_{\k,\Omega} \, P_\L(\k) + a_{\k,\Omega}\, P_\T(\k) - c_{\k,\Omega} \, \varepsilon \right].
\end{equation}
This is easiest to see by writing the matrix $M^{-1}_{\k,\Omega}$ in the basis spanned by $\hat{\k}$ and $\hat{z} \times \hat{\k}$ (note that the fully antisymmetric $\varepsilon$ tensor is rotationally invariant):
\begin{equation}
M^{-1} = \begin{pmatrix}
    a & c \\ -c & b
\end{pmatrix} \implies M = \frac{1}{ab + c^2} \begin{pmatrix}
    b & -c \\ c & a
\end{pmatrix}  ,
\end{equation}
where we have suppressed the $(\k,\Omega)$ indices for clarity.
Therefore, the object of interest to calculate the longitudinal correlation function for vortex lattice phonons is given by
\begin{equation}
\Tr[P_\L M M^\dagger] = \frac{1}{|ab + c^2|^2} \Tr[P_\L( b P_\L + a P_\T - c \varepsilon)(b^* P_\L + a^* P_\T - c^* \varepsilon)] = \frac{|a|^2 + |c|^2}{|ab + c^2|^2},
\label{eq:CLVLapp}
\end{equation}
where we have used that $P_\L^2 = P_\L$, $P_\T^2 = P_\T$, $\varepsilon^2 = - \mathbb{I}_2$, $P_\L P_\T = 0$ and $\Tr[P_\L \varepsilon] = 0 = \Tr[P_\T \varepsilon]$. 
Finally, plugging the explicit forms of $a_{\k,\Omega} =  c_{11}(k)k^2/n_v -\mu_v \Omega^2 - i \eta \Omega$, $b_{\k,\Omega} = c_{66}(k) k^2/n_v - \mu_v \Omega^2 - i \eta \Omega$ and $c_{\k,\Omega}= i \Omega n_s h$ from Eq.~\eqref{eq:MinvVLapp} to Eq.~\eqref{eq:CLVLapp}, we find
\begin{equation}
\Tr[P_\L(\k) M_{\k,\Omega} M^\dagger_{\k,\Omega}] = \frac{|c_{11}(k)k^2/n_v -\mu_v \Omega^2 - i \eta \Omega|^2 + (\Omega n_s h)^2}{|(c_{11}(k)k^2/n_v -\mu_v \Omega^2 - i \eta \Omega)(c_{66}(k) k^2/n_v - \mu_v \Omega^2 - i \eta \Omega) - (\Omega n_s h)^2|^2},
\end{equation}
which leads to Eq.~\eqref{eq:uuLongCorrVL} in the main text.

\section{Vortex liquid}
\label{app:liquid}

\subsection{Derivation of diffusive vortex dynamics}
In this appendix, we provide details on the magnetic noise calculations for the vortex liquid phase that was omitted in the main text.
To this end, we first derive the diffusion constant of the vortex liquid.
For this purpose, we shall derive the Fokker-Planck equations to the corresponding Langevin equations of \eqref{eq:eom_Vliquid}.
The Fokker-Planck equation models the evolution of the probability distribution from the Langevin equations, and allows us to read off the diffusion constant directly.

Recall that the Langevin equations in the vortex liquid are given as,
\begin{equation}
    M
    \begin{pmatrix}
        \partial_t u^x \\
        \partial_t u^y
    \end{pmatrix}
    =
    \begin{pmatrix}
        \bar\zeta^x \\
        \bar\zeta^y
    \end{pmatrix}
    ,\textrm{ where }
    M = \begin{pmatrix}
        \bar{\eta} & -n_s h \\
        n_s h & \bar{\eta}
    \end{pmatrix}
    \label{appeq:Langevin_liquid}
\end{equation}
and $\bar\zeta$ are Gaussian white noise with zero mean satisfying the FDT,  
\begin{equation}
    \langle{\bar \zeta^\alpha (t) \bar \zeta^\beta(t') \rangle} = 2\bar\eta k_B T \, \delta^{\alpha\beta}\,  \delta(t-t') \,.
    \label{appeq:LL_noise}
\end{equation}
Let us apply $M^{-1}$ on both sides of Eq.~\eqref{appeq:Langevin_liquid}.
We find
\begin{equation}
    \partial_t u^\alpha = \tilde \zeta^\alpha \,, \textrm{ where } \ \tilde\zeta^\alpha = \frac{\bar\eta ~ \bar \zeta^\alpha + n_s h~ \epsilon_{\alpha\beta} \bar \zeta^\beta}{\bar \eta^2 + n_s^2 h^2}  .
    \label{appeq:Langevin_liquid2}
\end{equation}
It follows from Eq.~\eqref{appeq:LL_noise} that $\langle{\tilde \zeta^\alpha(t) \rangle} = 0$, and $\langle \tilde \zeta^\alpha(t) \tilde \zeta^\beta(t') \rangle = 2\frac{\bar\eta k_B T}{\bar\eta^2 + (n_s h)^2} \delta^{\alpha\beta} \delta(t-t')$.
This result has two important consequences.
First, the noise correlations remain diagonal despite the presence of the Magnus force, implying that the vortex dynamics can still be mapped onto effective Brownian motion. 
Second, the absence of cross-correlations indicates that the $x$ and $y$ components of the displacement are dynamically decoupled and may be treated independently.

Having simplified the Langevin equations, we now derive the corresponding Fokker-Planck equation to model the evolution of the probability distribution due to the stochastic dynamics of Eq.~\eqref{appeq:Langevin_liquid}~\cite{Kamenev}. 
To this end, let us consider a smooth function $h(u)$, and ask how the expectation value of $h(u(t))$ evolves under a set of stochastic differential equations (such as Eqs.~\eqref{appeq:Langevin_liquid} and \eqref{appeq:Langevin_liquid2}).
\begin{equation}
\begin{split}
    \frac{\partial} {\partial t} \langle h(u(t)) \rangle &= \int du \, h(u) \partial_t P(u,t) \\
    &= \lim_{\Delta t\rightarrow 0} \frac{1}{\Delta t} \left[ \int du \int dv\, h(u) P(u,t+\Delta t | v, t) P(v,t) - \int dv\, h(v) P(v,t) \right] \\
    &= \lim_{\Delta t\rightarrow 0} \frac{1}{\Delta t} \left[ \int du \int dv\, \sum \frac{(u-v)^n}{n!} h^{(n)}(v) P(u,t+\Delta t | v, t) P(v,t) - \int dv\, h(v) P(v,t) \right] \,,
\end{split}
\end{equation}
where in the second line, $P(u,t+\Delta t | v, t)$ denotes the conditional probability that a particle at position $u$ at time $t$ moves to position $v$ at time $t + \Delta t$.
In the third line we have performed a Taylor expansion of $h(u)$ about $u = v$.
Since $\int du\, P(u,t+\Delta t | v, t) = 1$, we find that the zeroth order of the expansion cancels, and we have,
\begin{equation}
\begin{split}
    \frac{\partial} {\partial t} \langle h(u(t)) \rangle
    &= \lim_{\Delta t\rightarrow 0} \frac{1}{\Delta t} \left[ \int du \int dv\, h^{(1)}(v) \langle (u-v) \rangle + \frac{h^{(2)}(v)}{2} \langle (u-v)^2 \rangle \right] \,,
\end{split}
\label{appeq:KMexpansion}
\end{equation}
where $\langle (u-v)^n \rangle = \int dv\, P(v,t) (u-v)^n$.
Here, we have omitted the the higher than quadratic order in the Taylor expansion, which vanish in the limit of $\Delta t \rightarrow 0$, as we will show later.
Let us denote with 
$\lim_{\Delta t \rightarrow 0} \frac{\langle (u-v) \rangle}{\Delta t} = D^{(1)}(v)$ \,,  $\lim_{\Delta t \rightarrow 0} \frac{\langle (u-v)^2 \rangle}{\Delta t} = D^{(2)}(v)$.
Integrating the right hand side of Eq.~\eqref{appeq:KMexpansion} by parts,
We finally arrive at,
\begin{equation}
\begin{split}
    \int du\, h(u) \partial_t P(u,t) &= \int du\, h(u) \left\{-\partial_u D^{(1)}(u) P(u,t) + \frac12 \partial_u^2 D^{(2)}(u) P(u,t) \right\}\\
    &\Rightarrow \partial_t P(u,t) = - \partial_u D^{(1)} P(u,t) + \frac12 \partial_u^2 D^{(2)}(u)  .
\end{split}
\end{equation}
Consequently, to model the evolution of the probability distribution due to the stochastic differential equations of Eqs.~\eqref{appeq:Langevin_liquid}, we simply need to find $D^{(1,2)}$.

From Eq.~\eqref{appeq:Langevin_liquid2}, we immediately find that $D^{(1)}$ is zero.
To see this, let us find the change in $u$ from time $t$ to time $t+\Delta t$.
Since $\langle \partial_t u \rangle = \langle \zeta_i \rangle = 0$, it follows that $D^{(1)} = \lim_{\Delta t \rightarrow 0} \frac{\langle \Delta u \rangle}{\Delta t} = 0$.
Now let us find $D^{(2)}$.
\begin{equation}
    D^{(2)} = \lim_{\Delta t \rightarrow 0} \frac{\langle \Delta u^2 \rangle}{\Delta t} = \lim_{\Delta t \rightarrow 0} \frac{\langle (\int_t^{t+\Delta t} dt' \,\tilde \zeta_{t'})^2  \rangle}{\Delta t} = \lim_{\Delta t \rightarrow 0} \frac{\int_t^{t+\Delta t} dt'\, dt'' \,\frac{2\bar\eta k_B T}{\bar\eta^2 + (n_s h)^2} \delta(t'-t'')}{\Delta t} = \frac{2\bar\eta k_B T}{\bar\eta^2 + (n_s h)^2}  .
\end{equation}

Putting this altogether, we consequently find that the Fokker-Planck equations are given by,
\begin{equation}
    \partial_t P(u,t) = -\frac{2\bar\eta k_B T}{\bar\eta^2 + (n_s h)^2} \partial_u^2 P(u,t) \,.
\end{equation}

Consequently, we find that despite the presence of Magnus force, the overall vortex dynamics is still diffusive.
Accordingly, the vortex density-density correlation function takes the following form,

\begin{equation}
C_{n_v n_v}(\k,\Omega) = \langle n_v(\k,\Omega) n_v(-\k,-\Omega) \rangle = \frac{2 k_B T \chi_v D_v k^2}{\Omega^2 + (D_v k^2)^2},
\end{equation}
where $\chi_v = (\partial n_v / \partial \mu)_{T}$ denotes the vortex compressibility. The above equation is the Eq.~\eqref{eq:n_v_correlator} in the main text.

\subsection{Details on different noise regimes of the vortex liquid}

We now discuss the various regimes of the magnetic noise in the vortex liquid phase.
We start with the expression for noise for a liquid of Pearl vortices in 2D:
\begin{equation}
\cN_{zz}(\Omega) = \frac{ k_B T \sigma_v \Phi_0^2}{\pi D_v^2} I_2(z_0, \Lambda, \ell_\Omega). \text{ where } I_2(z_0, \Lambda, \ell_\Omega) = \int_0^\infty d\tilde{k} \, \frac{\tilde{k}^3  \, e^{- 2 \tilde{k} z_0/\Lambda}}{(1 + \tilde{k})^2 \left[ \left( \frac{\Lambda}{\ell_\Omega} \right)^4 + \tilde{k}^4 \right]}.
\label{eq:NPearl_VL_app}
\end{equation}
We note that the integral $I_2(z_0, \Lambda, \ell_\Omega)$ in Eq.~\eqref{eq:NPearl_VL_app} is determined by the hierarchy of three lengthscales --- $z_0$, $\Lambda$ and $\ell_\Omega = \sqrt{D_v/\Omega}$. 
Hence, we evaluate the integral in these different limits, leading to the results presented in Table~\ref{tab:VLscalings} in the main text. 

First, if $\Lambda, z_0 \ll \ell_\Omega$, then the integral can be approximated by:
\begin{equation}
I_2(z_0, \Lambda, \ell_\Omega) \approx \int_{\Lambda/\ell_\Omega}^\infty d\tilde{k} \, \frac{ \, e^{- 2 \tilde{k} z_0/\Lambda}}{\tilde{k}(1 + \tilde{k})^2 } \approx \begin{cases}
\displaystyle  \int_{\Lambda/\ell_\Omega}^{\Lambda/z_0} \frac{d\tilde{k}}{\tilde{k}} = \ln\left( \frac{\ell_\Omega}{z_0} \right), \text{ if } z_0 \gg \Lambda \\ \\
\displaystyle  \int_{\Lambda/\ell_\Omega}^{1} \frac{d\tilde{k}}{\tilde{k}} + \int_{1}^{\Lambda/z_0} \frac{d\tilde{k}}{\tilde{k}^3} = \ln\left( \frac{\ell_\Omega}{\Lambda} \right) + \frac{1}{2}\left[ 1 - \left( \frac{z_0}{\Lambda} \right)^2 \right] \approx  \ln\left( \frac{\ell_\Omega}{\Lambda} \right), \text{ if } z_0 \ll \Lambda
\end{cases}.
\end{equation}

Next, if $\Lambda, \ell_\Omega \ll z_0$, then the integral gets cut off at $\tilde{k} \sim \Lambda/z_0  \ll 1$, so we may approximate:
\begin{equation}
I_2(z_0, \Lambda, \ell_\Omega) \approx \left( \frac{\ell_\Omega}{\Lambda} \right)^4  \int_0^\infty d\tilde{k} \, \frac{\tilde{k}^3  \, e^{- 2 \tilde{k} z_0/\Lambda}}{\left[ 1+ \left(\tilde{k} \ell_\Omega/\Lambda\right)^4 \right]}  \xrightarrow[]{\tilde{k} = \Lambda k} \ell_\Omega^4 \int_0^\infty \, dk \, \frac{k^3 e^{-2 k z_0}}{1 + (k \ell_\Omega)^4} \approx \ell_\Omega^4 \int_0^\infty \, dk \, k^3 e^{-2 k z_0}=  \frac{3}{8}\left( \frac{\ell_\Omega}{z_0} \right)^{4}.
\end{equation}

Finally, if $\ell_\Omega, z_0 \ll \Lambda$, then the answer depends on the hierarchy of $\ell_\Omega$ and $z_0$. 
If $\ell_\Omega \ll z_0$, then 
\begin{equation}
I_2(z_0, \Lambda, \ell_\Omega) = \left( \frac{\ell_\Omega}{\Lambda} \right)^4  \int_0^\infty d\tilde{k} \, \frac{\tilde{k}^3  \, e^{- 2 \tilde{k} z_0/\Lambda}}{(1 + \tilde{k})^2 \left[ 1+ \left(\tilde{k} \ell_\Omega/\Lambda\right)^4 \right]}  \approx  \left( \frac{\ell_\Omega}{\Lambda} \right)^4  \int_0^\infty d\tilde{k} \, \frac{\tilde{k}^3  \, e^{- 2 \tilde{k} z_0/\Lambda}}{(1 + \tilde{k})^2} \approx  \left( \frac{\ell_\Omega}{\Lambda} \right)^4  \int_0^\infty d\tilde{k} \, \tilde{k} e^{- 2 \tilde{k} z_0/\Lambda} = \frac{\ell_\Omega^4}{4 z_0^2 \Lambda^2},
\end{equation}
where we have used that the exponential cuts off the integral at $\tilde{k} \sim \Lambda/z_0 \gg 1$ (but still $\ll \Lambda/\ell_\Omega$ where the $1+ (\tilde{k} \ell_\Omega/\Lambda)^4$ term in the denominator becomes important), while the $\tilde{k}^3$ prefactor suppresses it close to zero, so the majority of the contribution comes from $\tilde{k} \gtrsim 1$ such that we may approximate $(1 + \tilde{k})^2 \approx \tilde{k}^2$.

If $z_0 \ll \ell_\Omega$, we can neglect the exponential factor and again approximate $(1 + \tilde{k})^2 \approx \tilde{k}^2$, by noting that the dominant contribution comes from $1 \lesssim \tilde{k} \lesssim \Lambda/\ell_\Omega$.
Accordingly, 
\begin{equation}
I_2(z_0, \Lambda, \ell_\Omega) \approx  \int_0^\infty d\tilde{k} \, \frac{\tilde{k} }{ \left[ \left(\Lambda/\ell_\Omega\right)^4+ \tilde{k}^4 \right]} = \frac{1}{2 (\Lambda/\ell_\Omega)^2 } \arctan\left( \frac{\tilde{k}^2 \ell_\Omega^2}{\Lambda^2} \right)\bigg|_{0}^{\infty} = \frac{\pi}{4} \left( \frac{\ell_\Omega}{\Lambda} \right)^2.
\end{equation}

Next, we consider rod-like Abrikosov vortices in 3D, where the noise in the vortex liquid phase is given by
\begin{equation}
\cN_{zz}(\Omega) = \frac{ k_B T \sigma_v \Phi_0^2}{\pi D_v^2} I_3(z_0, \Lambda, \ell_\Omega). \text{ where } I_3(z_0, \Lambda, \ell_\Omega) = \int_0^\infty d\tilde{k} \, \frac{\tilde{k}^3  \, e^{- 2 \tilde{k} z_0/\Lambda}}{(1 + \tilde{k}^2)^2 \left[ \left( \frac{\Lambda}{\ell_\Omega} \right)^4 + \tilde{k}^4 \right]}.
\label{eq:NPearl_VL_app}
\end{equation}

We once again proceed case by case. 
If $\lambda_L, z_0 \ll \ell_\Omega$, then the integral can be approximated by:
\begin{equation}
I_3(z_0, \lambda_L, \ell_\Omega) \approx \int_{\lambda_L/\ell_\Omega}^\infty d\tilde{k} \, \frac{ \, e^{- 2 \tilde{k} z_0/\lambda_L}}{\tilde{k}(1 + \tilde{k}^2)^2 } \approx \begin{cases}
\displaystyle  \int_{\lambda_L/\ell_\Omega}^{\lambda_L/z_0} \frac{d\tilde{k}}{\tilde{k}} = \ln\left( \frac{\ell_\Omega}{z_0} \right), \text{ if } z_0 \gg \lambda_L \\ \\
\displaystyle  \int_{\lambda_L/\ell_\Omega}^{1} \frac{d\tilde{k}}{\tilde{k}} + \int_{1}^{\lambda_L/z_0} \frac{d\tilde{k}}{\tilde{k}^5} = \ln\left( \frac{\ell_\Omega}{\lambda_L} \right) + \frac{1}{4}\left[ 1 - \left( \frac{z_0}{\lambda_L} \right)^4 \right] \approx  \ln\left( \frac{\ell_\Omega}{\lambda_L} \right), \text{ if } z_0 \ll \lambda_L
\end{cases}.
\end{equation}

In the regime where $z_0 \gg \lambda_L$, the vortices are essentially screened point objects. 
Formally, the integrand in $I_3(z_0, \Lambda, \ell_\Omega)$ is cut of at $\tilde{k} \sim \lambda_L/z_0 \ll 1$, and we can approximate $1 + \tilde{k}^2 \approx 1$ in the denominator. 
In these regimes, therefore, the Pearl (point) or Abrikosov (line) nature of the vortices does not play any role. 
Thus for $\lambda_L, \ell_\Omega \ll z_0$, we have
\begin{equation}
I_3(z_0, \lambda_L, \ell_\Omega) \approx \ell_\Omega^4 \int_0^\infty \, dk \, k^3 e^{-2 k z_0}=  \frac{3}{8}\left( \frac{\ell_\Omega}{z_0} \right)^{4}.
\end{equation}

Finally, we consider the cases where $\lambda_L$ is the largest lengthscale. 
For $\ell_\Omega \ll z_0 \ll \lambda_L$, we find that
\begin{equation}
I_3(z_0, \lambda_L, \ell_\Omega) \approx \left( \frac{\ell_\Omega}{\lambda_L}\right)^4 \int_0^\infty d\tilde{k} \, \frac{\tilde{k}^3 e^{- 2 \tilde{k} z_0/\lambda_L}}{(1 + \tilde{k}^2)^2} \approx \left( \frac{\ell_\Omega}{\lambda_L}\right)^4   \int_0^{\lambda_L/z_0} d\tilde{k} \, \frac{\tilde{k}^3}{(1 + \tilde{k}^2)^2}  \approx \frac{1}{2}  \left( \frac{\ell_\Omega}{\lambda_L}\right)^4  \ln\left( \frac{\lambda_L}{z_0}\right).
\end{equation}
For $z_0 \ll \ell_\Omega \ll \lambda_L$, the exponential term becomes significantly different from one when $\tilde{k} \sim \lambda_L/z_0 \gg 1$, at which point the integral is already suppressed by the power law factors in the denominator. 
Hence, we may approximate
\begin{equation}
 I_3(z_0, \lambda_L, \ell_\Omega) \approx \int_0^\infty d\tilde{k} \, \frac{\tilde{k}^3  }{(1 + \tilde{k}^2)^2 \left[ \left( \frac{\lambda_L}{\ell_\Omega} \right)^4 + \tilde{k}^4 \right]} = \frac{ -1 + \left(\frac{\lambda_L}{\ell_\Omega} \right)^2
\left[ \pi - \left(\frac{\lambda_L}{\ell_\Omega}\right)^2 \right] + \left[\left(\frac{\lambda_L}{\ell_\Omega}\right)^4 - 1\right]\ln(\frac{\lambda_L}{\ell_\Omega})}{\left[1 + \left(\frac{\lambda_L}{\ell_\Omega}\right)^4\right]^2} \approx  \left(\frac{\ell_\Omega}{\lambda_L} \right)^4 \ln\left( \frac{\lambda_L}{\ell_\Omega} \right).
\end{equation}
Thus, we have derived all the scaling regimes presented in Table~\ref{tab:VLscalings} in the main text.

\end{widetext}
\end{document}